UNIVERSITY OF CALIFORNIA

Los Angeles

# Information Security and Quantum Mechanics: Security of Quantum Protocols

A dissertation submitted in partial satisfaction

of the requirements for the degree

Doctor of Philosophy in Physics

by

**Patrick Oscar Boykin**

2002

The dissertation of Patrick Oscar Boykin is approved.

---

Hong-Wen Jiang

---

David Cline, Committee Co-chair

---

Vwani P. Roychowdhury, Committee Co-chair

University of California, Los Angeles

2002

*to Julie,*

*with much love and many thanks*

TABLE OF CONTENTS

# LIST OF FIGURES

# ACKNOWLEDGMENTS

I am extremely grateful to Vwani Roychowdhury, who encouraged me to think about about a wider array of subjects, and introduced me to the study of information. Without the academic freedom which he afforded me, this work would not have been possible. I would also like to particularly thank Tal Mor, who introduced me to the topic of this work, quantum security protocols. Tal's eagerness to collaborate on research has been particularly helpful. Thanks to Farrokh Vatan and Somshubro Bandyopadhyay for frequent, stimulating discussions, some of the results of which are chapters in this work. I thank Professors Steve Kivelson, Hong-Wen Jiang and particularly David Cline, for serving on my thesis committee. I am grateful to Professor Rick Wesel, who allowed me to frequently interrupt his work to answer questions about information theory. Thanks to Celina Liebmann and Letty Mar, who regularly inconvenienced themselves in order to help me. I very much thank my co-authors for allowing me to reprint versions of papers here. Chapter 3 is a version of E. Biham, M. Boyer, P. O. Boykin, T. Mor, and V. Roychowdhury, "A Proof of the Security of Quantum Key Distribution", to appear in Journal of Cryptology. Chapter 5 is a version of S. Bandyopadhyay, P. O. Boykin, Vwani Roychowdhury, and F. Vatan, "A new proof for the existence of mutually unbiased bases", quant-ph/0103162, to appear in Algorithmica. Special thanks to my family for their personal support during my research and the writing of this thesis: my mother, my father, and my brother; and to Jesse Bridgewater, who offered never-ending motivational encouragement and stimulating discussion.

# VITA

| | |
|---|---|
| 1974 | Born, Raleigh, North Carolina, USA. |
| 1996 | B.S. (Mathematics) and B.S. (Physics), Georgia Institute of Technology. |
| 1996–1998 | Teaching Assistant, Physics Department, UCLA. |
| 1998 | M.S. (Physics), UCLA, Los Angeles, California. |
| 1998-2002 | Research Assistant, Electrical Engineering Department, UCLA. |

## PUBLICATIONS AND PRESENTATIONS

S. Bandyopadhyay, P. O. Boykin, Vwani Roychowdhury, and F. Vatan, "A new proof for the existence of mutually unbiased bases", quant-ph/0103162, to appear in Algorithmica.

E. Biham, M. Boyer, P. O. Boykin, T. Mor, and V. Roychowdhury, "A Proof of the Security of Quantum Key Distribution", to appear in Journal of Cryptology quant-ph/9912053.

E. Biham, M. Boyer, P. O. Boykin, T. Mor, and V. Roychowdhury, "A Proof of the Security of Quantum Key Distribution", Proceedings of the 32'nd Ann. ACM Symposium on the Theory of Computing (STOC'00), ACM Press, 2000, pp. 715-724.

P. O. Boykin, T. Mor, M. Pulver, V. Roychowdhury, and F. Vatan, "A new universal and fault-tolerant quantum basis", Information Processing Letters, vol. 75, pp. 101–107, 2000.

P. O. Boykin, T. Mor, M. Pulver, V. Roychowdhury, and F. Vatan, "On universal and fault-tolerant quantum computing: a novel basis and a new constructive proof of universality for Shor's basis", Proc. 40th Annual IEEE Symposium on the Foundations of Computer Science, pp. 486–494, 1999.

P. O. Boykin, T. Mor, V. Roychowdhury, F. Vatan, and R. Vrijen, "Algorithmic Cooling of Ensemble Computers", Proceedings of the National Academy of Science, vol. 99, no. 6, pp. 3388–3393, 2002

ABSTRACT OF THE DISSERTATION

# Information Security and Quantum Mechanics: Security of Quantum Protocols

by

## Patrick Oscar Boykin

Doctor of Philosophy in Physics

University of California, Los Angeles, 2002

Professor Vwani P. Roychowdhury, Co-chair

Professor David Cline, Co-chair

The problem of security of quantum key protocols is examined. In addition to the distribution of classical keys, the problem of encrypting quantum data and the structure of the operators which perform quantum encryption is studied. It is found that unitary bases are central to both encryption of quantum information, as well as the generation of states used in generalized quantum key distribution (which are called mutually unbiased bases). A one-to-one correspondence between certain unitary bases and mutually unbiased bases is found. Finally, a new protocol for making anonymous classical broadcasts is given along with a security proof. An experimental procedure to implement this protocol is also given. In order to prove these new results, some new bounds for accessible information of quantum sources are obtained.

# CHAPTER 1

# Introduction

## 1.1 Physics and Information Theory

Physics is the study of matter and energy. This study of matter and energy is almost always done by considering the states of objects or the environment. The physical laws concern themselves with the dynamics of physical states. These states are distinguishable by measurements to varying degrees. In the early part of the last century, a new branch of physics called quantum mechanics was discovered. This theory predicts startlingly different dynamics for some systems than do previous theories. A celebrated result from this new theory is the Heisenberg uncertainty relation, which states that one cannot precisely measure both the momentum and position of an object. The Heisenberg uncertainty relation comes from the fact that, in this new theory of quantum mechanics, the state of a physical object may be thought to be an eigenvector of an operator representing a physical quantity. A result of quantum mechanics is that not all quantities have compatible eigenvectors; hence, when a physical object has a well-defined momentum state, it cannot also be in an eigenstate of position.

Information theory was introduced by Claude Shannon. There are two main problems with which information theory concerns itself: channel capacity and source coding. Both problems can be couched in language with which a physicist would be very comfortable. A channel is a map from input states onto output states. A channel may be thought of as an interaction: a system is in an initial state; it then undergoes an interaction, which puts it into a final state. The capacity is a measure of correlation of the initial state with the final state, or how much information the final state has about the initial state. The source coding problem is also related to physics. A source is something that outputs states with a given probability. One does not know which state will come out at each time step, but one does know the probability distribution. The source coding problem seeks to quantify how much information is required, on average, to describe an output of the source. Though both of these problem statements might seem familiar to physicists, it may still come as a surprise that the quantity that plays a central role in answering both of these questions is also familiar: entropy.

## 1.2 Entropy

### 1.2.1 Boltzmann

Entropy is familiar to physicists as a quantity of fundamental importance to the theory of statistical mechanics and thermodynamics. One of the most celebrated physical laws is the Second Law of Thermodynamics, which states that entropy of a system can never decrease. Boltzmann first gave a functional form of entropy as:

**Definition 1.2.1** *Boltzmann entropy:*

$$S = k_b \ln \Omega$$

Where $\Omega$ is the statistical weight, or the number of microstates consistent with the observed properties of the macrostate. Further, Boltzmann added two postulates: each state consistent with the observed properties of the system is equally likely, and in equilibrium $\Omega$ is maximized.

**Definition 1.2.2** *The inverse of temperature is the partial derivative of entropy with*

*respect to energy:*

$$\frac{1}{T} \equiv \frac{\partial S}{\partial E}$$

Instead of definition 1.2.1 we can use the following:

**Definition 1.2.3** *Generalized Boltzmann entropy of a system with probability of being in state $r$ is $p_r$:*

$$S = -k_b \sum_r p_r \ln p_r$$

Since $S$ can never decrease, but may increase, the maximum of $S$ which is consistent with observed properties is an equilibrium. So, the probability distribution on the state of the system is that which maximizes $S$ under any set of constraints.

As an example, we begin by considering a system in which a state $r$ has energy $E_r$. Normalization and energy constraints are: $\sum_r p_r = 1$ and $\sum_r p_r E_r = E$ respectively. Using usual techniques to maximize given constraints, we have:

$$J = -k_b \sum_r p_r \ln p_r + \lambda_0 \sum_r p_r + \lambda_1 \sum_r p_r E_r$$

Finding the maximum for each $p_r$, we differentiate and set equal to zero:

$$\frac{\partial J}{\partial p_r} = -k_b(\ln p_r + 1) + \lambda_0 + \lambda_1 E_r = 0$$

Which gives:

$$p_r = e^{\frac{\lambda_0 - 1}{k_b}} e^{\lambda_1 E_r / k_b} \tag{1.1}$$

If we define[1] $Z \equiv \sum_r e^{\lambda_1 E_r / k_b}$, then we see that $p_r = \frac{1}{Z} e^{\lambda_1 E_r / k_b}$. Note that $\frac{\partial Z}{\partial \lambda_1} = EZ/k_b$. Putting what we know into the equation for entropy, we find that:

$$S = k_b \ln Z - \lambda_1 E \tag{1.2}$$

---
[1] this definition of $Z$ is called the partition function

And applying definition 1.2.2, we can compute $\lambda_1$:

$$\begin{aligned}
\frac{\partial S}{\partial E} &= \frac{k_b}{Z}\frac{\partial Z}{\partial E} - \lambda_1 - E\frac{\partial \lambda_1}{\partial E} \\
&= \frac{k_b}{Z}(\frac{\partial Z}{\partial E} - \frac{EZ}{k_b}\frac{\partial \lambda_1}{\partial E}) - \lambda_1 \\
&= \frac{k_b}{Z}(\frac{\partial Z}{\partial E} - \frac{\partial Z}{\partial \lambda_1}\frac{\partial \lambda_1}{\partial E}) - \lambda_1 \\
&= -\lambda_1
\end{aligned}$$

Thus we find that $\lambda_1 = \frac{-1}{T}$. Pulling it together: $p_r = \frac{1}{Z}e^{-\frac{E_r}{k_b T}}$.

All of the above is easily generalized to the case of continuous state variables (an integral takes the place of the sum), or the case of more constraints (such as a volume constraint). These sorts of approaches constitute the discipline of classical equilibrium statistical mechanics.

It might be noted that the generalized definition of entropy given in definition 1.2.3 only has to do with probability theory. There are no physical quantities (other than the normalization constant), only probabilities. Physics only entered the picture to give a constraint on the probabilities ($\sum_r p_r E_r = E$) and the definition of temperature (equation 1.2.2). We shall see that entropy will play a large role in results in information theory.

### 1.2.2 Shannon

Claude Shannon gave a definition of entropy[2] as:

**Definition 1.2.4** *Shannon entropy:*

$$H_s(X) = -\sum_{x_i} p(X = x_i) \log_2 p(X = x_i)$$

---

[2] We will denote the Shannon Entropy as $H_s$ in order to avoid confusion with a Hamiltonian $H$.

Note that this is the same as definition 1.2.1 up to a normalization. While the entropy was originally given as a function which yields a consistent theory of statistical mechanics and thermodynamics, it may be derived axiomatically as the function with the following properties[CT91]:

- Normalization: $H_s(\frac{1}{2}, \frac{1}{2}) = 1$

- Continuity: $H_s(p, 1-p)$ is a continuous function of $p$

- Grouping: $H_s(p_0, p_1, \ldots, p_{n-1}) = H_s(p_0 + p_1, p_2, \ldots, p_{n-1})$
  $+ (p_0 + p_1) H_s(\frac{p_0}{p_0+p_1}, \frac{p_1}{p_0+p_1})$

Only the function given in definition 1.2.4 satisfies all these criteria. The first two axioms should be clear. The grouping property states that the amount of disorder should partition in a somewhat straightforward way: the disorder of the entire system should be equal to the disorder of the system when the first two states are considered as one, combined with the probability of being in the first two states times their disorder.

In addition to entropy, Shannon defined a measure of correlation called mutual information.

**Definition 1.2.5** *The mutual information between two random variables $X, Y$ is:*

$$I(X;Y) \equiv H_s(X) - H_s(X|Y) = H_s(Y) - H_s(Y|X)$$

where $H_s(X|Y) \equiv \sum_{x_i} p(X = x_i) H_s(Y|X = x_i)$. Armed with these definitions, Shannon was able to show that the number of bits required to describe a source which outputs state $x_i$ with probability $p(X = x_i)$ is $H_s(X)$. A channel is now a map which maps input states in $X$ onto output states in $Y$ with a transition matrix: $p(Y = y_j | X = x_i)$. The channel capacity, which is the number of bits that the output has in common with the input, is: $I(X;Y) = H_s(Y) - H_s(Y|X)$.

In the following section, we will see how these information theoretic concepts due to Shannon may be applied to discover physics of information.

### 1.2.3 Boltzmann Meets Shannon

Physicists have long understood entropy as a measure of disorder, but quantified it in terms of its connections to energy and temperature. Shannon proved that entropy is indeed a precise quantification of disorder, as the number of bits necessary to describe a system. Landauer observed that physics has something to say about erasing information[Lan61]. If a bit is erased, the information required to describe the system is decreased by exactly 1 bit. Using the Clausius inequality, Landauer derived the thermodynamic cost of erasing information. Since $S = k_b \ln 2 H_s$, and $\Delta H_s = -1$, we have:

$$\Delta S - \frac{Q}{T} \geq 0$$
$$-k_b \ln 2 - \frac{Q}{T} \geq 0$$
$$-Q \geq k_b T \ln 2$$

Hence the laws of thermodynamics tell us that erasing 1 bit of energy requires removing $k_b T \ln 2$ of heat from the system.

## 1.3 Quantum Mechanics

So far we have described the relationship between statistical mechanics and information theory saying nothing of quantum mechanics. By now we can see how applicable information theory is to physics questions; one might wonder how the picture changes when we allow states to be quantum states. Schrodinger gave us the formula for the evolution of quantum states:

$$i\hbar \frac{\partial \psi}{\partial t} = H\psi \tag{1.3}$$

where $H$ is the Hamiltonian of the system and $\psi$ is any vector in the Hilbert space over which the Hamiltonian acts. In the so-called Heisenberg picture, we consider not the evolution of the state but the evolution of an operator:

$$|\psi(t)\rangle = U(t)|\psi(0)\rangle$$

which gives rise to a new dynamical equation:

$$\frac{\partial U}{\partial t} = \frac{-i}{\hbar} HU$$
$$U(0) = I$$

Using this new equation, one sees that $\frac{\partial (U^\dagger U)}{\partial t} = 0$. Therefore $U(t)^\dagger U(t) = I$, which is the definition of unitarity. Hence all evolution in quantum mechanics is unitary.

### 1.3.1 Quantum Statistical Mechanics

In order to consider how statistical mechanics changes in the quantum picture, we must define what we mean by a distribution of quantum states. According to the axioms of quantum mechanics, measurement outcomes are random. However, for any vector in a Hilbert space, there is always a basis where the probability distribution is a delta function. A state which cannot be represented as a vector in a Hilbert space $|\psi\rangle$ is called a mixed state and is represented by a density matrix:

$$\rho = \sum_i p_i |\psi_i\rangle\langle\psi_i|$$

The entropy of such states is given by:

**Definition 1.3.1** *Von Neumann entropy:*

$$S(\rho) = -Tr(\rho \log_2 \rho)$$

One may rewrite the Schrodinger equation for density matrices:

$$i\hbar \frac{\partial \rho}{\partial t} = [H, \rho] \tag{1.4}$$

It is interesting to note that the Schrodinger equation conserves energy for constant Hamiltonians:

$$\begin{aligned}
E &= Tr(\rho H) \\
\frac{\partial E}{\partial t} &= Tr(\frac{\partial \rho}{\partial t} H) = Tr(\frac{-i}{\hbar}[H, \rho]H) \\
&= \frac{-i}{\hbar} Tr(H\rho H - \rho H^2) \\
&= \frac{-i}{\hbar}(Tr(H\rho H) - Tr(\rho H^2)) \\
&= \frac{-i}{\hbar}(Tr(\rho H^2) - Tr(\rho H^2)) = 0
\end{aligned}$$

If we consider a quantum state in equilibrium, then $\frac{\partial \rho}{\partial t} = 0$. Hence $[H, \rho] = 0$, which means that $\rho$ has the same eigenbasis as $H$. If the eigenbasis of $H$ is $\{|\phi_0\rangle, |\phi_1\rangle, \ldots, |\phi_{n-1}\rangle\}$, with $H|\phi_i\rangle = E_i|\phi_i\rangle$ where $E_i$ is the energy of the $i^{th}$ state, then we may write $\rho$ as:

$$\rho = \sum_i p_i |\phi_i\rangle\langle\phi_i|$$

Since $\rho$ is diagonal, the Von Neumann entropy reduces to the Shannon entropy:

$$S(\rho) = H_s(p_0, p_1, \ldots, p_{n-1})$$

The energy is $Tr(\rho H) = \sum_i p_i E_i$. Aside from using a Hilbert space to represent the states of the system, this is the same as the Boltzmann entropy we considered in subsection 1.2.1. Using the same techniques of maximization under a constraint, the density matrix is obtained:

$$\rho = \frac{e^{-\frac{H}{k_b T}}}{Tr(e^{-\frac{H}{k_b T}})}$$

### 1.3.2 Deriving the Second Law of Thermodynamics

After seeing some consequences of evolution via the Schrodinger equation, such as the fact that energy is explicitly conserved, we might wonder what other results we can prove. The first thing to notice is that unitary evolution of a system does not its change entropy:

$$\begin{aligned}
S(U\rho U^\dagger) &= -Tr(U\rho U^\dagger \log_2(U\rho U^\dagger)) \\
&= -Tr(U\rho U^\dagger U(\log_2 \rho)U^\dagger) \\
&= -Tr(U\rho(\log_2 \rho)U^\dagger) \\
&= -Tr(\rho \log_2 \rho) = S(\rho)
\end{aligned}$$

Hence a closed system evolving according to the Schrodinger equation will keep the entropy of the state as a constant of the motion. How then, can entropy *increase*? When a system interacts with the environment, the Schrodinger equation will not apply, since this is not a Hamiltonian interaction unless we consider the system and the environment together.

In probability theory, there is the notion of marginalizing a variable of a probability distribution. For instance, if one has a probability distribution $p(x,y)$ and one is only concerned with the probabilities over $x$, the marginalized probability distribution is easily obtained: $p(x) = \sum_y p(x,y)$. Marginalization may also be done for density matrices, and it is called "tracing out" a subsystem. If we consider a system which has a basis $|x,y\rangle = |x\rangle|y\rangle$, any density matrix may be written as:

$$\begin{aligned}
\rho &= \sum_{x,x',y,y'} \alpha_{x,y,x',y'} |x,y\rangle\langle x',y'| \\
&= \sum_{x,y} \alpha_{x,y,x',y'} |x\rangle\langle x'| \otimes |y\rangle\langle y'|
\end{aligned}$$

Where we have used $\otimes$ to represent the tensor product. We can marginalize over the

$|y\rangle$ subsystem by "tracing out" $y$:

$$\begin{aligned}
\rho_x &= Tr_y(\rho) = \sum_y \langle y| \left( \sum_{x,x',y'',y'} \alpha_{x,y'',x',y'} |x\rangle\langle x'| \otimes |y''\rangle\langle y'| \right) |y\rangle \\
&= \sum_y \sum_{x,x',y'',y'} \alpha_{x,y'',x',y'} |x\rangle\langle x'| \langle y||y\rangle\langle y'||y\rangle \\
&= \sum_y \sum_{x,x',y'',y'} \alpha_{x,y'',x',y'} |x\rangle\langle x'| \delta_{y,y''} \delta_{y,y'} \\
&= \sum_{x,x'} \sum_y \alpha_{x,y,x',y} |x\rangle\langle x'| \\
&= \sum_{x,x'} \beta_{x,x'} |x\rangle\langle x'|
\end{aligned}$$

Where $\beta_{x,x'} = \sum_y \alpha_{x,y,x',y}$.

We need one more definition notion to prove the second law of thermodynamics:

**Definition 1.3.2** *Relative Von Neumann entropy:*

$$S(\rho|\sigma) \equiv tr(\rho \log_2 \rho) - tr(\rho \log_2 \sigma)$$

**Theorem 1.3.1** *Relative Von Neumann entropy is always positive:*

$$S(\rho|\sigma) \geq 0$$

Proof. See [Pre]. ∎ Using this definition, we can see that

$$\begin{aligned}
S(\rho_{AB}|\rho_A \otimes \rho_B) &= -S(\rho_{AB}) - tr(\rho_{AB} \log_2(\rho_A \otimes \rho_B)) \\
&= -S(\rho_{AB}) - tr\left(\rho_{AB}(\log_2((\rho_A \otimes 1_B)(1_A \otimes \rho_B)))\right) \\
&= -S(\rho_{AB}) + S(\rho_A) + S(\rho_B)
\end{aligned}$$

Since $S(\rho_{AB}|\rho_A \otimes \rho_B) \geq 0$, then we see that:

$$S(\rho_A) + S(\rho_B) \geq S(\rho_{AB})$$

Putting this all together we can see that the second law of thermodynamics is a result of the functional form of entropy.

Suppose that we have an environment and a system. Initially these two are independent: $\rho_{SE} = \rho_S \otimes \rho_E$. These two systems undergo a Hamiltonian interaction to become $\rho'_{SE}$. As we saw in the beginning of this chapter, Hamiltonian interactions produce unitary evolution. We also know that unitary evolution does not change entropy. Putting this all together:

$$\begin{aligned} S(\rho'_S) + S(\rho'_E) &\geq S(\rho'_{SE}) \\ &= S(U\rho_{SE}U^\dagger) \\ &= S(\rho_{SE}) \\ &= S(\rho_S) + S(\rho_E) \end{aligned}$$

Defining $\Delta S_i \equiv S(\rho'_i) - S(\rho_i)$, the above implies:

$$\Delta S_S + \Delta S_E \geq 0$$

As such, the entropy of systems (when considered independently) grows in time.

### 1.3.3 Quantum Information Theory

The two main problems of classical information theory, namely channel capacity and source coding, have also been considered in the quantum case. One of the earliest results answers the question of how much classical information is carried by quantum systems. Since there are an uncountably infinite number of bases of any Hilbert space, and uncountably infinite elements in any Hilbert space, one might wonder if quantum states might be able to hold more information than classical states. The answer, unfortunately, is no[Hol73]. In fact, the amount of information transmitted is always less than the $\log$ of the dimension of the system, just as in classical information theory.

More precisely, what Holevo showed is that if a source $A$ outputs quantum state $\rho_i$ with probability $p_i$, then the maximum mutual information between $i$ and any measurement is bounded by $\chi$, which Holevo defined:

**Definition 1.3.3** *Holevo $\chi$:*

$$\chi = S(\rho) - \sum_i p_i S(\rho_i)$$

which closely mirrors the classical case (see definition 1.2.5). What if the information is quantum, and not classical?

One can ask the following question: for a given quantum source, how many quantum bits, on average, is required to describe the output? This problem has been solved in very much the same way as Shannon solved it[Sch95]. The answer is very much the same: the number of qubits required is the entropy of the source, $S(\rho)$. While the question of source coding has been solved for quantum systems, the problem of channel coding for general quantum channels has not been solved. To prove the capacity result, Shannon considered correlated inputs over several channel usages. In the quantum case, in addition to correlated quantum inputs, one must also consider entangled inputs. The solution to the capacity of a quantum channel is still an open problem.

## 1.4 Secret Communications

Much of Shannon's work was the result of studies done during World War II, in which secret communications played a large role. Shannon sought to formalize many aspects of communication mathematically. If $M$ is the random variable for a message, $K$ is the random variable for a key, and $C$ is the random variable for the output of the encryption process, or cipher-text, then we may define informationally secure cryptography in the

following way[Sha49]:

$$I(M;C) = H(C) - H(C|M) = 0 \ . \tag{1.5}$$

The above relationship implies $p(c|m) = p(c)$, i.e., that the cipher-text, $c$, is independent of the message, $m$. Since one must be able to recover the message from the cipher-text given the key, one must also satisfy $I(M;C|K) = H(M)$. Hence, the secrecy condition combined with the recoverability condition imply that $H(K) \geq H(M)$ and $H(C) \geq H(M)$ for informationally secure cryptography. Of course, we see that one cannot "reuse" keys and keep security; that would mean using the same amount of key entropy on a larger message, and we have already shown that the size of the key is lower-bounded by the size of the message.

An example of informationally secure cryptography is the one time pad[Ver26]. The message $m$ is compressed to its entropy, and then a full-entropy random string of length $H(M)$ is chosen and called $k$. Then, the cipher-text is $c = m \oplus k$. Given $c$, one knows nothing of $m$, but given $c$ and $k$, one has $m$ exactly.

So, we have a proof that to get perfectly secure communications, one must first share a secret key as long as the message. If sharing secret keys were easy, why not simply secretly share the message? Of course, sharing keys does have advantages: for instance one may share secret keys in advance and subsequently use them to send secure communications when sharing keys is not possible. However, these sorts of perfectly secure systems are sufficiently unwieldy to prevent any sort of common usage.

If there was a way to know whether an eavesdropper had seen the key or not, one could tell whether or not it might be safe to use it. Alas, classical information theory offers no tools which makes this possible. If we apply quantum mechanics, we will see that we can accomplish this task.

## 1.5 Secrecy with Quantum States

Bennett and Brassard proposed a method to distribute secret key bits using polarization states of photons[BB84]. Their idea was simple: Alice sends polarized photons to Bob, either up, down, left or right. Bob measures randomly in one of two polarization bases: up-down or left-right. After Alice sends the photons to Bob, she announces which basis for each photon, but not which values she sent. They keep the photons where Bob's measurement basis coincided with Alice's transmission basis. The Heisenberg uncertainly principle says that the photon is either in a eigenstate of up-down or left-right, but not both[3]. So if an eavesdropper attempts to listen, she will presumably cause some errors, since she will project the state into eigenstates of the "wrong" basis about half the time. By testing a few of the photons, Alice and Bob should be able to detect an eavesdropper. If there is no eavesdropper, they use their shared secret bits as a key as described before. This scheme has become known as the BB84 protocol.

As simple as it sounds, it was more than 10 years before full proofs of security for the BB84 protocol were found[May96, BBB99]. In chapter 2, we will develop powerful tools for deriving security results and show the fundamental information vs. disturbance bound which is at work in the BB84 protocol. In chapter 3, we will prove a security result in the presence of noise. As we cannot know if noise is the result of the environment or an eavesdropper, steps must be taken to insure security. In chapter 4, we will see how the picture changes if one wants to encrypt not classical information, but quantum information. In chapter 5, we will see that there is a relationship between operators used to encrypt quantum states and the states used in quantum key distribution schemes. In chapter 6, we will describe a new quantum protocol and prove that it is secure. This protocol will allow any number of users to efficiently and securely make anonymous classical announcements. In chapter 7, we will see a prescription for an experimental implementation of the quantum protocol for an anonymous channel.

---
[3] because the operators for these observables do not commute

# CHAPTER 2

# Tools for Quantum Information Security Results

In this chapter, we will get into some of the fundamental results in quantum information security. Most of these results may be viewed as a sort of uncertainty principle for information. We can sum up the basic property as follows: gaining information in one basis necessarily causes errors in a conjugate basis.

We will also derive some new distinguishability measures. These are bounds on the amount of classical information that can be obtained from any measurement of a source of quantum states. We will extend some previous work[FG99] on such bounds which turns out be useful for quantum security results. These results are powerful because they only depend on the source and not on any measurement done. Later, we will apply these bounds on distinguishability to relate the amount of information eavesdroppers can obtain to the disturbance they cause in the quantum state.

First, we will generalize a main result in [FG99] to work with the case where the outcomes are not equally likely; then we will consider the case where the number of outcomes is unlimited. The result allows us to derive very general information vs. disturbance results.

## 2.1 Bound on Mutual Information for 1-bit Sources

Suppose there is a classical source $S$ which sends one of two signals; zero or one. Also suppose that $p_{s=1} \leq p_{s=0}$. Following [FG99], we first come up with a linear bound on

$H(p)$:

**Lemma 2.1.1** *For any $p' \leq 1/2$, $H(p) \geq H(p') - \frac{H(p')}{p'}|p - p'|$*

**P**roof. Consider two regions, $p \leq p'$ and $p > p'$. $H(p)$ is concave, which means that $H(\alpha x + (1-\alpha)y) \geq \alpha H(x) + (1-\alpha)H(y)$. Applying this with $x = p'$, $\alpha = p/p'$ and $y = 0$, we obtain: $H(p) \geq \frac{H(p')}{p'}p$, which is exactly what we need for $p \leq p'$. In the region $p > p'$ we want to show that $H(p) \geq H(p')(2 - \frac{p}{p'})$. Again using the concavity, set $y = p'$, $x = 1$ and $\alpha = \frac{p-p'}{1-p'}$ We see then that

$$H(p) \geq \frac{1-p'-p+p'}{1-p'}H(p') = \frac{1-p}{1-p'}H(p')$$

Since $p' \leq 1/2$, this implies that $\frac{1}{1-p'} \leq 2 \leq \frac{1}{p'}$.

$$\begin{aligned} H(p) &\geq \frac{1-p}{1-p'}H(p') \\ &\geq H(p')(2 - \frac{p}{1-p'}) \\ &\geq H(p')(2 - \frac{p}{p'}) \end{aligned}$$

∎

**Lemma 2.1.2** *The mutual information between the random variable $E$ and the random bit $S$ (with $p(s = 0) \geq p(s = 1)$) is bounded:*

$$I(E; S) \leq H(S)p(s = 0) \sum_e |p(e|s = 1) - p(e|s = 0)|$$

**Proof.** Using lemma 2.1.1 as a bound on $H(S|E)$ with $p' = p(s = 1)$, we can obtain the bound on mutual information:

$$\begin{aligned}
I(E;S) &= H(S) - H(S|E) \\
&= H(S) - \sum_e p_e H(S|E=e) \\
&\leq H(S) - \sum_e p_e (H(p(s=1)) - \frac{H(S)}{p(s=1)}|p(s=1|e) - p(s=1)|) \\
&= H(S) \sum_e |p(e|s=1) - p(e)| \\
&= H(S) \sum_e |p(e|s=1) - (p(s=0)p(e|s=0) + p(s=1)p(e|s=1))| \\
&= H(S)p(s=0) \sum_e |p(e|s=1) - p(e|s=0)|
\end{aligned}$$

∎

**Lemma 2.1.3** *If a source $S$ outputs quantum states $\rho_0$ and $\rho_1$ with probabilities $p_0$ and $p_1$ with $p_0 \geq p_1$, then mutual information between this source and the output of any measuring device $E$ is bounded: $I(E;S) \leq H(S)p(s=0)Tr|\rho_0 - \rho_1|$*

**Proof.** The source sends two states, $\rho_0$ and $\rho_1$. Eve does some POVM[Per93] on them. The probability that Eve gets outcome $x$ for her measurement given an input $s$ is: $p(e|s) = Tr(E_e \rho_s)$. This gives:

$$I(E;S) \leq H(S)p(s=0) \sum_e |Tr(E_e(\rho_0 - \rho_1))|$$

Since $\rho_0 - \rho_1$ is Hermitian, we can diagonalize it as $\sum_i \lambda_i |\psi_i\rangle\langle\psi_i|$. Taking this and applying the facts that $E_e$ are positive semi-definite and $\sum_e E_e = I$, we get:

$$\begin{aligned}
I(E;S) &\leq H(S)p(s=0)\sum_e |Tr(E_e(\rho_0 - \rho_1))| \\
&= H(S)p(s=0)\sum_e |Tr(E_e(\sum_i \lambda_i |\psi_i\rangle\langle\psi_i|))| \\
&= H(S)p(s=0)\sum_e |\sum_i \lambda_i \langle\psi_i|E_e|\psi_i\rangle| \\
&\leq H(S)p(s=0)\sum_e \sum_i |\lambda_i|\langle\psi_i|E_e|\psi_i\rangle \\
&= H(S)p(s=0)\sum_i |\lambda_i|\langle\psi_i|\sum_e E_e|\psi_i\rangle \\
&= H(S)p(s=0)\sum_i |\lambda_i| \\
&= H(S)p(s=0)Tr|\rho_0 - \rho_1|
\end{aligned}$$

∎

**Corollary 2.1.1** *If a source $S$ outputs quantum states $\rho_0$ and $\rho_1$, then mutual information between this source and the output of any measuring device $E$ is bounded: $I(E;S) \leq H(S)Tr|\rho_0 - \rho_1|$*

**Proof.** Consider two cases, the first where $p_0 \geq p_1$ and the second where $p_1 > p_0$. If $p_0 \geq p_1$, then using lemma 2.1.3 we have that $I(E;S) \leq H(S)p(s=0)Tr|\rho_0 - \rho_1|$. Since $p(s=0) \leq 1$, we get the result. If $p_1 > p_0$ then relabel the $\rho_1$ as $\rho_0$ and vice versa. Hence in the original labeling, lemma 2.1.3 becomes

$$I(E;S) \leq H(S)p(s=1)Tr|\rho_1 - \rho_0|$$

, and since $p(s=0) \leq 1$ we get the result. ∎

## 2.2 Bound On Information For Any Source

In lemma 2.1.3, we derived a bound for the amount of mutual information about a 1 bit function that can be obtained by an arbitrary measurement based on the trace norm of the density matrices. In this section, we extend our results to a source of any number of outputs. As we will see later in the chapter, this allows us to derive the fundamental information vs. disturbance results that are at work in quantum security protocols. Additionally, this result gives an important insight into the robustness of the trace norm as a metric bound for information.

This metric will be applied later in order to make a point about the great robustness of quantum security protocols.

**Lemma 2.2.1** *For any random variable $X'$ with each probability $p_i' \leq 1/2$:*

$$H(X) \geq H(X') - \sum_i \log(\frac{1}{p_i'})|p_i - p_i'|$$

**P**roof. $H(X) = -\sum_i p_i \log p_i$, so if we define $f(p_i) \equiv -p_i \log p_i$, we see that $H(X) = \sum_i f(p_i)$. See that $f$ is concave and has a maximum at $1/2$ and is zero at $p_i = 0, 1$; thus, we can use the same type of lower bound as lemma 2.1.1:

$$f(p_i) \geq f(p_i') - \frac{f(p_i')}{p_i'}|p_i - p_i'|$$

Plugging this into the definition of entropy:

$$\begin{aligned} H(X) &= \sum_i f(p_i) \\ &\geq \sum_i (f(p_i') - \frac{f(p_i')}{p_i'}|p_i - p_i'|) \\ &= H(X') - \sum_i \log(\frac{1}{p_i'})|p_i - p_i'| \end{aligned}$$

∎

**Lemma 2.2.2** *For any source S that outputs s with probability $p_s$ such that $p_s \leq 1/2$, the mutual information is bounded:*

$$I(S;E) \leq \sum_s p_s \log(\frac{1}{p_s}) \sum_e |p(e|s) - p(e)|$$

**P**roof. Make use of lemma 2.2.1:

$$\begin{aligned} I(S;E) &= H(S) - H(S|E) \\ &= H(S) - \sum_e p_e H(S|E=e) \\ &\leq H(S) - \sum_e p_e \left( H(S) - \sum_s \log(\frac{1}{p_s}) |p(s|e) - p_s| \right) \\ &= \sum_e p_e \sum_s \log(\frac{1}{p_s}) |p(s|e) - p_s| \\ &= \sum_e \sum_s p_s \log(\frac{1}{p_s}) |\frac{p(e)p(s|e)}{p_s} - p(e)| \\ &= \sum_e \sum_s p_s \log(\frac{1}{p_s}) |p(e|s) - p(e)| \end{aligned}$$

**Lemma 2.2.3** *If a source S outputs quantum states $\rho_i$ with probabilities $p_i$ with $p_i \leq 1/2$, then mutual information between this source and the output of any measuring device E is bounded:*

$$I(S;E) \leq \sum_s p_s \log(\frac{1}{p_s}) Tr|\rho_s - \sum_s p_s \rho_s|$$

**P**roof. Define the notation $\rho = \sum_s p_s \rho_s$. Starting from lemma 2.2.2, we use the definition of a POVM to replace $p(e|s)$ with $Tr(E_e \rho_s)$:

$$\begin{aligned} I(S;E) &\leq \sum_e \sum_s p_s \log(\frac{1}{p_s}) |p(e|s) - p(e)| \\ &= \sum_e \sum_s p_s \log(\frac{1}{p_s}) |Tr(E_e \rho_s) - Tr(E_e \rho)| \\ &= \sum_e \sum_s p_s \log(\frac{1}{p_s}) |Tr(E_e(\rho_s - \rho))| \end{aligned}$$

Using the same facts about POVMs from the previous sections, one can show that

$$\sum_e |Tr(E_e(\rho_s - \rho))| \leq Tr|\rho_s - \rho|$$

Hence, we have:

$$I(S;E) \leq \sum_s p_s \log(\frac{1}{p_s}) Tr|\rho_s - \rho|$$

∎

**Corollary 2.2.1** *If a source $S$ outputs one of $n$ quantum states $\rho_i$ with probability $1/n$, then mutual information between this source and the output of any measuring device $E$ is bounded: $I(S;E) \leq \log n \sum_s \frac{1}{n} |\rho_s - \rho|$.*

**P**roof. For all $n \geq 2$, then $1/n \leq 1/2$, hence lemma 2.2.3 applies:

$$\begin{aligned} I(S;E) &\leq \sum_s p_s \log(\frac{1}{p_s}) Tr|\rho_s - \rho| \\ &= \log n \sum_s \frac{1}{n} Tr|\rho_s - \rho| \end{aligned}$$

∎

## 2.3 Bounding the Trace Norm

As we have seen in the previous section, the trace norm distance between quantum states is a powerful tool for bounding mutual information. Now we look at some bounds on trace norm distances.

**Lemma 2.3.1** *The trace norm distance between two pure states is:*

$$||\psi\rangle\langle\psi| - |\phi\rangle\langle\phi|| = 2\sqrt{1 - |\langle\psi|\phi\rangle|^2}$$

**Proof.** Define $\langle \psi | \phi \rangle = \alpha$. Defining a new orthonormal basis we can write:

$$|e_0\rangle \equiv |\psi\rangle$$
$$|e_1\rangle \equiv \frac{1}{\sqrt{1-|\alpha|^2}}(|\phi\rangle - \alpha|\psi\rangle)$$

Inverting these equations we have:

$$|\psi\rangle = |e_0\rangle$$
$$|\phi\rangle = \alpha|e_0\rangle + \sqrt{1-|\alpha|^2}|e_1\rangle$$

Using this new basis, we find that:

$$||\psi\rangle\langle\psi| - |\phi\rangle\langle\phi|| = |(1-|\alpha|^2)|e_0\rangle\langle e_0| - (1-|\alpha|^2)|e_1\rangle\langle e_1|$$
$$-\sqrt{1-|\alpha|^2}(\alpha^*|e_1\rangle\langle e_0| + \alpha|e_0\rangle\langle e_1|)|$$

This is just a $2 \times 2$ matrix and we can compute the trace norm by taking the absolute value of the eigenvalues, which are:

$$\lambda = \pm\sqrt{1-|\alpha|^2}$$

∎

**Lemma 2.3.2** *The trace norm distance between any state and any pure state is bounded:*

$$|\rho - |\psi\rangle\langle\psi|| \leq 2\sqrt{1 - \langle\psi|\rho|\psi\rangle}$$

**Proof.** Let $\rho = \sum_i p_i |\phi_i\rangle\langle\phi_i|$ and apply $\sum_i p_i x_i \leq \sqrt{\sum_i p_i x_i^2}$:

$$|\rho - |\psi\rangle\langle\psi|| = |\sum_i p_i|\phi_i\rangle\langle\phi_i| - |\psi\rangle\langle\psi||$$
$$\leq \sum_i p_i ||\phi_i\rangle\langle\phi_i| - |\psi\rangle\langle\psi||$$
$$= \sum_i p_i \sqrt{1 - |\langle\psi|\phi_i\rangle|^2}$$
$$\leq \sqrt{\sum_i p_i(1 - |\langle\psi|\phi_i\rangle|^2)}$$
$$= 2\sqrt{1 - \langle\psi|\rho|\psi\rangle}$$

**Definition 2.3.1** *Purification of $\rho$: any pure state $|\psi\rangle$ in $\mathcal{H}_1 \otimes \mathcal{H}_2$ such that $Tr_2(|\psi\rangle\langle\psi|) = \rho$*

**Lemma 2.3.3** *The trace norm distance is reduced by partial trace:*

$$|\rho' - \sigma'| \leq |\rho - \sigma|$$

*Where $\rho$ and $\sigma$ are density matrices over states in $\mathcal{H}_1 \otimes \mathcal{H}_2$ and the partial trace is over one of the subsystems: $\rho' = Tr_2(\rho)$ and $\sigma' = Tr_2(\sigma)$.*

**P**roof. See [Per93].

## 2.4 Security of Quantum Key Distribution

We now have the tools necessary in order to derive an information theoretic analogy to the Heisenberg uncertainty principle. This result is the basis for quantum security results that we will examine in chapters 3 and 6.

Quantum key distribution is related to the problems we considered in the previous section. Figure 2.1 gives a general attack that Eve might perform. From her perspective, she has access to a source (the system she has used to interact with the states sent by Alice) and she can make any measurement to get information about what was sent. The intuition about quantum mechanics is that measurements will disturb the system. We will make this a precise statement about information and disturbance.

### 2.4.1 Security of Quantum Keys

**Theorem 2.4.1** *If Alice sends $n$-qubit states to Bob, each with equal probability, the Information Eve can get about the state sent is bounded by the square root of the*

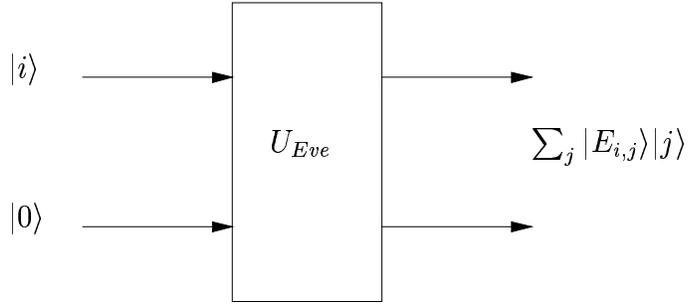

Figure 2.1: Most general attack by an eavesdropper

*probability that the Eve would have caused errors in the Fourier transformed basis:*

$$I(A;E) \leq 4n\sqrt{P_{\bar{e}}}$$

**P**roof. We will use lemmas 2.3.3 and 2.3.2 and corollary 2.2.1. Starting from corollary 2.2.1 we see that: $I(A;E) \leq n \sum_i \frac{1}{2^n} |\rho_i - \rho|$. Our approach will be to bound this by introducing a purification[1] for $\rho_i$ (the state that Eve holds when Alice sends $i$). Using the purification and lemma 2.3.3 we can bound the original trace norm distance.

To attack the state sent to Bob, Eve attaches a probe in a fixed state (say the $|0\rangle$ state) and applies a unitary operator. She then passes Bob his part, and does some generalized measurement on what she still holds. We can characterize this formally:

$$|0\rangle_E |i\rangle_A \xrightarrow{U} \sum_j |E_{i,j}\rangle |j\rangle$$

We represent the Fourier transformed states as:

$$|\bar{i}\rangle \equiv \frac{1}{\sqrt{2^n}} \sum_j (-1)^{i \cdot j} |j\rangle$$

Applying this to Eve's attack, we obtain:

$$|0\rangle_E |\bar{i}\rangle_A \xrightarrow{U} \sum_j |\bar{E}_{i,j}\rangle |\bar{j}\rangle$$

---
[1] see definition 2.3.1

where $|\bar{E}_{i,j}\rangle \equiv \frac{1}{2^n} \sum_{i',j'} (-1)^{i \cdot i}(-1)^{j \cdot j'} |E_{i,j}\rangle$.

From the axioms of quantum mechanics, we know that if Alice sends $|i\rangle$ the probability that Bob will measure $|j\rangle$ is $P(j|i) = \langle E_{i,j}|E_{i,j}\rangle$. Similarly, if Alice sends $|\bar{i}\rangle$ Bob will measure $|\bar{j}\rangle$ with probability $\bar{P}(j|i) = \langle \bar{E}_{i,j}|\bar{E}_{i,j}\rangle$.

We are now prepared to compute the probability that there are no errors in Fourier-transformed basis:

$$\begin{aligned}
P(e) &\equiv \sum_i p(i) \bar{P}(j = i \oplus e | i) \\
&= \frac{1}{2^n} \sum_i \langle \bar{E}_{i,j} | \bar{E}_{i,j} \rangle \\
&= \frac{1}{2^{3n}} \sum_{i,i',i'',j',j''} (-1)^{i \cdot (i' \oplus i'' \oplus j' \oplus j'')} (-1)^{e \cdot (j' \oplus j'')} \langle E_{i',j'} | E_{i'',j''} \rangle \\
&= \frac{1}{2^{2n}} \sum_{i',i'',j',j''} \left( \sum_i \frac{1}{2^n} (-1)^{i \cdot (i' \oplus i'' \oplus j' \oplus j'')} \right) (-1)^{e \cdot (j' \oplus j'')} \langle E_{i',j'} | E_{i'',j''} \rangle \\
&= \frac{1}{2^{2n}} \sum_{i',i'',j',j''} \delta_{i' \oplus j', i'' \oplus j''} (-1)^{e \cdot (j' \oplus j'')} \langle E_{i',j'} | E_{i'',j''} \rangle
\end{aligned}$$

For notational ease, we define a new variable $k = i' \oplus j'$; due to the delta function, all the terms that do not also have $k = i'' \oplus j''$ are zero. We restrict our case to the zero error probability:

$$\begin{aligned}
P(0) &= \frac{1}{2^{2n}} \sum_{i',i'',j',j''} \delta_{i' \oplus j', i'' \oplus j''} (-1)^{0 \cdot (j' \oplus j'')} \langle E_{i',j'} | E_{i'',j''} \rangle \\
P(0) &= \frac{1}{2^{2n}} \sum_{i',i'',k} \langle E_{i', i' \oplus k} | E_{i'', i'' \oplus k} \rangle
\end{aligned}$$

When Eve's states are considered without Bob, her state will look like $\rho_i = \sum_j |E_{i,j}\rangle\langle E_{i,j}|$. Now we will define a purification for Eve's states that will allow us to compute the trace norm easily. Using the purification from [BBB99], we assume that Eve holds $|\phi_i\rangle \equiv \sum_j |E_{i,j}\rangle_1 |i \oplus j\rangle_2$, which is a purification of her state $\rho_i$. We also define the Fourier transform of these states: $|\bar{\phi}_j\rangle \equiv \frac{1}{\sqrt{2^n}} \sum_i (-1)^{j \cdot i} |\phi_i\rangle$. The Fourier transform is unitary, so see that $|\phi_i\rangle = \frac{1}{\sqrt{2^n}} \sum_j (-1)^{j \cdot i} |\bar{\phi}_j\rangle$. It should be noted that our purification $|\phi_i\rangle$ for Eve's states is not orthonormal or normalized. In fact, this is a property of which we will make use in order to get a bound. We now calculate the norm of the $|\bar{\phi}_0\rangle$ and see that it is proportional to the probability that there was no error, $P(0)$:

$$\begin{aligned}
\langle \bar{\phi}_0 | \bar{\phi}_0 \rangle &= \frac{1}{2^n} \sum_{i,j} \langle \phi_i | \phi_j \rangle \\
&= \frac{1}{2^n} \sum_{i,j} \sum_{k,l} \langle E_{i,k} | E_{j,l} \rangle \langle i \oplus k | j \oplus l \rangle \\
&= \frac{1}{2^n} \sum_{i,j} \sum_{k'} \langle E_{i,i \oplus k'} | E_{j,j \oplus k'} \rangle \\
&= 2^n P(0)
\end{aligned}$$

Where we have re-labeled the states in the last step. In fact, $\langle \bar{\phi}_e | \bar{\phi}_e \rangle = 2^n P(e)$ but we do not need this result. We are now ready to prove the theorem. Define $\rho_i' \equiv |\phi_i\rangle\langle\phi_i|$ and $\rho' \equiv \frac{1}{2^n} \sum_i \rho_i$. Since $Tr_2(\rho_i') = \rho_i$ and $Tr_2(\rho') = \rho$ we may apply lemma 2.3.3. We will see that we may introduce an intermediate pure state to make the bounding of the information easier. The pure state we will use is $\frac{|\bar{\phi}_0\rangle\langle\bar{\phi}_0|}{\langle\bar{\phi}_0|\bar{\phi}_0\rangle}$. Starting with corollary

2.2.1:

$$
\begin{aligned}
I(A;E) &\leq n \sum_i \frac{1}{2^n}|\rho_i - \rho| \\
&\leq n \sum_i \frac{1}{2^n}|\rho_i' - \rho'| \\
&= n \sum_i \frac{1}{2^n}|\rho_i' - \frac{|\bar{\phi}_0\rangle\langle\bar{\phi}_0|}{\langle\bar{\phi}_0|\bar{\phi}_0\rangle} + \frac{|\bar{\phi}_0\rangle\langle\bar{\phi}_0|}{\langle\bar{\phi}_0|\bar{\phi}_0\rangle} - \rho'| \\
&\leq n \sum_i \frac{1}{2^n}(|\rho_i' - \frac{|\bar{\phi}_0\rangle\langle\bar{\phi}_0|}{\langle\bar{\phi}_0|\bar{\phi}_0\rangle}| + |\frac{|\bar{\phi}_0\rangle\langle\bar{\phi}_0|}{\langle\bar{\phi}_0|\bar{\phi}_0\rangle} - \rho'|) \\
&\leq n \sum_i \frac{1}{2^n}\left(2\sqrt{1 - \frac{\langle\bar{\phi}_0|\rho_i'|\bar{\phi}_0\rangle}{\langle\bar{\phi}_0|\bar{\phi}_0\rangle}} + 2\sqrt{1 - \frac{\langle\bar{\phi}_0|\rho'|\bar{\phi}_0\rangle}{\langle\bar{\phi}_0|\bar{\phi}_0\rangle}}\right) \\
&= 2n\left(\sqrt{1 - \frac{\langle\bar{\phi}_0|\rho'|\bar{\phi}_0\rangle}{\langle\bar{\phi}_0|\bar{\phi}_0\rangle}} + \sum_i \frac{1}{2^n}\sqrt{1 - \frac{\langle\bar{\phi}_0|\rho_i'|\bar{\phi}_0\rangle}{\langle\bar{\phi}_0|\bar{\phi}_0\rangle}}\right) \\
&\leq 2n\left(\sqrt{1 - \frac{\langle\bar{\phi}_0|\rho'|\bar{\phi}_0\rangle}{\langle\bar{\phi}_0|\bar{\phi}_0\rangle}} + \sqrt{1 - \frac{\langle\bar{\phi}_0|(\sum_i \frac{1}{2^n}\rho_i')|\bar{\phi}_0\rangle}{\langle\bar{\phi}_0|\bar{\phi}_0\rangle}}\right) \\
&= 4n\sqrt{1 - \frac{\langle\bar{\phi}_0|\rho'|\bar{\phi}_0\rangle}{\langle\bar{\phi}_0|\bar{\phi}_0\rangle}}
\end{aligned}
$$

Now we compute $\langle\bar{\phi}_0|\rho'|\bar{\phi}_0\rangle$:

$$
\begin{aligned}
\langle\bar{\phi}_0|\rho'|\bar{\phi}_0\rangle &= \sum_i \frac{1}{2^n}|\langle\bar{\phi}_0|\phi_i\rangle|^2 \\
&= \sum_i \frac{1}{2^n}|\langle\bar{\phi}_0|\frac{1}{\sqrt{2^n}}\sum_j (-1)^{i\cdot j}|\bar{\phi}_j\rangle|^2 \\
&\geq |\frac{1}{2^n}\langle\bar{\phi}_0|\frac{1}{\sqrt{2^n}}\sum_j \sum_i (-1)^{i\cdot j}|\bar{\phi}_j\rangle|^2 \\
&= |\frac{1}{2^n}\langle\bar{\phi}_0|\frac{1}{\sqrt{2^n}}\sum_j 2^n \delta_{j,0}|\bar{\phi}_j\rangle|^2 \\
&= \frac{1}{2^n}|\langle\bar{\phi}_0|\bar{\phi}_0\rangle|^2
\end{aligned}
$$

Putting this together:

$$\begin{aligned} I(A;E) &\leq 4n\sqrt{1 - \frac{\langle \bar{\phi}_0 | \rho' | \bar{\phi}_0 \rangle}{\langle \bar{\phi}_0 | \bar{\phi}_0 \rangle}} \\ &\leq 4n\sqrt{1 - \frac{1}{2^n}\langle \bar{\phi}_0 | \bar{\phi}_0 \rangle} \\ &= 4n\sqrt{1 - P(0)} = 4n\sqrt{P_{\bar{e}}} \end{aligned}$$

Where $P_{\bar{e}}$ is the probability that there is an error in the Fourier transformed basis, which proves the theorem. ∎

The previous theorem is what gives security to quantum key distribution schemes; however, we have only shown that QKD schemes are secure if the errors caused in the Fourier-transformed basis are extremely small. Ideally, we would like Eve's information to be exponentially close to zero. We consider in detail the case where there are errors on the channel between Alice and Bob in chapter 3. We show that, as long as the errors are not too large, they can all be corrected with Eve only gaining exponentially little information. It is the result in chapter 3 that proves the security of realistic quantum key distribution.

### 2.4.2 Security of Functions of Messages

According to theorem 2.4.1, if the fidelity Bob would have had in the Fourier transformed basis is exponentially close to unity, then Eve's information is exponentially low. It does not address the question of what information Eve might get about a *function* of a message encrypted with that key. Suppose Eve only wants to know if the message has a particular value. This function only has exponentially little information about the message itself. Could Eve learn this information? The next theorem will show that this, too, is impossible.

**Theorem 2.4.2** *If Alice sends the $n$ qubit state $|k\rangle$ to Bob, with $k$ chosen uniformly at*

*random, and after Bob has received the state Alice announces* $a = m \oplus k$, *then the information Eve can get about any function of* $m$, $F(M)$, *is bounded by the square root of the probability that the Eve would have caused errors in the Fourier transformed basis:*

$$I(F(M); E|A) \leq H(F(K)) 4\sqrt{P_{\bar{e}}}$$

**P**roof. This proof will follow closely the proof of theorem 2.4.1 and use the same tools. The state consistent with a function value $i$ is:

$$\sigma_i{}^a \equiv \frac{1}{q_i} \sum_{k: f(a \oplus k) = i} p_k \rho_k$$

with $q_i \equiv \sum_{k: f(a \oplus k) = i} p_k$. Note that since $p_k = \frac{1}{2^n}$, then the probability of an announcement $a = m \oplus k$ is also $\frac{1}{2^n}$. As such, $q_i$ does not depend on $m$ and is only related to the number of inputs to the function $f$ which have a given output. The averaged state is:

$$\sigma^a \equiv \sum_i q_i \sigma_i{}^a$$
$$= \sum_i \sum_{k: f(k \oplus a) = i} p_k \rho_k$$

Since each input has one and only one output and $p_k = \frac{1}{2^n}$:

$$\sigma^a = \sum_k \frac{1}{2^n} \rho_k = \rho$$

The definition of mutual information[CT91] means that:

$$I(F(M); E|A) = \sum_a p_a I(F(M); E|A = a)$$

Using lemma 2.2.3

$$\sum_a p_a I(F(M); E|A=a)$$
$$\leq -\sum_a p_a \sum_i q_i \log q_i |\sigma_i{}^a - \sigma^a|$$
$$= -\sum_i q_i \log q_i \sum_a p_a |\sigma_i{}^a - \rho|$$
$$= -\sum_i q_i \log q_i \sum_a p_a |\sigma_i{}^a - \frac{|\bar{\phi}_0\rangle\langle\bar{\phi}_0|}{\langle\bar{\phi}_0|\bar{\phi}_0\rangle} + \frac{|\bar{\phi}_0\rangle\langle\bar{\phi}_0|}{\langle\bar{\phi}_0|\bar{\phi}_0\rangle} - \rho|$$
$$\leq -\sum_i q_i \log q_i \sum_a p_a \left(|\sigma_i{}^a - \frac{|\bar{\phi}_0\rangle\langle\bar{\phi}_0|}{\langle\bar{\phi}_0|\bar{\phi}_0\rangle}| + |\frac{|\bar{\phi}_0\rangle\langle\bar{\phi}_0|}{\langle\bar{\phi}_0|\bar{\phi}_0\rangle} - \rho|\right)$$
$$= -\sum_i q_i \log q_i \sum_a p_a \left(2\sqrt{1 - \frac{\langle\bar{\phi}_0|\sigma_i{}^a|\bar{\phi}_0\rangle}{\langle\bar{\phi}_0|\bar{\phi}_0\rangle}} + 2\sqrt{1 - \frac{\langle\bar{\phi}_0|\rho|\bar{\phi}_0\rangle}{\langle\bar{\phi}_0|\bar{\phi}_0\rangle}}\right)$$
$$\leq -\sum_i q_i \log q_i \left(2\sqrt{1 - \frac{\langle\bar{\phi}_0|\sum_a p_a \sigma_i{}^a|\bar{\phi}_0\rangle}{\langle\bar{\phi}_0|\bar{\phi}_0\rangle}} + 2\sqrt{1 - \frac{\langle\bar{\phi}_0|\rho|\bar{\phi}_0\rangle}{\langle\bar{\phi}_0|\bar{\phi}_0\rangle}}\right)$$

We can simplify the quantity $\sum_a p_a \sigma_i{}^a$ by remembering that $p_a = 1/2^n$ and $q_i$ is independent of $a$:

$$\sum_a \frac{1}{2^n}\sigma_i{}^a = \sum_a \frac{1}{2^n} \frac{\sum_{k:f(k\oplus a)=i} \frac{1}{2^n}\rho_k}{q_i}$$
$$= \frac{1}{q_i}\sum_a \frac{1}{2^n} \sum_{m:f(m)=i} \frac{1}{2^n}\rho_{a\oplus m}$$
$$= \frac{1}{q_i}\sum_{m:f(m)=i} \frac{1}{2^n} \sum_a \frac{1}{2^n}\rho_{a\oplus m}$$

In the last sum, we sum over all $a$ with equal weight; hence, the $m$ dependence disappears:

$$\sum_a \frac{1}{2^n}\sigma_i{}^a = \frac{1}{q_i}\sum_{m:f(m)=i} \frac{1}{2^n} \sum_a \frac{1}{2^n}\rho_{a\oplus m}$$
$$= \frac{1}{q_i}(\sum_{m:f(m)=i} \frac{1}{2^n})\rho$$
$$= \rho$$

Putting this back into the information bound:

$$\sum_a p_a I(F(M); E|A=a)$$

$$\leq -\sum_i q_i \log q_i \left(2\sqrt{1 - \frac{\langle \bar{\phi}_0 | \sum_a p_a \sigma_i{}^a | \bar{\phi}_0 \rangle}{\langle \bar{\phi}_0 | \bar{\phi}_0 \rangle}} + 2\sqrt{1 - \frac{\langle \bar{\phi}_0 | \rho | \bar{\phi}_0 \rangle}{\langle \bar{\phi}_0 | \bar{\phi}_0 \rangle}}\right)$$

$$= -\sum_i q_i \log q_i (4\sqrt{1 - \frac{\langle \bar{\phi}_0 | \rho | \bar{\phi}_0 \rangle}{\langle \bar{\phi}_0 | \bar{\phi}_0 \rangle}})$$

$$= 4H(Q)\sqrt{1 - \frac{\langle \bar{\phi}_0 | \rho | \bar{\phi}_0 \rangle}{\langle \bar{\phi}_0 | \bar{\phi}_0 \rangle}}$$

$$\leq H(F(K)) 4\sqrt{P_{\bar{e}}}$$

Which proves the result. ∎

## 2.5  Summary

By developing bounds on entropy, we are able to apply these to bound the amount of information that measurements can get from a quantum source. Modeling eavesdropping in quantum key distribution as a quantum source, we are able to bound information that an eavesdropper can get. Since this bound is a function of the errors that would be caused in a Fourier transformed basis, Alice and Bob can use their measurements to estimate this figure. Therefore, Alice and Bob can bound information that Eve has about the information they share. In addition to showing security of such information, we show for the first time that any function of messages encrypted with this secret information is secure. This is a very strong statement about the robustness of quantum security.

What lies next is to In the following chapter, we examine the case where there are errors on the channel. We then show a similar but stronger result, namely that Eve must cause more errors than Alice and Bob could have corrected in the Fourier transformed basis.

# CHAPTER 3

# Security of Quantum Key Distribution

## 3.1 Introduction

Quantum key distribution [BBB92, BB84] uses the power of quantum mechanics to suggest the distribution of a key that is secure against an adversary with unlimited computation power. Such a task is beyond the ability of classical information processing. The extra power gained by the use of quantum bits (quantum two-level systems) is due to the fact that the state of such a system cannot be cloned. On the other hand, the security of conventional key distribution is based on the (unproven) existence of various one-way functions, and mainly on the difficulty of factoring large numbers, a problem which is assumed to be difficult for a classical computer, and is proven to be easy for a hypothetical quantum computer [Sho97].

Various proofs of security were previously obtained against collective attacks [BM97b, BM97a, BBB98], and we continue this line of research here to prove the ultimate security of quantum key distribution (QKD), against any attack (under the conventional assumptions of theoretical QKD, as explained below). Note that the eavesdropper is assumed to have unlimited technology (e.g., a quantum memory, a quantum computer), while the legitimate users use practical tools (or more precisely, simplifications of practical tools).

To prove security against such a super-strong eavesdropper we develop some important technical tools and we reach some surprising results: we obtain a new *infor-* *mation versus disturbance* result, where the power of quantum information theory is manifested in an intuitive and clear way. We show explicitly how the randomness of the choice of bases, and the randomness of the choice of test-bits provides the desired security of QKD. We adopt and generalize sophisticated tools invented in [BBB98], and developed more in chapter 2: a "purification" which simplifies Eve's states; a bound on accessible information (using Trace-Norm-Difference of density matrices) which avoids any complicated optimization of Eve's possible measurements; a connection between Eve's accessible information and the error-rate she induces. We add some more simplifications (which were not required in the analysis of collective attacks in [BBB98]): a reduction to a scheme in which all qubits are used by Alice and Bob, and a usage of a symmetry of the problem under investigation.

Recently there have been a few security results announced [May96, May, LC99, BBB99, Lo, SP00] [1]. This proof differs from other proofs in that uses tools similar to those in chapter 2 to find explicit bounds on information which are a function of the errors that the eavesdropper would cause. Other proofs, specifically [LC99, Lo, SP00], make use of an ingenious reduction to allow the proof of security in the case of exponentially few errors, to apply to a case where an error correction code is used. The current work may be elucidating to security proofs for protocols which might not have such a mapping onto quantum error codes.

We follow the standard assumptions of QKD (assumption 3 is discussed in [BMS, BLM00] in much details): 1) Alice and Bob share an unjammable classical channel. This assumption is usually replaced by the demand that Alice and Bob share a short secret key to be used for authenticating a standard classical channel (hence the protocol is then a quantum key expansion protocol). 2) Eve cannot attack Alice's and Bob's labs. She can only attack the quantum channel and listen to all transmissions on the classical channel. 3) Alice sends quantum bits (two level systems).

---

[1]This chapter is a adaptation of [BBB99], of which I was a coauthor

We prove the security of the Bennett-Brassard-84 (BB84) protocol [BB84], against any attack allowed by the rules of quantum physics. We prove the security for instances in which the error rate in the transmission from Alice to Bob is up to 7.56%.

### 3.1.1 The BB84 Protocol and the Used-Bits-BB84 Protocol

Quantum cryptography [Wie83, BB84] was described in several places, some of which are also introducing the notations in a more expository way, and a reader who is unfamiliar with the basics of quantum information processing is referred for instance to the Appendix in [BBC93b].

In the BB84 protocol Alice and Bob use four possible quantum states in two bases (using "spin" notations, and connecting them to "computation basis" notations): (i) $|0_z\rangle \equiv |0\rangle$; (ii) $|1_z\rangle \equiv |1\rangle$; (iii) $|0_x\rangle = \frac{1}{\sqrt{2}}(|0\rangle + |1\rangle)$; and (iv) $|1_x\rangle = \frac{1}{\sqrt{2}}(|0\rangle - |1\rangle)$. We shall refer to these states as the BB84 states. By comparing bases after Alice transmit such a state and Bob receives it, a common key can be created in instances when Alice and Bob used the same basis.

We prove here the security of a simplified protocol in which only the relevant bits are discussed (we call it the "used-bits-BB84"). The proof of the security of the original BB84 protocol follows immediately, due to a simple reduction, as we show in section 3.6.

Let us describe the used-bits protocol in detail, splitting it into *creating the sifted key* and *creating the final key from the sifted key*. This simplified protocol assumes that Bob has a quantum memory.

I. Creating the sifted key:

1. Alice and Bob choose a large integer $n \gg 1$. The protocol uses $2n$ bits.

2. Alice randomly selects two $2n$-bit strings, $b$ and $i$ which are then used to create qubits: The string $b$ determines the basis $0 \equiv z$, and $1 \equiv x$ of the qubits. The string $i$ determines the value (0 or 1) of each of the $2n$ qubits (in the appropriate bases).

   Alice generates $2n$ qubits according to her selection, and sends them to Bob via a quantum communication channel.

3. Bob tells Alice when he receives the qubits.

4. Alice publishes the bases she used, $b$; this step should be performed only after Bob received all the qubits.

   Bob measures the qubits in Alice's bases to obtain a $2n$-bit string $j$.

   We shall refer to the resulting $2n$-bit string as the sifted key, and it would be the same for Alice and Bob, i.e. $j = i$, if natural errors and eavesdropping did not exist.

II. Creating the final key from the sifted key:

1. Alice chooses at random a $2n$-bit string $s$ which has exactly $n$ zeroes and $n$ ones. There are $\binom{2n}{n}$ such strings to choose from.

2. From the $2n$ bits, Alice selects a subset of $n$ bits, determined by the zeros in $s$, to be the test bits. Alice publishes the values of these test bits (given by a string $i_T$). The values of Bob's bits on the test bits are given by $j_T$.

   The other $n$ bits are the information bits (given by a string $i_I$). They are used for deriving a final key via error correction codes (ECC) and privacy amplification (PA) techniques.

   Alice shall send the ECC and PA information to Bob, hence Bob needs to correct his errors and use PA to obtain a key equal to Alice's.

3. Bob verifies that the error rate $p_{test} = |i_T \oplus j_T|/n$ in the test bits is lower than some agreed error-rate $p_{allowed}$, and aborts the protocol if the error rate is larger.

4. Bob also publishes the values of his test bits ($j_T$). This is not crucial for the protocol, but it is done to simplify the proof.

5. Alice selects a linear ECC with $2^k$ code words of $n$ bits, and a minimal Hamming distance $d$ between any two words: an $(n, k, d)$ code, and publishes it along with the ECC parities on the information bits. The strategy is that Alice announces the parity check matrix of an ECC, i.e., $r = n - k$ parity check strings of n bits: $v_s$, $s = 1, \ldots, r$. She then announces $r$ bits which are the parities of her string $i_I$ with respect to the parity check matrix, which is $v_s \cdot i_I$ for all $s$. Bob doesn't announce anything. The condition on the ECC is that it corrects $t \geq (p_{allowed} + \epsilon_{rel})n$ errors, for some positive $\epsilon_{rel}$. If an ECC corrects has $d \geq 2t + 1$ it will always correct $t$ errors, and thus $d \geq 2(p_{allowed} + \epsilon_{rel})n + 1$ is sufficient for all codes. For Random Linear Codes $d \geq (p_{allowed} + \epsilon_{rel})n + 1$ is also sufficient as noted in [May].

6. Bob performs the correction on the information bits.

7. Alice selects a privacy amplification function (PA) and publishes it. The PA strategy is to publish $m$ $n$-bit strings and use the parities of the bits masked by these strings as the secret key. That is she announces *privacy-amplification-strings* $v_s$, where $s = r + 1, \ldots, r + m$, of $n$ bits each. The final secret key bits are $v_s \cdot i$. This strategy is similar to error correction except that the parities are kept secret.

   The PA strings must be chosen such that the minimal distance $\hat{v}$, between any string in their span and any string in the span of their union with the ECC parity-check-strings, is at least $\hat{v} \geq 2(p_{allowed} + \epsilon_{sec})\, n$. Note that, by definition, the minimal distance of the space spanned by the ECC and PA strings, $d^\perp$, is less than the above distance, hence if we demand $d^\perp \geq 2(p_{allowed} + \epsilon_{sec})\, n$, the above criterion is automatically satisfied.

8. Bob performs the PA on the corrected information bits. The result obtained is the final key.

### 3.1.2 Eavesdropping

Eve attacks the qubits in two steps. First she lets all qubits pass through a device that tries to probe their state. Then, after receiving all the classical data, she measures the probe. She can gain nothing by measuring the probe earlier, since such a measurement is a special case of applying a unitary operation (it is the application of a measurement gate). Thus we can split Eve's attack into her transformation and her measurement.

> Eve's transformation: The qubits can be attacked by Eve while they are in the channel between Alice and Bob. Eve can perform any attack allowed by the laws of physics, the most general one being any unitary transformation on Alice's qubits and Eve's probe. We are generous to Eve, allowing her to attack all the bits together (in practice, she usually needs to send the preceding qubit toward Bob before she has access to the next one).
>
> Without loss of generality we assume that all the noise on the qubits, is caused by Eve, and can be used by her in any way she likes.
>
> Eve's measurement: Eve keeps the probe in a quantum memory. After Eve receives *a*ll the classical information from Alice and Bob, including the bases of all bits, the choice of test bits, the test bits values, the ECC, the ECC parities, and the PA, she tries to guess the final key using her best strategy of measurement.

Eve's goal is to learn as much information as possible on the final key without causing Alice and Bob to abort the protocol due to a failure of the test. The task of finding Eve's optimal operation in these two steps is very difficult. Luckily, to prove security that task need not be solved, and it is enough to find bounds on Eve's optimal information (via any operation she could have done): In order to analyze her optimal transformation we find bounds for *any* transformation she could perform, and in order to analyze her optimal measurement we find bounds for *any* measurement she could perform.

### 3.1.3 Security and Reliability

The issue of the security criterion is non-trivial since the obvious security criterion (that Eve's information given that the test passed, is small) does not work.

To be more precise, let $\mathcal{A}$ be a random variable presenting Alice's final key, $\mathcal{B}$ be a random variable presenting Bob's final key, and $\mathcal{E}$ a random variable representing a string in Eve's hands as result of her measurements. Let $\mathcal{T}$ be a random variable presenting if the test passed or failed. What one would like to obtain as a security criterion is $I(\mathcal{A}; \mathcal{E} \mid \mathcal{T} = \text{pass}) \leq A_{\text{info}} \, e^{-\beta_{\text{info}} n}$ with $A$ and $\beta$ (with any subscript) positive constants.

Unfortunately the above bound is *not* satisfied in quantum cryptography. Given that the test is passed, Eve can still have full information. Consider the *swap attack*: Eve takes Alice's qubits and puts them into a quantum memory. She sends random BB84 states to Bob. Eve measures the qubits she kept after learning their bases, hence gets full information on Alice's final key. In this case, Bob will almost always abort the protocol because it is very unlikely that his bits will pass the test. However, even in the rare event when the test is passed, Eve still has full information on Alice's key. So, given the test is passed (a rare event), information is still $m$ bits, and the above criterion cannot be satisfied.

Another potential security criterion says the following: "if Eve tries an attack that gives her non-negligible information on a final key she has to be extremely lucky in order to pass the test." As observed in an earlier version of [May], this criterion is also inappropriate. Consider the half-SWAP attack in which Eve does nothing with probability half, and performs the SWAP attack with probability half. This half-SWAP attack gives information of exactly m/2, and it passes the test with probability larger than half. Obviously these two cases, getting a non-negligible information, and passing the test with high probability, will not happen in the same event, hence this example motivates a more precise definition of security (first used in [May]).

In order to prove security we show that the event where the test is passed *and* Eve obtains meaningful information on the key is extremely unlikely.

Formally, the security criterion is:

$$P\left[(\mathcal{T} = \text{pass}) \wedge (I_{Eve} \geq A_{\text{info}}\, e^{-\beta_{\text{info}} n})\right] \leq A_{\text{luck}}\, e^{-\beta_{\text{luck}} n} \qquad (3.1)$$

where $\mathcal{T}$ is the test outcome and $I_{Eve} \equiv I(\mathcal{A}; \mathcal{E} | i_T, c_T, b, s)$ is the information Eve has on the key, after the particular protocol values $(i_T, j_T, b, s)$ are announced by Alice and Bob. The event in which the test is passed includes all the cases such that $c_T = i_T \oplus j_T$ satisfies $|c_T| \leq n p_{allowed}$. Note that Alice and Bob can increase the number of bits $n$ as they like to increase security.

We show that the final $m$-bit key is reliable: the keys distiled by Alice and Bob are identical except for some exponentially small probability $A_{\text{rel}}\, e^{-\beta_{\text{rel}} n}$.

### 3.1.4 Structure of the Chapter

The rest of the paper contains three main steps: In Section 3.2 we reduce the problem to a simpler problem of optimizing over all attacks symmetric to the bit values 0 and 1. In Section 3.3 we analyze the information bits in the bases actually used by Alice and Bob, and we prove our main *information versus disturbance* theorem: the eavesdropper information on the final key is bounded by the following probability: the probability of error if the *other* bases were used by Alice and Bob (this probability is well defined). We then obtain in Section 3.4 a bound on

$$\sum_{i_T, c_T, b, s} P(\mathcal{T} = \text{pass}, i_T, c_T, b, s) \, I(\mathcal{A}; \mathcal{E} | i_T, c_T, b, s)$$

and prove that this bound is exponentially small with $n$. This expression could also serve as a security criterion, since it immediately follows (from this bound) that the security criterion 3.1 is satisfied as shown in section 3.7.

Various technical details and Lemmas are proven in the appendices, so that the proof can be read more smoothly.

## 3.2 Eve's Attack

In the used-bits BB84 protocol Alice sends a string $i$ encoded in the bases of her choice $b$, and Bob measures a string $j$ using the same set of bases. Eve prepares a probe in a known state, say $|0\rangle$. Eve applies a unitary transformation $U$ on *a*ll the qubits and her probe and then she sends the disturbed qubits to Bob, while leaving her probe in her hands. The unitary transformation $U$ is written in the basis $b$, $U(|0\rangle|i\rangle) = \sum_j |E'_{i,j}\rangle|j\rangle$, with $|E'_{i,j}\rangle$ the unnormalized states of Eve's probes if Alice sent $|i\rangle$, and Bob received $|j\rangle$.

Later on Eve obtains *all* classical information sent by Alice and Bob. Eve learns $b$ (the bases) and $s$ (which bits are the test bits and which are the information bits). She also learns the values of the test bits $i_T$ and $j_T$. We also use $i_I$ and $j_I$ to denote the values of the information bits. Then, Eve's attack (written in a basis $b$ chosen by Alice)

looks like:

$$U[(|0\rangle)_{Eve}(|i_T\rangle|i_I\rangle)_{Alice}] \equiv \sum_{j_T,j_I} |E'_{i_T,i_I,j_T,j_I}\rangle|j_T\rangle|j_I\rangle \ . \quad (3.2)$$

Once the additional data regarding the bases and the values of the test bits is given to Eve, this data modifies her probes' states. We define $|\psi_{i_I}\rangle$ to be the state of Eve+Bob if Alice chose a bases $b$, a sample $s$, and values $i_T i_I$, Eve's attack is $U$, and Bob received $j_T$ in his measurement on the test bits. After renormalization Eve and Bob's state is:

$$|\psi_{i_I}\rangle = \frac{1}{\sqrt{p(j_T|i_T,i_I,b,s)}}[\langle j_T|]U[(|0\rangle)_{Eve}(|i_T\rangle|i_I\rangle)_{Alice}] \quad (3.3)$$

Since $\langle\psi_{i_I}|\psi_{i_I}\rangle = 1$, using the above definitions the normalization is fixed[2]: $p(j_T|i_I,i_T,b,s) = \sum_{j_I}||E'_{i_I,i_T,j_I,j_T}||^2$, which makes use of the norm notation: $||E_{i,j}||^2 \equiv \langle E_{i,j}|E_{i,j}\rangle$.

For a given $i_T$ and $j_T$, we define:

$$|E_{i_I,j_I}\rangle \equiv \frac{1}{\sqrt{p(j_T|i_T,i_I,b,s)}}|E'_{i_T,i_I,j_T,j_I}\rangle$$

Eve's states *for a given* classical data regarding $(i_T, j_T, s, b)$. Then,

$$|\psi_{i_I}\rangle = \sum_{j_I} |E_{i_I,j_I}\rangle|j_I\rangle \ . \quad (3.4)$$

We now present a symmetrized attack (which is symmetrical to Alice sending 0 or 1 for the bit values). This is required in order to obtain Eq.(3.11,3.12). Recall that the choice of 0/1 is random. As a result, any attack chosen by Eve can be replaced by an equivalent attack which is as good, with $i$ replaced by $i \oplus k$ and with $j$ replaced by $j \oplus k$, for any $k$. Thus, any attack chosen by Eve can also be replaced by an equivalent symmetric attack which is as good, in which $k$ is chosen at random. The symmetrization does not change the average induced error-rate. For an arbitrary attack, the symmetrization can improve Eve's final information on the common secret key (or leave it the same). Thus, if the *optimal attack* is asymmetric, there is also an equivalent symmetric attack which is as good, hence also optimal. Thus, the optimal attack can be assumed to be symmetric, without loss of generality (WLG), and we therefore need to bound Eve's information only for attacks symmetric to 0/1.

---

[2]$p(x)$ is the usual probability theory notation: $Prob(X = x)$

The symmetrization is performed using bit-wise operations: Given any transformation of Eve, it is symmetrized as follows: For each qubit $|q_l\rangle$ sent by Alice, Eve adds a qubit in a state $|w_l\rangle = H|0\rangle = \frac{1}{\sqrt{2}}(|0_z\rangle + |1_z\rangle)$, and performs a pseudo-controlled-NOT transform on this bit as the control and $|q_j\rangle$ as the target: if $|w_j\rangle = 0$ leave $|q_j\rangle$ as is; otherwise negate it (i.e., rotate the spin by 180 degrees) in both $x$ basis and $z$ basis. After the application of the (possibly asymmetric) attack $U$, Eve performs the inverse of this pseudo-controlled-NOT transform, and let Alice's qubit continue to Bob. The gate which does the negation properly in both $x$ and $z$ bases is the Control-$(\sigma_x \sigma_z)$ on this ancillary qubit and the data qubit. For any attack $U$ chosen by Eve, she applies the symmetrization and the resulting attack is $U^{sym}$. The 0/1 symmetrization ensures that the errors are independent of the values 0 or 1 that Alice sends (in either basis). The overall attack on all qubits is then described by $U^{sym}(|0\rangle|i\rangle) = \sum_j |E'^{sym}_{i,j}\rangle|j\rangle$ with $E'^{sym}$ which can be written using $E'$ as follows:

$$|E'^{sym}_{i,j}\rangle = \frac{1}{\sqrt{2^{2n}}} \sum_m (-1)^{(i \oplus j) \cdot m} |m\rangle |E'_{i \oplus m, j \oplus m}\rangle. \tag{3.5}$$

If Eve measures $|m\rangle$ at any stage of the attack (and receives a specific value, say $k$), the resulting attack is one in which $i$ and $j$ of the original attack are replaced by $i \oplus k$ and $j \oplus k$ as previously described, hence an attack which is equivalent to the original attack.

To prove that the 0/1 symmetrization does not change the average error-rate is obvious since Eve can always project onto one particular $m$ (and destroy the symmetry) by measuring $|m\rangle$, and any such projection leads to the same attack (up to a shift of $i$ and $j$ by $m$). Since Eve can perform that measurement later on (when she does not hold Bob's qubits anymore), her measurement cannot affect Bob's outcomes due to causality argument. It is also obvious that the symmetric attack cannot be worse (for Eve) in terms of Eve's information, since she can always measure $m$. Clearly, the symmetrization can only increase Eve's information since she does not have to measure $m$ but can also do other things.

The property of symmetric attacks that we are using is:

$$\langle E'^{sym}_{i \oplus m, i \oplus m \oplus c} | E'^{sym}_{i \oplus m, i \oplus m \oplus c} \rangle = \langle E'^{sym}_{i, i \oplus c} | E'^{sym}_{i, i \oplus c} \rangle .$$

More explicitly, we calculate:

$$\begin{aligned}
p(j_T | i_I, i_T, b, s) &= \sum_{j_I} ||E'^{sym}_{i_I, i_T, j_I, j_T}||^2 \\
&= \frac{1}{2^{2n}} \sum_{j_I, m_I, m_T} ||E'_{i_I \oplus m_I, i_T \oplus m_T, j_I \oplus m_I, j_T \oplus m_T}||^2 \\
&= \frac{1}{2^{2n}} \sum_{i'_I, j'_I, m_T} ||E'_{i'_I, i_T \oplus m_T, j'_I, j_T \oplus m_T}||^2 .
\end{aligned} \quad (3.6)$$

Hence, $p(j_T | i_I, i_T, b, s) = p(j_T | i_T, b, s)$ is independent of $i_I$, and using standard probability theory we then get
$p(i_I | i_T, j_T, b, s) = p(i_I | i_T, b, s)$. Thus, for any symmetrized attack

$$p(i_I | i_T, j_T, b, s) = 1/2^n \quad (3.7)$$

because of independence: $p(i_I | i_T, j_T, b, s) = p(i_I | i_T, b, s)$ as shown above, and $p(i_I | i_T, b, s) = p(i_I)$ because Alice chooses $i$, $b$ and $s$ independently of each other. In Subsections 3.3.3 and 3.3.4 we make use of this property $p(i_I | i_T, j_T, b, s) = p(i_I | i_T, b, s) = 1/2^n$.

## 3.3 Information vs. Disturbance

In this section we analyze the information bits alone (for a given symmetric attack $U^{sym}$, a given input $i_T$ and outcome $j_T$ on the test bits, and given bases $b$ and choice of test bits $s$). Our result here applies for any $U^{sym}$, hence in particular *for the optimal one*. The optimization over Eve's measurement is avoided by using the fact that trace-norm of the difference of two density matrices provides an upper bound on the accessible information one could obtain *via any measurement* when having the two density matrices as the possible inputs.

### 3.3.1 Eve's State

When Alice sends a state $|i_I\rangle$ for the information bits (written in the basis actually used by her and Bob for these bits), the state of Eve and Bob together, $|\psi_{i_I}\rangle = \sum_{j_I} |E_{i_I,j_I}\rangle |j_I\rangle$ is fully determined by Eve's attack and by the data regarding the test bits. Eve's state in that case is fully determined by tracing-out Bob's subsystem $|j_I\rangle$ from the Eve-Bob state, and it is

$$\rho^{i_I} = \sum_{j_I} |E_{i_I,j_I}\rangle\langle E_{i_I,j_I}|,$$

calculated given $i_T$ and $j_T$. This state in Eve's hands is a mixed state.

### 3.3.2 Purification and a Related Basis

We can "purify" the state while giving more information to Eve by assuming she keeps the state

$$|\phi_i\rangle = \sum_{j_I} |E_{i_I,j_I}\rangle |i_I \oplus j_I\rangle$$

where we introduce another subsystem for the "purification". [Note that an index $\{\ \}_I$ for $\phi_i$ is not required since the purified state is only defined on the information bits]. The term purification means different things in different papers, thus we explain it a bit more: a mixed state can also be obtained from a pure state in an enlarged system (the original system plus an ancilla), once the ancilla is traced out; the pure state of the enlarged system is called a purification of the mixed state. In a more general case, the state in the enlarged system is not necessarily pure, and then we refer to it as a "lift-up" [BBB98] of the state of the original system.

The resulting purified state (i.e., any purification or any lift-up of Eve's states) is at least as informative to Eve as $\rho^{i_I}$ is. This is because the density matrix is exactly the same as it was if Eve ignores the $i_I \oplus j_I$ register of $\phi$. Thus, any information Eve can obtain from her mixed state is bounded by the information she could get if the purified state was available to her. Note that the overlap between these purified states satisfies

$$\begin{aligned}\langle\phi_l|\phi_{l\oplus k}\rangle &= \sum_j \sum_{j'} \langle E_{l_I,j_I}|E_{l_I\oplus k_I,j'_I}\rangle \langle l_I\oplus j_I|l_I\oplus k_I\oplus j'_I\rangle \\ &= \sum_j \langle E_{l_I,j_I}|E_{l_I\oplus k_I,j_I\oplus k_I}\rangle .\end{aligned}$$

For the space spanned by the purified states $\phi_i$, we define a basis $|\eta\rangle$, and show that it is possible to compute a bound on Eve's information on the information bits, once the purified states are expressed in this basis.

**Definition 3.3.1**

$$|\eta_i\rangle = \frac{1}{2^n}\sum_l (-1)^{i\cdot l}|\phi_l\rangle \; ; \; d_i^2 = \langle\eta_i|\eta_i\rangle \; ; \; \hat{\eta}_i = \eta_i/d_i$$

Using the above definitions and $(1/2^n)\sum_l(-1)^{(i\oplus j)\cdot l} = \delta_{ij}$, Eve's purified state can be rewritten as:

$$|\phi_i\rangle = \sum_l (-1)^{i\cdot l}|\eta_l\rangle = \sum_l (-1)^{i\cdot l} d_i|\hat{\eta}_l\rangle \tag{3.8}$$

Note that $\langle\eta_i|\eta_i\rangle = \frac{1}{2^{2n}}\sum_l\sum_k(-1)^{i\cdot k}\langle\phi_l|\phi_{l\oplus k}\rangle$, hence the length of the vectors $\eta_i$ is the average over all $l$, of the Fourier transform of the overlap $\langle\phi_l|\phi_{l\oplus k}\rangle$. In terms of Eve's states we get

$$d_i^2 = \langle\eta_i|\eta_i\rangle = \frac{1}{2^{2n}}\sum_l\sum_k(-1)^{i\cdot k}\sum_j\langle E_{l,j}|E_{l\oplus k,j\oplus k}\rangle . \tag{3.9}$$

### 3.3.3 Eve's State and Probability of Errors Induced on Information Bits

In this subsection we show that the probability of any error string Eve would have induced if the conjugate basis was used for the information bits, is a simple function of $d_i$'s (of Definition 3.3.1), hence a function of the overlap.

First, we discuss the error rate in the basis $b_I, b_T$ actually used by Alice and Bob. For any attack

$$P(j_I = i_I \oplus c_I \mid i_I, i_T, j_T, b, s) = \langle E_{i_I, i_I \oplus c_I} | E_{i_I, i_I \oplus c_I} \rangle . \tag{3.10}$$

For any symmetrized attack satisfying the 0/1 symmetry, the error distribution in the information bits is

$$\begin{aligned} P(c_I | i_T, j_T, b, s) \\ &= \sum_{i_I} P(i_I | i_T, j_T, b, s) P(j_I = i_I \oplus c_I \mid i_I, i_T, j_T, b, s) \\ &= \frac{1}{2^n} \sum_i \langle E_{i_I, i_I \oplus c_I} | E_{i_I, i_I \oplus c_I} \rangle \end{aligned} \tag{3.11}$$

the average probability of an error syndrome $c$ for the information bits (when the test bits, basis and sequence are given). The first equality is derived using standard probability theory and the second is due to Eq. (3.7) and Eq. (3.10).

Due to the linearity of quantum mechanics, given Eve's attack in one basis we can write Eve's attack in any other basis, and in particular, in a basis $\bar{b}_I, b_T$, where the $x/z$ bases of each information qubit are interchanged. We refer to this basis as the "conjugate basis", but note that it is only conjugate on the information bits. For an input string $i = i_I, i_T$ in the conjugate basis ($\bar{b}_I, b_T$) and an output string $j = j_I, j_T$, the error distribution for the information bits is

$$P(c_I|i_T, j_T, \bar{b}_I, b_T, s)$$
$$= \sum_{i_I} p(i_I|i_T, j_T, \bar{b}_I, b_T, s) P(j_I = i_I \oplus c_I|i_I, i_T, j_T, \bar{b}_I, b_T, s)$$
$$= \frac{1}{2^n} \sum_i \langle E^o_{i_I, i_I \oplus c_I} | E^o_{i_I, i_I \oplus c_I} \rangle \quad (3.12)$$

with $|E^o_{k,l}\rangle = \frac{1}{2^n} \sum_{i,j} (-1)^{i \cdot k} (-1)^{j \cdot l} |E_{i,j}\rangle$ (see section 3.8.1). The independence of $(i_I, j_T)$ and $(i_I, i_T, b_I, b_T, s)$ is used here exactly as in Eq. (3.11). The following lemma shows that the probability of an error syndrome $c$ (if the conjugate bases were used) equals the coefficients $d_c$ (calculated for the purifications in the basis actually in use $b_I$) when writing the purification of Eve's states using the basis $|\eta_c\rangle$.

**Lemma 3.3.1**

$$P(c_I \mid i_T, j_T, \bar{b}_I, b_T, s) = d^2_{c_I} . \quad (3.13)$$

**Proof 3.3.1** *See section 3.8.1.*

### 3.3.4 Bounds on Eve's Information

In this subsection we improve upon a result based on [BBB98]. Eve's information on a particular bit of the final key (even if all other bits of the final key are given to her) is bounded. We take into consideration the error-correction data that is given to Eve, and we do it more efficiently than in [BBB98], hence we obtain a much better threshold for the allowed error-rate.

Let us first discuss one-bit final key, defined to be the parity of substring of the input $i_I$. The substring is defined using a mask $v$, meaning that the secret key is $v \cdot i_I$ (so $v$ tells us the subset of bits whose parity is the final key). Bob first correct his errors using the error correction code data, hence he learns the string $i_I$ of Alice. Eve does not know $i_I$, but she learns the error correcting code $\mathcal{C}$ used by Alice and Bob as well as $v$ and the parity bits $\xi$ sent by Alice to help Bob correct the sequence he received. All the possible inputs $i_I$ that have the correct parities $\xi$ for the code $\mathcal{C}$ form a set denoted $\mathcal{C}_\xi$.

When the purification of Eve's state is given by $|\phi_i\rangle$ the density matrix is $\rho^i = |\phi_i\rangle\langle\phi_i|$. In order to guess the key $b = v \cdot i$, Eve must now distinguish between two *ensembles* of states: the ensemble of [equally likely, due to Eq.(3.7)] states $\rho^i$ with $i_I \in \mathcal{C}_\xi$ and key $b = v \cdot i_I = 0$, and the ensemble of (equally likely) states $\rho^i$ with $i_I \in \mathcal{C}_\xi$ and key $b = v \cdot i_I = 1$. For $b \in \{0, 1\}$ these ensembles are represented by the following density matrices:

$$\rho_b = \frac{1}{2^{n-(r+1)}} \sum_{\substack{i_I \in \mathcal{C}_\xi \\ i_I \cdot v = b}} \rho^i$$

(the lift-ups of the states really known to Eve) and Eve's goal is to distinguish between them. A good measure for their distinguishability is the optimal mutual information (known as the accessible information) that one could get if one needs to guess the bit $b$ by performing an optimal measurement to distinguish between $\rho_0$ and $\rho_0$ when the two are given with equal probability (half). We call this Shannon Distinguishability ($SD$) to emphasize that it is a distinguishability measure, and $SD \equiv \text{opt}\{I(\mathcal{A}_j; \mathcal{E}|i_T, j_T, b, s)\}$ where the optimization (maximization) is over all possible measurements.

In the same way that $v$ acts as a mask and the secret bit is $v \cdot i_I$, the error-correction data also acts as masks: the $r$ "parity-check strings" $v_1, v_2, \ldots v_r$, and the parities: $\{v_1 \cdot i_I, v_2 \cdot i_I, \ldots v_r \cdot i_I\}$ are given to Eve. Let us assume (WLG) that these parity-check strings are linearly independent. Eve also knows the parity of any linear combination of the $r$ parity strings, e.g., $(v_1 \oplus v_2) \cdot i_I$. As result, a total of $2^r$ parity strings and parity bits are known to Eve. Let us take $s$ to be an index running from 0 to $2^r - 1$, so we call the set of all these $2^r$ parity strings $S_\mathbf{s}$, and $v_s \in S_\mathbf{s}$ means that $v_s$ is in this set.

Let $\hat{v}$ be the minimum Hamming distance between $v$ and any (error correction) parity string $v_s$. [The minimal Hamming weight of $v \oplus v_s$ when the minimum is over all strings $v_s \in S_\mathbf{s}$]. Then, for Eve's purified states $|\phi_i\rangle = \sum_l (-1)^{i \cdot l} d_l |\hat{\eta}_l\rangle$, we obtain that

**Lemma 3.3.2** The Shannon distinguishability between the parity 0 and the parity 1 of the information bits over any PA string, $v$, is bounded above by the following inequality:

$$SD_v \leq \alpha + \frac{1}{\alpha} \sum_{|l| \geq \frac{\hat{v}}{2}} d_l^2 \ , \qquad (3.14)$$

where $\alpha$ is any positive constant, and $SD_v$ is the optimal mutual information that Eve can obtain regarding the parity bit defined by the string $v$ (given the test and unused bits).

The proof is given in section 3.9

This gives an upper bound for Eve's information about the bit defined by this privacy amplification string $v$. In order to prove security in case of $m$ bits in the final key, we start by proving security of each bit when we assume that Eve is given the ECC information and in addition, she is also given the values of all the other bits of the key. This is like using a code with $r + m - 1$ independent parity check strings, or like using less code words. Since $r$ does not appear in the above bound, replacing $r$ by $r + m - 1$ leaves the same result as before, $SD_v \leq \alpha + \frac{1}{\alpha} \sum_{|l| \geq \frac{\hat{v}}{2}} d_l^2$ , as a bound on Eve's information on (any) one bit of the final key Although looks identical to Eq.(3.14), there is a difference between the two bounds since the additional parity strings of the privacy amplification causes a decrease in $\hat{v}$, which is now the minimal Hamming distance between a particular parity string of the privacy amplification, and any parity string of the error correction together with the other privacy amplification strings (and their linear superpositions).

### 3.3.5 Eve's Information versus the Induced Disturbance

We have already shown in Eq.(3.13) that $P(c_I \mid i_T, j_T, \bar{b}_I, b_T, s) = d_{c_I}^2$. Thus,

$$SD_v \leq \alpha + \frac{1}{\alpha} \sum_{|c_I| \geq \frac{\hat{v}}{2}} P(c_I | i_T, j_T, \bar{b}_I, b_T, s) \ . \tag{3.15}$$

This equation bounds the information of Eve using the probability of the error syndromes in the other basis, and it completes the "information versus disturbance" result of our proof. Previous security proofs (for simpler attacks), such as [FGG97, BM97a, BBB98] are also based on various "information versus disturbance" arguments, since the non-classicality of QKD is manifested via such arguments.

The result is expressed using classical terms: Eve's information is bounded using the probability of error strings with large Hamming weight. If only error strings with low weight have non-zero probability, Eve's information becomes zero. Such a result is a "low weight" property and it resembles a similar result with this name which was derived by Yao [Yao95] for the security analysis of quantum oblivious transfer. Henceforth we no longer concern ourselves with the delicate issues of quantum mechanics.

From this point on we want to use standard information theory and probability notations. Shannon Distinguishability is the optimal mutual information between Eve's bits ($\mathcal{E}$) and Alice's $j^{th}$ bit ($\mathcal{A}_j$) (when all other PA bits are given together with the ECC data and test data). Therefore, $I(\mathcal{A}_j; \mathcal{E}|i_T, j_T, b_I, b_T, s) \leq \alpha + \frac{1}{\alpha} \sum_{|c_I| \geq \frac{\hat{v}}{2}} P(c_I|i_T, j_T, \bar{b}_I, b_T, s)$.

When summing over the $m$ bits of the key, the total information Eve receives on the final $m$-bit key is bounded by

$$\frac{I(\mathcal{A}; \mathcal{E}|i_T, j_T, b_I, b_T, s)}{m} \leq \alpha + \frac{1}{\alpha} \sum_{|c_I| \geq \frac{\hat{v}}{2}} P(c_I|i_T, j_T, \bar{b}_I, b_T, s) \tag{3.16}$$

as proven in section 3.10

If $\alpha = \sqrt{\sum_{|c_I| \geq \frac{\hat{v}}{2}} P(c_I|i_T, j_T, \bar{b}_I, b_T, s)}$, then

$$I(\mathcal{A}; \mathcal{E}|i_T, j_T, b_I, b_T, s) \leq 2m \sqrt{\sum_{|c_I| \geq \frac{\hat{v}}{2}} P(c_I|i_T, j_T, \bar{b}_I, b_T, s)},$$

however, to derive the security criterion we need not fix $\alpha$ yet.

## 3.4 Completing the Security Proof

In this section we analyze the attack on the test and information together (3.2). For these states, we bound a weighted average of Eve's information:

$$\sum_{i_T, c_T, b, s} P(\mathcal{T} = \text{pass}, i_T, c_T, b, s) \, I(\mathcal{A}; \mathcal{E}|i_T, c_T, b, s)$$

We show that the above bound is small and then we show that security is achieved. Note that $c_T$ replaces $j_T$ from this point forward (when $i_T$ is given). Recall that $c_T = i_T \oplus j_T$, so once $c_T$ is known $j_T$ is uniquely given.

### 3.4.1 Exponentially Small Bound on Eve's Information

We generalize here previous proofs [BMS96, BM97b, BBB98] that information on parity bits is exponentially small, to be applicable for the joint attack.

The maximum error rate that still passes the test is $p_{allowed}$ (or $p_a$). Also recall that $\mathcal{T}$ denotes the random variable for the test. Making use of Eq. 3.16 we get:

**Lemma 3.4.1**

$$\sum_{i_T, c_T} P(\mathcal{T} = \text{pass}, i_T, c_T|b, s) I(\mathcal{A}; \mathcal{E} \mid i_T, c_T, b, s)$$
$$\leq m \left\{ \alpha + \frac{1}{\alpha} P \left[ (|c_I| > \frac{\hat{v}}{2}) \wedge (\frac{|c_T|}{n} \leq p_a) \mid \bar{b}_I, b_T, s \right] \right\}$$

**Proof 3.4.1** *See appendix 3.8.2.*

For an $\epsilon$ (called earlier $\epsilon_{sec}$) such that $\hat{v} \geq 2n(p_a + \epsilon)$ we get the following bound:

$$\sum_{i_T, c_T} P(\mathcal{T} = \text{pass}, i_T, c_T | b, s) I(\mathcal{A}; \mathcal{E} \mid i_T, c_T, b, s)$$
$$\leq m \left\{ \alpha + \frac{1}{\alpha} P \left[ (\frac{|c_I|}{n} > p_a + \epsilon) \wedge (\frac{|c_T|}{n} \leq p_a) \mid \bar{b}_I, b_T, s \right] \right\}$$

Thus far, there is nothing that causes the bound on the right hand side to be a small number. The result above is true even if Eve is told in advance the bases of Alice and Bob (the string $b$), or if she is told in advance which are the test bits and which are the used bits (the string $s$), two cases in which Eve easily obtains full information.

Only Eve's lack of knowledge regarding the random $b$ and $s$ provides an exponentially small number at the right hand side. Since Eve must fix her attack *before* she knows the basis or order, we compute the average information for a fixed attack over all bases and orders. Averaging over $b$ means that we sum over all $b$'s and multiply each term by the constant $P(b) = 1/2^{2n}$. The averaging over $b$ removes the dependence on the particular basis:

**Lemma 3.4.2**

$$\sum_{i_T, c_T, b} P(\mathcal{T} = \text{pass}, i_T, c_T, b \mid s) I(\mathcal{A}; \mathcal{E} \mid i_T, c_T, b, s)$$
$$\leq m \left\{ \alpha + \frac{1}{\alpha} \frac{1}{2^{2n}} \sum_b P[(\frac{|c_I|}{n} > p_a + \epsilon) \wedge (\frac{|c_T|}{n} \leq p_a) | b, s] \right\}$$

**Proof 3.4.2** *Summing over $b$ i.e. $b_I, b_T$ is the same as summing over $\bar{b}_I, b_T$.*

By averaging over all values of the sample strings $s$ and basis choices $b$,

**Lemma 3.4.3**

$$\sum_{i_T, c_T, b, s} P(\mathcal{T} = \text{pass}, i_T, c_T, b, s) I(\mathcal{A}; \mathcal{E} \mid i_T, c_T, b, s)$$
$$\leq m \left\{ \alpha + \frac{1}{\alpha} P[(\frac{|c_I|}{n} > p_a + \epsilon) \wedge (\frac{|c_T|}{n} \leq p_a)] \right\}$$

**Proof 3.4.3** *This follows from the definition of the probability of an event: it is calculated by averaging over all values of $b$ and $s$.*

By assigning a value to the free parameter $\alpha$ we get:

**Lemma 3.4.4**

$$\sum_{i_T, c_T, b, s} P(\mathcal{T} = \text{pass}, i_T, c_T, b, s) I(\mathcal{A}; \mathcal{E} \mid i_T, c_T, b, s)$$
$$\leq 2m \sqrt{P[(\frac{|c_I|}{n} > p_a + \epsilon) \wedge (\frac{|c_T|}{n} \leq p_a)]}.$$

**Proof 3.4.4** *Let $h(p_a, \epsilon) \equiv P[(\frac{|c_I|}{n} > p_a + \epsilon) \wedge (\frac{|c_T|}{n} \leq p_a)]$ for short. The right-hand side of lemma 3.4.3 is $m \left\{ \alpha + \frac{1}{\alpha} h(p_a, \epsilon) \right\}$. Replacing $\alpha$ by the particular (and optimal) value $\sqrt{h(p_a, \epsilon)}$ leaves*

$$\sum_{i_T, c_T, b, s} P(\mathcal{T} = \text{pass}, i_T, c_T, b, s) I(\mathcal{A}; \mathcal{E} \mid i_T, c_T, b, s) \leq 2m \sqrt{h(p_a, \epsilon)}$$

The current bound can be dealt with the help of a random sampling theorem (Hoeffding's law of large numbers [Hoe63]). For a long string, the test bits and the information bits should have similar number of errors if the test is picked at random. The probability that they have different numbers of errors should go to zero exponentially fast as shown in the following lemma.

**Lemma 3.4.5** For any $\epsilon > 0$, $h(p_a, \epsilon) \leq e^{-\frac{1}{2} n \epsilon^2}$.

**Proof 3.4.5** *See section 3.8.3.*

As a Corollary we get

$$\sum_{i_T, c_T, b, s} P(\mathcal{T} = \text{pass}, i_T, c_T, b, s) I(\mathcal{A}; \mathcal{E} \mid i_T, c_T, b, s)$$

$$\leq 2m\sqrt{e^{-\frac{1}{2}n\epsilon^2}} = Ae^{-\beta n} \;,$$

with $A = 2m$ and $\beta = \epsilon^2/4$.

Using $I_{Eve} \equiv I(\mathcal{A}; \mathcal{E} | i_T, c_T, b, s)$ and the above bound we obtain the security criterion:

$$P\left[(\mathcal{T} = \text{pass}) \wedge (I_{Eve} \geq A_{\text{info}}\, e^{-\beta_{\text{info}} n})\right] \leq A_{\text{luck}}\, e^{-\beta_{\text{luck}} n} \quad (3.17)$$

with $A_{\text{info}} = A_{\text{luck}} = \sqrt{A}$ and $\beta_{\text{info}} = \beta_{\text{luck}} = \beta/2$. The above follows from standard probability and information theory as shown in section 3.7.

Note that Lemma 3.4.5 also provides the proof that once the test passes there are no more than $(p_a + \epsilon_{\text{rel}})n$ errors in the information string (except for exponentially small probability $e^{-\frac{1}{2}n\epsilon_{\text{rel}}^2}$). Thus $A_{rel} = 1$ and $\beta_{rel} = \epsilon_{rel}^2/2$, in the reliability criterion.

The above bound of Eve's information is exponentially small, but it assumes that we have a desired ECC. If we restrict ourselves to linear codes, then the properties required of the ECC are: (1) It can correct all the errors (except for exponentially small probability $e^{-\frac{1}{2}n\epsilon_{\text{rel}}^2}$; see Lemma 3.4.5) in the information string, and (2) The minimum distance of the code words in the span of the dual code and the PA strings (hence, the augmented dual code is of dimension $r + m$) should have a minimum distance of $|\hat{v}| \geq 2n(p_{allowed} + \epsilon_{\text{sec}})$. In fact, it can be shown that for random linear codes (RLC's), requirement (1) can be satisfied with only an exponentially small probability of error if the minimum distance is $\geq n(p_{allowed} + \epsilon_{\text{rel}})$. We show in section 3.11 that the two above-mentioned requirements can be satisfied and one can generate an $m$-bit secret key, if one picks an $(n, n - r)$ RLC, where $r$ and $m$ satisfy the following:

$$H_2(p_a + \epsilon_{\rm rel} + 1/n) < r/n$$

$$H_2(2p_a + 2\epsilon_{\rm sec}) + H_2(p_a + \epsilon_{\rm rel} + 1/n) < 1 - R_{secret},$$

where $R_{secret} \equiv m/n$. In the limit of large $n$ and $\epsilon$'s close to zero, $p_{allowed} < 7.56\%$ satisfies the bound and hence this is our threshold.

## 3.5 Summary

We proved the security of the Bennett-Brassard (BB84) protocol for quantum key distribution. Our proof is based on a novel information-versus-disturbance result, on the optimality of symmetric attacks, on laws of large numbers, and on various techniques that simplifies the analysis of the problem.

## 3.6 Security of BB84

In the paper we prove that used-bits-BB84 is secure. Let us now present the original BB84 protocol and prove, by reduction, that its security follows immediately from the security of the used-bits-BB84 protocol.

The differences between the protocols are only in the first part:

I. Creating the sifted key:

1. Alice and Bob choose a large integer $n \gg 1$, and a number $\delta_{\rm num}$, such that $1 \gg \delta_{\rm num} \gg 1/\sqrt{(2n)}$. The protocol uses $n'' = (4 + \delta_{\rm num})n$ bits.

2. Alice randomly selects two $n''$-bit strings, $b$ and $i$, which are then used to create qubits: The string $b$ determines the basis $0 \equiv z$, and $1 \equiv x$ of the qubits. The string $i$ determines the value (0 or 1) of each of the $n''$ qubits (in the appropriate bases).

3. Bob randomly selects an $n''$-bit string, $b'$, which determines Bob's later choice of bases for measuring each of the $n''$ qubits.

4. Alice generates $n''$ qubits according to her selection of $b$ and $i$, and sends them to Bob via a quantum communication channel.

5. After receiving the qubits, Bob measures in the basis determined by $b'$.

6. Alice and Bob publish the bases they used; this step should be performed only after Bob received all the qubits.

7. All qubits with different bases are discarded by Alice and Bob. Thus, Alice and Bob finally have $n' \approx n''/2$ bits for which they used the same bases. The $n'$-bit string would be identical for Alice and Bob if Eve and natural noise do not interfere.

8. Alice selects the first $2n$ bits from the $n'$-bit string, and the rest of the $n'$ bits are discarded. If $n' < 2n$ the protocol is aborted.

   We shall refer to the resulting $2n$-bit string as the sifted key.

The second part of the protocol is identical to the second part of the used-bits-BB84 protocol. To prove that BB84 is secure let us modify BB84 by a few steps in a way that each step can only be helpful to Eve, and the final protocol is the used-bits-BB84.

Recall that Alice and Bob choose their strings of basis $b$ and $b'$ in advance. Recall the the two strings are random. Thus, the first modification below has no influence at all on the security or the analysis of the BB84 protocol. Note that after the first modification Alice knows the un-used bits in advance. The second modification is done in a way that Eve can only gain, hence security of the resulting protocol provides the security of BB84. The third modification is only "cosmetic", in order to derive precisely the used-bits-BB84 protocol. This modification changes nothing in terms of Eve's ability.

- Let Bob have a quantum memory. Let Alice choose $b'$ instead of Bob at step 3. When Bob receives the qubits at step 5, let him keep the qubits in a memory, and tell Alice he received them. In step 6, let Alice announce $b'$ to Bob, and Bob measure in bases $b'$.

  Bob immediately knows which are the used and the un-used bits (as follows directly from announcing $b$ and $b'$). Steps 7 and 8 are now combined since Alice and Bob know all the un-used bits already, and they ignore them, to be left with $2n$ bits.

- Let Alice generate and send to Bob only the used bits in step 4, and let her ask Eve to send the un-used bits (by telling her which these are, and also the preparation data for the relevant subsets, that is—$b_{un-used}$ and $i_{un-used}$). Knowing which are the used bits, and knowing their bases $b_{un-used}$ and values $i_{un-used}$ can only help Eve in designing her attack $U'$.

  Since Bob never uses the values of the unused bits in the protocol (he only ignores them), he doesn't care if Eve doesn't provide him these bits or provide them to him without following Alice's preparation request.

  After Bob receives the used and unused bits, let him give Eve the unused qubits (without measuring them), and ask her to measure them in bases $b'_{un-used}$. Having these qubits can only help Eve in designing her optimal final measurement.

  Since Bob never use the values of the unused bits in the rest of the protocol, he doesn't care if Eve doesn't provide him these values correctly or at all.

- Since Alice and Bob never made any use of the unused bits, Eve could have them as part of her ancilla to start with, and Alice could just create $2n$ bits, send them to Bob, and then tell him the bases.

  The protocol obtained after this reduction, is a protocol in which Eve has full control on her qubits and on the unused qubits. Alice and Bob have control on the preparation and measurement of the used bits only. This is the used-bits BB84, for which we prove security in the text.

One important remark is that the exponentially small probability that $n' < 2n$ in Step 8 (so that the protocol is aborted due to insufficient number of bits in the sifted key) now becomes a probability that the reduction fails.

Another important remark is that the issue of high loss rate of qubits (e.g., due to losses in transmission or detection) can also be handled via the same reduction. Thus, our proof applies also to a more practical BB84 protocol where high losses are allowed.

By the way, another practical aspect is imperfect sources (in which the created states are not described by a two-level system). This subject is the issue of recent subtlety regarding the security of practical schemes [BMS, BLM00], and it is not discussed in this current work.

## 3.7 Satisfying the Security Criterion

So far we have not shown that the security criterion is satisfied by bounding the following:

$$\sum_{i_T, c_T, b, s} P(\mathcal{T} = pass, i_T, c_T, b, s) I(\mathcal{A}; \mathcal{E} | i_T, c_T, b, s) \leq e^{2(\alpha - \beta n)} \quad (3.18)$$

We now show that when the above bound is satisfied, as shown in the paper, then the security criterion is satisfied:

$$Prob(\text{Test Passes and } I_{Eve} \geq e^{\alpha-\beta n}) \leq e^{\alpha-\beta n} \quad (3.19)$$

Where $I_{Eve} \equiv I(\mathcal{A}; \mathcal{E}|i_T, c_T, b, s)$.

To show the above break the sum into the parts where Eve has large information and the part where she has small. Then standard bounding techniques are used:

$$\sum_{i_T, c_T, b, s} P(\mathcal{T} = pass, i_T, c_T, b, s) I(\mathcal{A}; \mathcal{E}|i_T, c_T, b, s)$$

$$= \sum_{\substack{i_T, c_T, b, s \\ s.t.\ I_{Eve} < I'}} P(\mathcal{T} = pass, i_T, c_T, b, s) I(\mathcal{A}; \mathcal{E}|i_T, c_T, b, s)$$

$$+ \sum_{\substack{i_T, c_T, b, s \\ s.t.\ I_{Eve} \geq I'}} P(\mathcal{T} = pass, i_T, c_T, b, s) I(\mathcal{A}; \mathcal{E}|i_T, c_T, b, s)$$

$$\geq \sum_{\substack{i_T, c_T, b, s \\ s.t.\ I_{Eve} \geq I'}} P(\mathcal{T} = pass, i_T, c_T, b, s) I(\mathcal{A}; \mathcal{E}|i_T, c_T, b, s)$$

$$\geq \left( \sum_{\substack{i_T, c_T, b, s \\ s.t.\ I_{Eve} \geq I'}} P(\mathcal{T} = pass, i_T, c_T, b, s) \right) I'$$

The above steps follow from non-negativity of probability and mutual information. We are really already done:

$$\left( \sum_{\substack{i_T, c_T, b, s \\ s.t.\ I_{Eve} \geq I'}} P(\mathcal{T} = pass, i_T, c_T, b, s) \right) I'$$

$$\leq \sum_{i_T, c_T, b, s} P(\mathcal{T} = pass, i_T, c_T, b, s) I(\mathcal{A}; \mathcal{E}|i_T, c_T, b, s)$$

So far $I'$ is a free parameter. We can set it to any value we like, namely

$$I' = \sqrt{\sum_{i_T, c_T, b, s} P(\mathcal{T} = pass, i_T, c_T, b, s) I(\mathcal{A}; \mathcal{E}|i_T, c_T, b, s)}$$

:

$$Prob(\text{Test Passes and } I_{Eve} \geq I')$$

$$= \sum_{\substack{i_T, c_T, b, s \\ s.t.\ I_{Eve} \geq I'}} P(\mathcal{T} = pass, i_T, c_T, b, s)$$

$$\leq \sqrt{\sum_{i_T, c_T, b, s} P(\mathcal{T} = pass, i_T, c_T, b, s) I(\mathcal{A}; \mathcal{E}|i_T, c_T, b, s)}$$

If we assume that $\sum_{i_T, c_T, b, s} P(\mathcal{T} = pass, i_T, c_T, b, s) I(\mathcal{A}; \mathcal{E}|i_T, c_T, b, s) \leq e^{2(\alpha - \beta n)}$ then we have:

$$Prob(\text{Test Passes and } I_{Eve} \geq e^{\alpha - \beta n}) \leq e^{\alpha - \beta n} \quad (3.20)$$

Thus, the bounds that we have shown satisfy the security criterion.

## 3.8 A Few Technical Lemmas

### 3.8.1 A Proof of Lemma 3.3.1

From now on we assume the attack 0/1 symmetric (obtained by 0/1 symmetrization) and consider only the information bits; we consequently drop the superscript $Sym$ and the subscript $I$ and write $E_{i,j}$ to mean $E^{Sym}_{i_I, j_I}$. Those terms depend on $i_T, j_T, b_I, b_T$ and $s$ where $b_I$ fixes the bases on the information bits and $b_T$ those on test bits. Had we expanded according the conjugate basis $\overline{b_I}$ on information bits ($b_T$ unchanged) namely the basis $|k\rangle^o$

$$|k\rangle^o = \frac{1}{2^n} \sum_i (-1)^{i \cdot k} |i\rangle$$

we would have obtained the terms $|E^o_{k,l}\rangle$ with

$$|E^o_{k,l}\rangle = \frac{1}{2^n} \sum_{i,j} (-1)^{i \cdot k} (-1)^{j \cdot l} |E_{i,j}\rangle$$

$$P(c_I \mid i_T, j_T, \overline{b_I}, b_T, s) = \frac{1}{2^n} \sum_k \langle E^o_{k, k \oplus c_I} | E^o_{k, k \oplus c_I}\rangle =$$

$$= \frac{1}{2^n} \sum_k \sum_{i,j} \sum_{i',j'} \frac{1}{2^{2n}} (-1)^{(i\oplus i')\cdot k \oplus (j\oplus j')\cdot(k\oplus c)} \langle E_{i,j}|E_{i',j'}\rangle$$

$$= \frac{1}{2^{3n}} \sum_{i,i',j,j'} \left(\sum_k (-1)^{k\cdot(i\oplus i'\oplus j\oplus j')}\right)(-1)^{c\cdot(j\oplus j')}\langle E_{i,j}|E_{i',j'}\rangle$$

The sum over $k$ is non zero only when $i \oplus i' = j \oplus j' \stackrel{\Delta}{=} h$,

and then it is $2^n$, so

$$= \frac{1}{2^{2n}} \sum_{i,j,h} (-1)^{c\cdot h} \langle E_{i,j}|E_{i\oplus h,j\oplus h}\rangle$$

$$= \langle \eta_c|\eta_c\rangle = d_c^2$$

where the last equalities are due to the calculation of the norm of $\eta$ in Eq. (3.9).

### 3.8.2 A Proof of Lemma 3.4.1

We first recall that $c_T \equiv i_T \oplus j_T$ and $c_I \equiv i_I \oplus j_I$. From equation

$$\frac{I(\mathcal{A};\mathcal{E}|i_T,j_T,b_I,b_T,s)}{m} \leq \alpha + \frac{1}{\alpha} \sum_{|c_I|\geq \frac{\hat{v}}{2}} P(c_I|i_T,j_T,\bar{b}_I,b_T,s)$$

and $P(\mathcal{T} = \text{pass}, i_T, j_T \mid b, s) = P(i_T, j_T \mid b, s)$ if $|i_T \oplus j_T|/n \leq p_a$ and $= 0$ otherwise, we deduce that

$$\frac{1}{m} \sum_{i_T,j_T} P(\mathcal{T} = \text{pass}, i_T, j_T \mid b, s) I(\mathcal{A};\mathcal{E} \mid i_T, j_T, b, s)$$

is bounded above by

$$\alpha + \frac{1}{\alpha} \sum_{\substack{|c_I|\geq \frac{\hat{v}}{2} \\ |i_T\oplus j_T|/n \leq p_a}} P(c_I \mid i_T, j_T, \bar{b}_I, b_T, s) P(i_T, j_T \mid b, s)$$

Now we show that $P(i_T, j_T \mid b_I, b_T, s) = P(i_T, j_T \mid \bar{b}_I, b_T, s)$.

$$P(i_T, j_T | b_T, b_I, s)$$
$$= \sum_{i_I, j_I} P(i_I, i_T | b_I, b_T, s) P(j_T, j_I | i_I, i_T, b_I, b_T, s)$$
$$= \sum_{i_I, j_I} \frac{1}{2^{2n}} P(j_T, j_I | i_I, i_T, b_I, b_T, s)$$
$$= \frac{1}{2^{2n}} \sum_{i_I, j_I} \langle E_{i_I, i_T, j_I, j_T} | E_{i_I, i_T, j_I, j_T} \rangle$$

now look at the other:

$$P(i_T, j_T | \bar{b}_I, b_T, s)$$
$$= \sum_{i_I, j_I} P(i_I, i_T | \bar{b}_I, b_T, s) P(j_T, j_I | i_I, i_T, \bar{b}_I, b_T, s)$$
$$= \sum_{i_I, j_I} \frac{1}{2^{2n}} P(j_T, j_I | i_I, i_T, \bar{b}_I, b_T, s)$$
$$= \frac{1}{2^{4n}} \sum_{\substack{i_I, j_I, k \\ l, m, n}} (-1)^{i_I \cdot (k \oplus m) + j_I \cdot (l \oplus n)} \langle E_{m, i_T, n, j_T} | E_{k, i_T, l, j_T} \rangle$$

Since, $P(i_I, i_T | \bar{b}_I, b_T, s) = P(i_I, i_T | b_I, b_T, s) = \frac{1}{2^{2n}}$ because Alice chooses all those parameters independently, the sums over $i_I$ and $j_I$ give delta functions (removing one factor of $1/2^{2n}$) and leave you with:

$$P(i_T, j_T | \bar{b}_I, b_T, s) = \frac{1}{2^{2n}} \sum_{k, l} \langle E_{k, i_T, l, j_T} | E_{k, i_T, l, j_T} \rangle$$
$$= P(i_T, j_T | b_T, b_I, s)$$

Using the above: $P(c_I \mid i_T, j_T, \bar{b}_I, b_T, s) P(i_T, j_T \mid b, s)$

$$= P(c_I \mid i_T, j_T, \bar{b}_I, b_T, s) P(i_T, j_T \mid \bar{b}_I, b_T, s)$$
$$= P(c_I, i_T, j_T \mid \bar{b}_I, b_T, s)$$

and the terms inside brackets sum up to

$$P\left[\left(|c_I| \geq \frac{\hat{v}}{2}\right) \wedge \left(\frac{|c_T|}{n} \leq p_a\right) \mid \bar{b}_I, b_T, s\right]$$

which proves the lemma.

### 3.8.3 A Proof of Lemma 3.4.5

Let
$$h(p_a, \epsilon) = \sum_b p(b) h_b(p_a, \epsilon)$$
with
$$h_b(p_a, \epsilon) = P\left[\left(\frac{|c_I|}{n} > p_a + \epsilon\right) \wedge \left(\frac{|c_T|}{n} \leq p_a\right) \mid b\right]$$

This $h_b(p_a, \epsilon)$ is the probability that the information bits have $\epsilon$ more than the allowed error rate, when the test bits have less than the allowed error rate averaged over all choices of test and information bits, for a particular basis $b$, and is given by

$$\sum_c P\left[\left(\frac{|c_I|}{n} > p_a + \epsilon\right) \wedge \left(\frac{|c_T|}{n} \leq p_a\right) \mid c, b\right] P(c \mid b)$$

where $c$ is over all possible error strings on all bits, test and information. Note that in principle $P(c \mid b)$, can be calculated but we shall soon see that there is no need for it.

Now we must note that

$$P\left[\left(\frac{|c_I|}{n} > p_a + \epsilon\right) \wedge \left(\frac{|c_T|}{n} \leq p_a\right) \mid c, b\right]$$

does not depend on the attack. And in fact, in the aforementioned expression, the basis $b$ is superfluous. Once the error string $c$ is fixed, the values $\frac{|c_I|}{n}$ and $\frac{|c_T|}{n}$ depend uniquely on the random string $s$. In fact $\frac{|c_I|}{n}$ is the average of a random sampling without replacement of $n$ bits taken amongst the $2n$ bits $c$ whose mean $\mu$ is $\frac{|c|}{2n}$. From Hoeffding [Hoe63] we know that

$$P\left[\frac{|c_I|}{n} - \mu \geq \frac{\epsilon}{2} \mid c, b\right] \leq e^{-\frac{1}{2}n\epsilon^2} \qquad (3.21)$$

By definition $|c| = |c_I| + |c_T|$ and so

$$\mu = \frac{|c|}{2n} = \frac{|c_I|}{2n} + \frac{|c_T|}{2n}$$

Replacing $\mu$ by its value in (3.21) and simplifying, equation (3.21) becomes

$$P\left[\frac{|c_I|}{n} \geq \frac{|c_T|}{n} + \epsilon \mid c, b\right] \leq e^{-\frac{1}{2}n\epsilon^2} \qquad (3.22)$$

Now, since

$$(\frac{|c_I|}{n} > p_a + \epsilon) \wedge (\frac{|c_T|}{n} \leq p_a) \implies \frac{|c_I|}{n} \geq \frac{|c_T|}{n} + \epsilon$$

we deduce from (3.22) that

$$P\left[\left(\frac{|c_I|}{n} > p_a + \epsilon\right) \wedge \left(\frac{|c_T|}{n} \leq p_a\right) \mid c, b\right] \leq e^{-\frac{1}{2}n\epsilon^2}$$

and consequently,

$$h_b(p_a, \epsilon) = P\left[\left(\frac{|c_I|}{n} > p_a + \epsilon\right) \wedge \left(\frac{|c_T|}{n} \leq p_a\right) \mid b\right] \leq e^{-\frac{1}{2}n\epsilon^2}$$

and

$$h(p_a, \epsilon) \leq e^{-\frac{1}{2}n\epsilon^2}$$

## 3.9 Eve's Information Versus the disturbance

In this appendix we do not prove Lemma 3.3.2 immediately. We prove it later on, in the second subsection (the tight bound). For simplicity of the presentation, we first prove another Lemma which leads to a loose bound (with an additional factor of $2^r$), for which the derivation is simpler. The bulk of the loose bound was derived in [BBB98], and the tight bound is an improvement over that derivation. The loose bound lead to a much worse threshold for $p_{allowed}$ (less than 1%, instead of 7.56% derived from the tight bound), and this is the motivation for deriving the tight bound. One can skip directly to the second subsection if desired.

Both the loose and the tight bound are derived using the fact that the Shannon distinguishability between the parity 0 density matrix, $\rho_0$, and the parity 1 density matrix, $\rho_1$, is bounded by the trace norm of $\rho_0 - \rho_1$, and using the fact that the one can easily calculate this trace-norm when the purified states are given by Eq. 3.8.

### 3.9.1 The Loose Bound (BBBGM)

We have already defined a purification of Eve's state: $|\phi_{i_I}\rangle = \sum_l (-1)^{i_I \cdot l}|\eta_l\rangle$ The density matrix for such a $|\phi_{i_I}\rangle$ is

$$\rho^{i_I} = |\phi_{i_I}\rangle\langle\phi_{i_I}| = \sum_{l,l'}(-1)^{i_I \cdot (l \oplus l')} d_l d_{l'} |\hat{\eta}_l\rangle\langle\hat{\eta}_{l'}| \tag{3.23}$$

Recall that the final key is computed as $v \cdot i_I$. Eve does not know $i_I$, but she knows from the announced syndrome that $i_I$ is in the coset $\mathcal{C}_\xi$ for $\xi \equiv \xi^{Alice}$. Hence, in order to know the key, Eve must distinguish between the states $i_I = i_\xi \oplus c$ in $\mathcal{C}_\xi$ that give parity zero and the states $i_I = i_\xi \oplus c$ in $\mathcal{C}_\xi$ that give parity one. For $b \in \{0, 1\}$ the reduced density matrix is

$$\begin{aligned}\rho_b &= \frac{1}{2^{n-(r+1)}} \sum_{\substack{c \in \mathcal{C} \\ v(i_\xi \oplus c) = b}} \rho^{i_\xi \oplus c} = \\ &= \frac{1}{2^{n-(r+1)}} \sum_{\substack{c \in \mathcal{C} \\ v(i_\xi \oplus c) = b}} \sum_{l,l'} (-1)^{(i_\xi \oplus c)(l \oplus l')} d_l d_{l'} |\hat{\eta}_l\rangle\langle\hat{\eta}_{l'}|\end{aligned}$$

where the sum is over $c$ that satisfy both the condition of being a code word, and the condition of leading to the particular parity $b$ for the PA.

**Lemma.** Let $\mathcal{C}$ be any linear code in $\{0, 1\}^n$ and $a \in \{0, 1\}^n$ be such that $a \notin \mathcal{C}^\perp$ then

$$\sum_{c \in \mathcal{C}} (-1)^{c \cdot a} = 0 \tag{3.24}$$

**Proof.**— Let $\{w_1, \ldots, w_k\}$ be a basis of $\mathcal{C}$. Define $t \in \{0, 1\}^k$ by $t_\alpha = w_\alpha \cdot a$, $1 \leq \alpha \leq k$; $a \notin \mathcal{C}^\perp$ means that $t$ is not the zero string. Let now $h : \{0, 1\}^k \to \mathcal{C}$ be defined by $h(s) = \sum_{1 \leq \alpha \leq k} s_\alpha w_\alpha$; then $h(s) \cdot a = \sum s_\alpha w_\alpha \cdot a = \sum s_\alpha t_\alpha = s \cdot t$ and so

$$\sum_{c \in \mathcal{C}} (-1)^{c \cdot a} = \sum_s (-1)^{h(s) \cdot a} = \sum_s (-1)^{s \cdot t} = 0$$

**Lemma** The Shannon distinguishability between the parity 0 and the parity 1 of the information bits over any PA string, $v$, is bounded above by the following inequality:

$$SD_v \leq 2^r \left[\alpha + \frac{1}{\alpha} \sum_{|l| \geq \frac{\hat{v}}{2}} d_l^2 \right], \tag{3.25}$$

where $\hat{v}$ is the minimum weight of $v \oplus v_s$ for any $v_s \in S_\mathbf{s}$, and $\alpha$ is any positive constant.

**Proof.—** The Shannon distinguishability between the parity 0 and the parity 1 is bounded by the trace norm of $\rho_0 - \rho_1$ ([BBB98, FG99]). Let us calculate the required bound:

$$\begin{aligned}
\rho_0 - \rho_1 &= \frac{1}{2^{n-(r+1)}} \sum_{c \in \mathcal{C}} (-1)^{(i_\xi \oplus c)v} \sum_{l,l'} (-1)^{(i_\xi \oplus c)(l \oplus l')} d_l d_{l'} |\hat{\eta}_l\rangle\langle\hat{\eta}_{l'}| \\
&= \frac{1}{2^{n-(r+1)}} \sum_{l,l'} \left( \sum_{c \in \mathcal{C}} (-1)^{(i_\xi \oplus c)(l \oplus l' \oplus v)} \right) d_l d_{l'} |\hat{\eta}_l\rangle\langle\hat{\eta}_{l'}| \\
&= \frac{1}{2^{n-(r+1)}} \sum_{l,l'} (-1)^{i_\xi (l \oplus l' \oplus v)} \left( \sum_{c \in \mathcal{C}} (-1)^{c(l \oplus l' \oplus v)} \right) d_l d_{l'} |\hat{\eta}_l\rangle\langle\hat{\eta}_{l'}|
\end{aligned}$$

From equation (3.24) we know the sum over $\mathcal{C}$ is zero except when $l \oplus l' \oplus v \in \mathcal{C}^\perp = S_\mathbf{s}$, i.e. when $l' = l \oplus v \oplus v_\mathbf{s}$ for some $v_\mathbf{s} \in S_\mathbf{s}$. As a consequence:

$$\rho_0 - \rho_1 = 2 \sum_{v_s \in S_\mathbf{s}} (-1)^{i_\xi \cdot v_s} \sum_l d_l d_{l \oplus v \oplus v_s} |\hat{\eta}_l\rangle\langle\hat{\eta}_{l \oplus v \oplus v_s}|$$

The trace norm of this matrix serves as a bound on the information Eve receives.

$$SD_v \leq \frac{1}{2} Tr|\rho_0 - \rho_1|$$

Using the above and making use of the triangle inequality for the Trace norm, the following is obtained:

$$
\begin{aligned}
SD_v &\leq Tr\Big| \sum_{v_s \in S_\mathbf{s}} (-1)^{i\xi \cdot v_s} \sum_l d_l d_{l \oplus v \oplus v_s} |\hat{\eta}_m\rangle\langle\hat{\eta}_{m \oplus v \oplus v_s}| \Big| \\
&= \frac{1}{2} Tr\Big| \sum_{v_s \in S_\mathbf{s}} (-1)^{i\xi \cdot v_s} \sum_l d_l d_{l \oplus v \oplus v_s} (|\hat{\eta}_l\rangle\langle\hat{\eta}_{l \oplus v \oplus v_s}| + |\hat{\eta}_{l \oplus v \oplus v_s}\rangle\langle\hat{\eta}_l|) \Big| \\
&\leq \sum_{v_s \in S_\mathbf{s}} \sum_l d_l d_{l \oplus v \oplus v_s} \Big( \frac{1}{2} Tr\big| (|\hat{\eta}_l\rangle\langle\hat{\eta}_{l \oplus v \oplus v_s}| + |\hat{\eta}_{l \oplus v \oplus v_s}\rangle\langle\hat{\eta}_l|) \big| \Big) \\
&= \sum_{v_s \in S_\mathbf{s}} \sum_l d_l d_{l \oplus v \oplus v_s}
\end{aligned}
$$

Now we will concern ourselves with bounding each of the terms $\sum_l d_l d_{l \oplus w_s}$, where $w_s = v \oplus v_s$.

$$
\begin{aligned}
\sum_l d_l d_{l \oplus w_s} &= \sum_{|l| > \frac{|w_s|}{2}} d_l d_{l \oplus w_s} + \sum_{|l| \leq \frac{|w_s|}{2}} d_l d_{l \oplus w_s} \\
&= \sum_{|l| > \frac{|w_s|}{2}} d_l d_{l \oplus w_s} + \sum_{|l' \oplus w_s| \leq \frac{|w_s|}{2}} d_{l' \oplus w_s} d_{l'}
\end{aligned}
$$

If $|l' \oplus w_s| \leq \frac{|w_s|}{2}$ then $|w_s| = |l' \oplus w_s \oplus l'| \leq |l' \oplus w_s| + |l'| \leq \frac{|w_s|}{2} + |l'|$ and so

$|l'| \geq \frac{|w_s|}{2}$. Therefore,

$$
\begin{aligned}
\sum_{|l|>\frac{|w_s|}{2}} d_l d_{l\oplus w_s} + \sum_{|l'\oplus w_s|\leq \frac{|w_s|}{2}} d_{l'\oplus w_s} d_{l'} &\leq \sum_{|l|\geq \frac{|w_s|}{2}} d_l d_{l\oplus w_s} + \sum_{|l'|\geq \frac{|w_s|}{2}} d_{l'\oplus w_s} d_{l'} \\
&= 2 \sum_{|l|\geq \frac{|w_s|}{2}} d_l d_{l\oplus w_s} \\
&= \frac{1}{\alpha} \sum_{|l|\geq \frac{|w_s|}{2}} 2 d_l (\alpha d_{l\oplus w_s}) \\
&\leq \frac{1}{\alpha} \sum_{|l|\geq \frac{|w_s|}{2}} [d_l^2 + \alpha^2 d_{l\oplus w_s}^2] \\
&= \alpha \sum_{|l|\geq \frac{|w_s|}{2}} d_{l\oplus w_s}^2 + \frac{1}{\alpha} \sum_{|l|\geq \frac{|w_s|}{2}} d_l^2
\end{aligned}
$$

where the last three steps are true for any real $\alpha$, and real $d_l, d_{l\oplus w_s}$.

Due to the fact that the $d_l^2$ form a probability distribution, any sum of them is less than or equal to unity.

$$
\begin{aligned}
\sum_l d_l d_{l\oplus w_s} &\leq \alpha + \frac{1}{\alpha} \sum_{|l|\geq \frac{|w_s|}{2}} d_l^2 \\
&\leq \alpha + \frac{1}{\alpha} \sum_{|l|\geq \frac{\hat{v}}{2}} d_l^2
\end{aligned}
$$

where $\hat{v} = \min_{v_s} |v \oplus v_s|$ (remember that $w_s = v \oplus v_s$). Summing over all $v_s \in S_\mathbf{s}$ now leaves:

$$
SD_v \leq 2^r \left[ \alpha + \frac{1}{\alpha} \sum_{|l|\geq \frac{\hat{v}}{2}} d_l^2 \right] \tag{3.26}
$$

**Q**ED

The BBBGM result gives an upper bound for Eve's information about the bit defined by this privacy amplification string $v$. To prove security in case of $m$ bits in the final key, we prove security of each bit as follows: for each bit in the key we assume that Eve is given the ECC information and in addition, she is also given all the other bits in the key. This is like using a code with more parity check strings $2^{r+m-1}$ (or less code words), hence the previous result holds with

$$SD_v \leq 2^{r+m-1} \left[ \alpha + \frac{1}{\alpha} \sum_{|l| \geq \frac{\hat{v}}{2}} d_l^2 \right]. \tag{3.27}$$

Following the proof of the above Lemma, one can see that it is not a tight bound since we sum over $2^r$ terms while most of them are much smaller than the term (terms) with the minimal $\hat{v}$.

### 3.9.2 Eve's Information on One Bit – Tight Bound

We now show an improved technique, by defining a basis for the purification of the code words (instead of a basis for all the purification).

We will now make a finer analysis of Eve's state after she learns the parity matrix and the syndrome $\xi = \xi^{Alice}$. We start again from the equality:

$$|\phi_{i_I}\rangle = \sum_l (-1)^{i_I \cdot l} |\eta_l\rangle \tag{3.28}$$

First, any $l \in \{0,1\}^n$ has a unique representation $l = m \oplus n$ with $m \in S_{\mathbf{s}}^c$ and $n \in S_{\mathbf{s}}$. Next, for any $i_I \in \mathcal{C}_\xi$ we have $i_I = i_\xi \oplus c$ for some $c \in \mathcal{C}$ and thus for any $n \in S_{\mathbf{s}}$ we get $i_I \cdot n = (i_\xi \oplus c) \cdot n = i_\xi \cdot n$ [because $n \in S_{\mathbf{s}} = \mathcal{C}^\perp$]. Putting those two remarks together we get:

$$\begin{aligned}
|\phi_{i_I}\rangle &= \sum_{m\in S_{\mathbf{s}}^c}\sum_{n\in S_{\mathbf{s}}}(-1)^{i_I\cdot(m\oplus n)}|\eta_{m\oplus n}\rangle \\
&= \sum_{m\in S_{\mathbf{s}}^c}(-1)^{i_I\cdot m}\sum_{n\in S_{\mathbf{s}}}(-1)^{i_I\cdot n}|\eta_{m\oplus n}\rangle \\
&= \sum_{m\in S_{\mathbf{s}}^c}(-1)^{i_I\cdot m}\sum_{n\in S_{\mathbf{s}}}(-1)^{i_\xi\cdot n}|\eta_{m\oplus n}\rangle \\
&= \sum_{m\in S_{\mathbf{s}}^c}(-1)^{i_I m}|\eta'_m\rangle
\end{aligned}$$

where $\eta'_m$ is of course defined for each $m \in S_{\mathbf{s}}$ by

$$|\eta'_m\rangle = \sum_{n\in S_{\mathbf{s}}}(-1)^{i_\xi\cdot n}|\eta_{m\oplus n}\rangle \tag{3.29}$$

Now, since $\langle\eta_{m_1\oplus n_1}|\eta_{m_2\oplus n_2}\rangle = 0$ except when $m_1\oplus n_1 = m_2\oplus n_2$, which implies $m_1 = m_2$, the $\eta_m$'s are orthogonal. If $d'_m$ is the length of $\eta'_m$, we can then write

$$\eta'_m = d'_m \hat{\eta}'_m$$

with the $\hat{\eta}'_m$'s normalized and orthogonal and

$$d'^2_m = \sum_{n\in S_{\mathbf{s}}} d^2_{m\oplus n}$$

and the density matrix for $|\phi_{i_I}\rangle$ reduces to:

$$\begin{aligned}
\rho^{i_I} &= |\phi_{i_I}\rangle\langle\phi_{i_I}| = \\
&= \sum_{m,m'\in S_{\mathbf{s}}^c}(-1)^{i_I(m\oplus m')}d'_m d'_{m'}|\hat{\eta}'_m\rangle\langle\hat{\eta}'_{m'}|
\end{aligned}$$

Recall that the final key is computed as $b = v\cdot i_I$. Of course, Eve does not know $i_I$, but she knows from the announced syndrome $\xi = \xi^{Alice}$ that $i_I \in \mathcal{C}_\xi = \{i_\xi \oplus c \mid c \in \mathcal{C}\}$ and wants to determine $b$. For $b \in \{0,1\}$ the reduced density matrix is

$$\begin{aligned}
\rho_b &= \frac{1}{2^{n-(r+1)}}\sum_{\substack{c\in\mathcal{C}\\(i_\xi\oplus c)v=b}}\rho^{i_\xi\oplus c} = \\
&= \frac{1}{2^{n-(r+1)}}\sum_{\substack{c\in\mathcal{C}\\(i_\xi\oplus c)v=b}}\sum_{m,m'\in S_{\mathbf{s}}^c}(-1)^{(i_\xi\oplus c)(m\oplus m')}d'_m d'_{m'}|\hat{\eta}'_m\rangle\langle\hat{\eta}'_{m'}|
\end{aligned}$$

**Lemma 3.9.1** 3.3.2 The Shannon distinguishability between the parity 0 and the parity 1 of the information bits over any PA string, $v$, is bounded above by the following inequality:

$$SD_v \leq \alpha + \frac{1}{\alpha} \sum_{|l| \geq \frac{\hat{v}}{2}} d_l^2 , \qquad (3.30)$$

where $\hat{v}$ is the minimum weight of $v \oplus v_s$ for any $v_s \in S_\mathbf{s}$, and $\alpha$ is any positive constant.

**P**roof: The Shannon distinguishability between the parity 0 and the parity 1 is bounded by the trace norm of $\rho_0 - \rho_1$:

$$\begin{aligned}
\rho_0 - \rho_1 &= \frac{1}{2^{n-(r+1)}} \sum_{c \in \mathcal{C}} (-1)^{(i_\xi \oplus c)v} \sum_{m,m' \in S_\mathbf{s}^c} (-1)^{(i_\xi \oplus c)(m \oplus m')} d'_m d'_{m'} |\hat{\eta}'_m\rangle \langle \hat{\eta}'_{m'}| \\
&= \frac{1}{2^{n-(r+1)}} \sum_{m,m' \in S_\mathbf{s}^c} \left( \sum_{c \in \mathcal{C}} (-1)^{(i_\xi \oplus c)(m \oplus m' \oplus v)} \right) d'_m d'_{m'} |\hat{\eta}'_m\rangle \langle \hat{\eta}'_{m'}| \\
&= \frac{1}{2^{n-(r+1)}} \sum_{m,m' \in S_\mathbf{s}^c} (-1)^{i_\xi(m \oplus m' \oplus v)} \left( \sum_{c \in \mathcal{C}} (-1)^{c \cdot (m \oplus m' \oplus v)} \right) d'_m d'_{m'} |\hat{\eta}'_m\rangle \langle \hat{\eta}'_{m'}|
\end{aligned}$$

Applying equality (3.24) the sum indexed by $c$ is zero except when $m \oplus m' \oplus v \in \mathcal{C}^\perp = S_\mathbf{s}$. But $m \oplus m' \oplus v \in S_\mathbf{s}^c$ because $m, m'$ and $v \in S_\mathbf{s}^c$. This implies $m \oplus m' \oplus v \in S_\mathbf{s} \cap S_\mathbf{s}^c = \{0\}$ and thus $m' = m \oplus v$. Of course, with $m \oplus m' \oplus v = 0$, the sum indexed by $c$ is $2^k = 2^{n-r}$ and the coefficient $(-1)^{i_\xi(m \oplus m' \oplus v)}$ is 1. Therefore $\rho_0 - \rho_1$ takes the very simple form:

$$\rho_0 - \rho_1 = 2 \sum_{m \in S_\mathbf{s}^c} d'_m d'_{m \oplus v} |\hat{\eta}'_m\rangle \langle \hat{\eta}'_{m \oplus v}|$$

As usual, the trace norm of this matrix serves as a bound on the information Eve receives. It is

$$SD_v \leq \frac{1}{2} Tr|\rho_0 - \rho_1|$$

First note that $v$ is in $S_{\mathbf{s}}^c$ and $S_{\mathbf{s}}^c$ is closed under addition. Further the set $S_{\mathbf{s}}^c$ is the same as $v \oplus S_{\mathbf{s}}^c$. Then the set defined by $m \in S_{\mathbf{s}}^c$ is identical to the set $m \oplus v \in S_{\mathbf{s}}^c$. We will use this identity to obtain the following inequality:

$$
\begin{aligned}
SD_v &\leq Tr| \sum_{m \in S_{\mathbf{s}}^c} d'_m d'_{m \oplus v} |\hat{\eta}'_m\rangle\langle\hat{\eta}'_{m \oplus v}|| \\
&= \frac{1}{2} Tr| \sum_{m \in S_{\mathbf{s}}^c} d'_m d'_{m \oplus v} |\hat{\eta}'_m\rangle\langle\hat{\eta}'_{m \oplus v}| + \sum_{m \oplus v \in S_{\mathbf{s}}^c} d'_m d'_{m \oplus v} |\hat{\eta}'_m\rangle\langle\hat{\eta}'_{m \oplus v}|| \\
&= \frac{1}{2} Tr| \sum_{m \in S_{\mathbf{s}}^c} d'_m d'_{m \oplus v} |\hat{\eta}'_m\rangle\langle\hat{\eta}'_{m \oplus v}| + \sum_{m \in S_{\mathbf{s}}^c} d'_{m \oplus v} d'_m |\hat{\eta}'_{m \oplus v}\rangle\langle\hat{\eta}'_m|| \\
&= \frac{1}{2} Tr| \sum_{m \in S_{\mathbf{s}}^c} d'_m d'_{m \oplus v} (|\hat{\eta}'_m\rangle\langle\hat{\eta}'_{m \oplus v}| + |\hat{\eta}'_{m \oplus v}\rangle\langle\hat{\eta}'_m|)| \\
&\leq \frac{1}{2} \sum_{m \in S_{\mathbf{s}}^c} d'_m d'_{m \oplus v} Tr ||\hat{\eta}'_m\rangle\langle\hat{\eta}'_{m \oplus v}| + |\hat{\eta}'_{m \oplus v}\rangle\langle\hat{\eta}'_m|| \\
&= \sum_{m \in S_{\mathbf{s}}^c} d'_m d'_{m \oplus v}
\end{aligned}
$$

Now we wish to give a bound in terms of the original $d$'s. Let us define

$$\Gamma_{\hat{v}} = \{m \in S_{\mathbf{s}}^c \mid |m \oplus n| \geq \hat{v}/2 \; \forall n \in S_{\mathbf{s}}\}$$

where $\hat{v}$ was defined in the statement of the lemma. We claim that for any $m \in S_{\mathbf{s}}^c$, either $m \in \Gamma_{\hat{v}}$ or $m \oplus v \in \Gamma_{\hat{v}}$. Indeed, if it were not so, there would be $n_1 \in S_{\mathbf{s}}$ and $n_2 \in S_{\mathbf{s}}$ such that $|m \oplus n_1| < \hat{v}/2$ and $|m \oplus v \oplus n_2| < \hat{v}/2$. But then $|n_1 \oplus n_2 \oplus v| = |m \oplus n_1 \oplus m \oplus n_2 \oplus v| < \hat{v}/2 + \hat{v}/2$ which, since $n_1 \oplus n_2 \in S_{\mathbf{s}}$, contradicts the definition of $\hat{v}$.

We now use the claim to break up the sum bounding $SD_v$ and prove the lemma.

$$
\begin{aligned}
SD_v &\leq \sum_{m \in S_{\mathbf{s}}^c} d'_m d'_{m \oplus v} \\
&\leq \left( \sum_{\substack{m \in S_{\mathbf{s}}^c \\ m \in \Gamma_{\hat{v}}}} d'_m d'_{m \oplus v} + \sum_{\substack{m \in S_{\mathbf{s}}^c \\ m \oplus v \in \Gamma_{\hat{v}}}} d'_m d'_{m \oplus v} \right) \\
&= \left( \sum_{\substack{m \in S_{\mathbf{s}}^c \\ m \in \Gamma_{\hat{v}}}} d'_m d'_{m \oplus v} + \sum_{\substack{m \oplus v \in S_{\mathbf{s}}^c \\ m \in \Gamma_{\hat{v}}}} d'_m d'_{m \oplus v} \right) \\
&= 2 \sum_{m \in \Gamma_{\hat{v}}} d'_m d'_{m \oplus v} \\
&= \frac{2}{\alpha} \sum_{m \in \Gamma_{\hat{v}}} (\alpha d'_{m \oplus v})(d'_m) \\
&\leq \alpha \sum_{m \in \Gamma_{\hat{v}}} {d'_{m \oplus v}}^2 + \frac{1}{\alpha} \sum_{m \in \Gamma_{\hat{v}}} {d'_m}^2 \\
&\leq \alpha \sum_{\substack{m \in S_{\mathbf{s}}^c, n \in S_{\mathbf{s}} \\ |m \oplus n| \geq \frac{\hat{v}}{2}}} d^2_{m \oplus n \oplus v} + \frac{1}{\alpha} \sum_{\substack{m \in S_{\mathbf{s}}^c, n \in S_{\mathbf{s}} \\ |m \oplus n| \geq \frac{\hat{v}}{2}}} d^2_{m \oplus n} \\
&= \alpha \sum_{|l| \geq \frac{\hat{v}}{2}} d^2_{l \oplus v} + \frac{1}{\alpha} \sum_{|l| \geq \frac{\hat{v}}{2}} d^2_l
\end{aligned}
$$

Due to the fact that the $d^2_l$ form a probability distribution, any sum of them is less than or equal to unity.

$$SD_v \leq \alpha + \frac{1}{\alpha} \sum_{|l| \geq \frac{\hat{v}}{2}} d^2_l \tag{3.31}$$

**Q**ED

Note that the number of parity check strings $r$ doesn't appear in the final expression, and this might seem surprising. However, it does appear there implicitly, since increasing $r$ by one increases the number of parity check strings from $2^r - 1$ to $2^{r+1} - 1$, hence potentially decreases $\hat{v}$.

## 3.10 Security of the Entire Key

We give a proof that bit-wise security implies security of the entire string. This is first shown classically, and then making use of Shannon Distinguishability, the same bound holds for quantum bits.

### 3.10.1 Classical Information Theory

**Lemma 3.10.1** *For independent random variables $\mathcal{A}_i$, $i \in (1, 2, \ldots, m)$ and random variable $\mathcal{E}$*

$$I(\mathcal{A}_i; \mathcal{E}|\mathcal{A}_1, \mathcal{A}_2, \ldots, \mathcal{A}_{i-1}) \leq I(\mathcal{A}_i; \mathcal{E}|\mathcal{A}_1, \mathcal{A}_2, \ldots, \mathcal{A}_{i-1}, \mathcal{A}_{i+1}, \ldots, \mathcal{A}_m)$$

**P**roof: First we define a few sets:

$$\begin{aligned}
\mathcal{A}_{<i} &\equiv \{\mathcal{A}_1, \mathcal{A}_2, \ldots, \mathcal{A}_{i-1}\} \\
\mathcal{A}_{>i} &\equiv \{\mathcal{A}_{i+1}, \mathcal{A}_{i+2}, \ldots, \mathcal{A}_m\} \\
\mathcal{A}_{\neq i} &\equiv \{\mathcal{A}_1, \mathcal{A}_2, \ldots, \mathcal{A}_{i-1}, \mathcal{A}_{i+1}, \ldots, \mathcal{A}_m\}
\end{aligned}$$

Of course, $\mathcal{A}_{\neq i} = \mathcal{A}_{<i} \cup \mathcal{A}_{>i}$. In this notation the lemma says: $I(\mathcal{A}_i; \mathcal{E}|\mathcal{A}_{\neq i}) - I(\mathcal{A}_i; \mathcal{E}|\mathcal{A}_{<i}) \geq 0$

$$\begin{aligned}
&I(\mathcal{A}_i; \mathcal{E}|\mathcal{A}_{\neq i}) - I(\mathcal{A}_i; \mathcal{E}|\mathcal{A}_{<i}) \\
&= H(\mathcal{A}_i|\mathcal{A}_{\neq i}) - H(\mathcal{A}_i|\mathcal{E}, \mathcal{A}_{\neq i}) - H(\mathcal{A}_i|\mathcal{A}_{<i}) + H(\mathcal{A}_i|\mathcal{E}, \mathcal{A}_{<i}) \\
&= (H(\mathcal{A}_i|\mathcal{E}, \mathcal{A}_{<i}) - H(\mathcal{A}_i|\mathcal{E}, \mathcal{A}_{\neq i})) - (H(\mathcal{A}_i|\mathcal{A}_{<i}) - H(\mathcal{A}_i|\mathcal{A}_{\neq i})) \\
&= (H(\mathcal{A}_i|\mathcal{E}, \mathcal{A}_{<i}) - H(\mathcal{A}_i|\mathcal{E}, \mathcal{A}_{<i}, \mathcal{A}_{>i})) \\
&\quad - (H(\mathcal{A}_i|\mathcal{A}_{<i}) - H(\mathcal{A}_i|\mathcal{A}_{<i}, \mathcal{A}_{>i})) \\
&= I(\mathcal{A}_i; \mathcal{A}_{>i}|\mathcal{E}, \mathcal{A}_{<i}) - I(\mathcal{A}_i; \mathcal{A}_{>i}|\mathcal{A}_{<i})
\end{aligned}$$

Due to the independence of $\mathcal{A}_i$, $I(\mathcal{A}_i; \mathcal{A}_{>i}|\mathcal{A}_{<i}) = 0$. Since any information is non-negative, $I(\mathcal{A}_i; \mathcal{A}_{>i}|\mathcal{E}, \mathcal{A}_{<i}) \geq 0$. Hence $I(\mathcal{A}_i; \mathcal{A}_{>i}|\mathcal{E}, \mathcal{A}_{<i}) - I(\mathcal{A}_i; \mathcal{A}_{>i}|\mathcal{A}_{<i}) \geq 0$
**Q**ED

**Theorem 3.10.1** *For independent random variables $\mathcal{A}_i$, $i \in (1, 2, \ldots, m)$ and random variable $\mathcal{E}$*
$I(\mathcal{A}_1, \mathcal{A}_2, \ldots, \mathcal{A}_m; \mathcal{E}) \leq m \, max_i(I(\mathcal{A}_i; \mathcal{E}|\mathcal{A}_1, \mathcal{A}_2, \ldots, \mathcal{A}_{i-1}, \mathcal{A}_{i+1}, \ldots, \mathcal{A}_m))$

**P**roof: Here we simply apply the chain rule for mutual information[CT91] and we then apply the above lemma. We will use the same notions introduced in the previous proof.

$$\begin{aligned}
I(\mathcal{A}_1, \mathcal{A}_2, \ldots, \mathcal{A}_m; \mathcal{E}) &= \sum_i I(\mathcal{A}_i; \mathcal{E}|\mathcal{A}_{<i}) \\
&\leq \sum_k I(\mathcal{A}_k; \mathcal{E}|\mathcal{A}_{\neq k}) \\
&\leq \sum_k max_i(I(\mathcal{A}_i; \mathcal{E}|\mathcal{A}_{\neq i})) \\
&= m \, max_i(I(\mathcal{A}_i; \mathcal{E}|\mathcal{A}_{\neq i}))
\end{aligned}$$

**Q**ED

**Lemma 3.10.2** *For independent random variables $\mathcal{A}_i$, $i \in (1, 2, \ldots, m)$ and random variable $\mathcal{E}$*
$I(\mathcal{A}_1, \mathcal{A}_2, \ldots, \mathcal{A}_m; \mathcal{E}) \leq m \, max_{i, a_{\neq i}}(I(\mathcal{A}_i; \mathcal{E}|\mathcal{A}_{\neq i} = a_{\neq i}))$. *Where $a_{\neq i}$ is a set of outcomes for all $\mathcal{A}$ except $i$.*

**Proof:** We must simply prove $I(\mathcal{A}_i; \mathcal{E}|\mathcal{A}_{\neq i}) \leq max_{a_{\neq i}} I(\mathcal{A}_i; \mathcal{E}|\mathcal{A}_{\neq i} = a_{\neq i})$ and then apply the previous theorem.

$$\begin{aligned}
I(\mathcal{A}_i; \mathcal{E}|\mathcal{A}_{\neq i}) &= \sum_{a_{\neq i}} P(\mathcal{A}_{\neq i} = a_{\neq i}) I(\mathcal{A}_i; \mathcal{E}|\mathcal{A}_{\neq i} = a_{\neq i}) \\
&\leq \sum_{a_{\neq i}} P(\mathcal{A}_{\neq i} = a_{\neq i}) max_{a'_{\neq i}} I(\mathcal{A}_i; \mathcal{E}|\mathcal{A}_{\neq i} = a'_{\neq i}) \\
&= max_{a_{\neq i}} I(\mathcal{A}_i; \mathcal{E}|\mathcal{A}_{\neq i} = a_{\neq i})
\end{aligned}$$

**QED**

### 3.10.2 Quantum Connection

We have used classical information theory to prove the above identities. In the quantum setting, Eve has a quantum system that may depend on Alice's bits, $\mathcal{A}_i$. The classical formulas are all valid once a particular measurement on the system (POVM) is fixed by Eve, so that:

$$I(\mathcal{A}_1, \mathcal{A}_2, \ldots, \mathcal{A}_m; \mathcal{E}^M) \leq m \, max_{i, a_{\neq i}} I(\mathcal{A}_i; \mathcal{E}^M|\mathcal{A}_{\neq i} = a_{\neq i}) \quad (3.32)$$

where $\mathcal{E}^M$ is the random variable obtained by Eve's output from her measurement $M$. In particular the above is true for any measurement, $\tilde{M}$, that Eve may consider optimal to learn the bits of Alice's key, $\mathcal{A}_i$, all at once.

Now we need the definition of Shannon Distinguishability:

$$SD^{i, a_{\neq i}} \equiv sup_M I(\mathcal{A}_i; \mathcal{E}^M|\mathcal{A}_{\neq i} = a_{\neq i}) \quad (3.33)$$

Note, a measurement that achieves (or nearly achieves) this upper bound may not be optimal for eavesdropping on the entire key, but that is of no consequence to the proof. Therefore, $I(\mathcal{A}_i; \mathcal{E}^M|\mathcal{A}_{\neq i} = a_{\neq i}) \leq SD^{i, a_{\neq i}}$ for all $M$ and in particular

$$I(\mathcal{A}_i; \mathcal{E}^{\tilde{M}}|\mathcal{A}_{\neq i} = a_{\neq i}) \leq SD^{i, a_{\neq i}} \quad (3.34)$$

Hence we have a bound for total mutual information for any measurement Eve might consider optimal:

$$I(\mathcal{A}_1, \mathcal{A}_2, \ldots, \mathcal{A}_m; \mathcal{E}^{\tilde{M}}) \leq m \, max_{i, a \neq i} SD^{i, a \neq i} \qquad (3.35)$$

## 3.11 Existence of Codes for Both Reliability and Security

Choosing a code which is good when $n$ is large (for constant error rate) is not a trivial problem in ECC. A Random Linear Code (RLC) is one such code, however, it does not promise us that the distances are as required, but only gives the desired distances with probability as close to one as we want. With RLC, we find that the threshold below which a secure key can be obtained is $p_{allowed} \leq 7.56\%$.

In order to correct $t$ errors with certainty, a code must have a minimal Hamming distance between the code words $d \geq 2t + 1$ so that all original code words, even when distorted by $t$ errors, can still be identified correctly. For any $c_T$ which passes the test, we are promised (due to Lemma 3.4.5) that the probability of having $t = |c_I| > n(p_{allowed} + \epsilon_{\text{rel}})$ errors is smaller than $h = 2e^{-(1/2)n\epsilon_{\text{rel}}^2}$.

Thus, we need to choose a RLC that promises a Hamming distance at least $d$ such that $p_{allowed} + \epsilon_{\text{rel}} < t/n = \frac{d-1}{2n}$, and then the $t$ errors are corrected except for a probability smaller than $h_1 = 2e^{-(1/2)n\epsilon_{\text{rel}}^2}$.

For any $n$, $r = n - k$, and for $\delta$ such that $H_2(\delta) < r/n$, an arbitrary *r*andom linear code $(n, k, d)$ satisfies $d/n \geq \delta$, except for a probability (see [Gal63], Theorem 2.2)

$$\text{Prob}(d/n < \delta) \leq \frac{c(\delta)}{\sqrt{n}} 2^{n(H_2(\delta) - r/n)} = g_1 \qquad (3.36)$$

where $c(\delta) = \frac{1}{1 - 2\delta} \sqrt{\frac{1-\delta}{2\pi\delta}}$.

If we choose $\delta = 2(p_{allowed} + \epsilon_{\text{rel}}) + 1/n$ then we are promised that the errors are corrected, except for probability that the error rate is larger than expected or a bad code was chosen.

Using such a code, $\epsilon_{\rm rel}$ is now a function of $\delta$ so that $\epsilon_{\rm rel} = \delta/2 - 1/(2n) - p_{allowed}$ and therefore,

$$h_1 = 2e^{-(n/4)(\delta - \frac{1}{n} - 2p_{allowed})^2} \tag{3.37}$$

and almost all such codes correct all the errors.

Therefore, the code is reliable except for a probability $g_1 + h_1$.

The above result can be improved [May] by taking RLC with distance $d - 1 \geq n(p_{allowed} + \epsilon_{\rm rel})$ (without the factor of 2), since such a code can also correct $t = n(p_{allowed} + \epsilon_{\rm rel})$ errors except for an exponentially small fraction $f_1$ of the possible errors. We get

$$f_1 = 2e^{-(n/4)(\delta - \frac{1}{n} - p_{allowed})^2} \tag{3.38}$$

and it is exponentially small (in the limit of large $n$) for any $\delta > p_{allowed}$.

Recall that we choose $\epsilon_{\rm sec}$ such that $|v| \geq 2n(p_{allowed} + \epsilon_{\rm sec})$. Let $|v|$ be the minimal distance between one PA string and any other parity check string (or linear combination) taken from ECC and PA. Clearly, the Hamming weight of the dual code of the ECC, once the PA is also added, provides a lower bound on $|v|$. Thus, it is sufficient to demand $d^\perp \geq 2n(p_{allowed} + \epsilon_{\rm sec})$ in order to prove security. Choosing a RLC for the ECC and PA, one cannot be completely sure that the distance indeed satisfies the constraint, but this shall be true with probability exponentially close to one. We use the dual code $(n, r^\perp, d^\perp)$, where $r^\perp = n - r - m$. Such codes satisfy $d^\perp/n \geq \delta^\perp$, except for a fraction of

$$\text{Prob}(d^\perp/n < \delta^\perp) \leq \frac{c(\delta^\perp)}{\sqrt{n}} 2^{n(H_2(\delta^\perp) - (n-r-m)/n)} = g_2 \tag{3.39}$$

With $\delta^\perp = 2(p_{allowed} + \epsilon_{\rm sec})$.

Assuming that Eve gets full information when the code fails we get:

$$\sum_{i_T,c_T,b,s} P(\mathcal{T} = pass, i_T, c_T, b, s) I(\mathcal{A}; \mathcal{E}|i_T, c_T, b, s) \le m \left(2\sqrt{2e^{-\frac{1}{2}n\epsilon_{\text{sec}}^2}} + g_2\right)$$

(3.40)

Since the first term is exponentially small we only need look at $g_2$. We also need to worry about the reliability so we need $g_1$ and $f_1$ to be exponentially small as well. All of them are exponentially small if the following conditions are met:

$$H_2(\delta) - r/n < 0$$
$$H_2(\delta^\perp) + r/n + m/n - 1 < 0$$

Or written another way:

$$H_2(p_{allowed} + \epsilon_{\text{rel}} + 1/n) < r/n$$
$$H_2(2p_{allowed} + 2\epsilon_{\text{sec}}) + H_2(p_{allowed} + \epsilon_{\text{rel}} + 1/n) < 1 - R_{secret}$$

Where $R_{secret} \equiv m/n$. In the limit of large $n$ and $\epsilon$'s close to zero, $p_{allowed} < 7.56\%$ satisfies the bound and hence this is our threshold.

Asymptotically, any $R_{secret} < 1 - H_2(2p_a) - H_2(p_a)$ is secure and reliable for the given ECC+PA. Note, as $p_a$ goes to zero, $R_{secret}$ goes to 1, which means all the information bits are secret.

This threshold is based on the property of the code, and other codes might give worse thresholds. It is possible to replace the RLC by a code that can be decoded and encoded efficiently (e.g., Reed-Solomon concatenated code), and add random PA strings. The Hamming distance between the PA check-strings and the ECC check-strings is still bounded below in the same way as for the RLC (see [May]).

A better threshold can be obtained by using privacy-distillation instead of the standard ECC+PA approach.

Note that any probability of failure in the classical transmission can be added in the same way that $g_2$ is added. This is important to prove security in the case where a fault-tolerant classical transmission is not 100% reliable. It shows an important advantage over the proof of [LC99] which is based on fault tolerant quantum ECC.

# CHAPTER 4

# Encryption of Quantum States

## 4.1 Introduction

In chapter 3 we saw how by using quantum states, or specifically qubits, two parties could share a key with perfect security. In this chapter we will consider how two parties could send *quantum* states with perfect security. The question we ask is: how many classically secure key bits do Alice and Bob need in order to encrypt a quantum bit with perfect security, and what operations will they perform? We consider informationally secure encryption protocols, where any potential eavesdropper, Eve, will have no information about the original quantum state, even if she manages to steal or intercept the entire encrypted quantum data. This scenario is very different from the well-known scheme of quantum cryptography, which in the usual sense[BB84, Ben92] is really a secure expansion of an existing classical key, using a quantum channel and a pre-selected set of quantum states. The resulting secure bits might then be used for an encryption algorithm on classical data. For the tasks targeted in this chapter, we need a method to make sure that even if the eavesdropper takes the quantum data, she will still learn nothing about the quantum information. In this case, the eavesdropper may not care about passing any tests, and may remove the qubits and replace them with qubits in any state.

We provide a simple method to get informationally secure encryption of any quantum state using a classical secret key. This could have several interesting applications.

For example, if we imagine a scenario where good quantum memories are expensive, one might rent quantum storage. Security in such a public-storage model would be a high priority. We assume the user cannot store quantum data herself, but can store classical data. Methods of using trusted centers for quantum cryptography have been developed[BHM96]. Our method would allow a user to encrypt her quantum data using a classical key and allow a potentially malicious center to store the data, and yet she would know that the center could learn nothing about her stored quantum data. Additionally, the untrusted center could act as a quantum communication provider. Several other applications which involve adaptations of classical cryptographic protocols, such as quantum secret sharing using classical key, are outlined later in the chapter.

## 4.2 Encryption of Quantum Data

Alice has a quantum state that she intends either to send to Bob, or to store in a quantum memory for later use. Eve may intercept the state during transmission or may access the quantum memory. Alice wants to make sure that *even if* Eve receives the entire state, she learns nothing. Toward this end, any encryption algorithm must be a unitary operation, or more specifically a set of unitary operations which may be chosen with some distribution. It must be unitary because one must be able to undo the encryption, and any quantum operation that is reversible is unitary[Pre].

The most general scheme is to have a set of $M$ operations, $\{U_k\}$, $k = 1, \ldots, M$, where each element $U_k$ is a $2^n \times 2^n$ unitary matrix. This set of unitary operations is assumed to be known to all, but the classical key, $k$, which specifies the $U_k$ that is applied to the $n$-bit quantum state, is secret. The key is chosen with some probability $p_k$ and the input quantum state is encrypted by applying the corresponding unitary operation $U_k$. In the decryption stage, $U_k^\dagger$ is applied to the quantum state to retrieve the original state.

The input state, $\rho$, is called the message state, and the output state, $\rho_c$, is called the cipher-state. The protocol is secure if for every input state, $\rho$, the output state, $\rho_c$, is the totally mixed state:

$$\rho_c = \sum_k p(k) U_k \rho U_k^\dagger = \frac{1}{2^n} I . \tag{4.1}$$

The reason that $\rho_c$ must be the totally mixed state is two fold. First, for security all inputs must be mapped to the same output density matrix (because $\rho_c$ must be independent of the input). Second, the output must be the totally mixed state because the totally mixed state is clearly mapped to itself by all encryption sets.

To see that this is secure, we note that Eve could prepare an n-bit totally mixed state on her own. Since two processes that output the same density matrices are indistinguishable[Per93], anything that can be learned from $\rho_c$ can also be learned from the totally mixed state.

The *design criterion is to find such a distribution of unitary operations $\{p_k, U_k\}$ that will map all inputs to the totally mixed state.* A construction of such a map is given next.

## 4.3 A Quantum One Time Pad

The algorithm is simple: for each qubit, Alice and Bob share two random secret bits. We assume these bits are shared in advance. If the first bit is 0 she does nothing, else she applies $\sigma_z$ to the qubit. If the second bit is 0 she does nothing, else she applies $\sigma_x$. Now she sends the qubit to Bob. She continues this protocol for the rest of the bits.

We now show that this quantum one time pad protocol is secure. First note that this bit-wise protocol can be expressed in terms of our general quantum encryption setup by choosing $p_k = 1/2^{2n}$ and $U_k = X^\alpha Z^\beta$ ($\alpha, \beta \in \{0,1\}^n$), where $X^\alpha = \bigotimes_{i=1}^{n} \sigma_x^{\alpha(i)}$ and $Z^\beta = \bigotimes_{i=1}^{n} \sigma_z^{\beta(i)}$. Thus $X^\alpha$ corresponds to applying $\sigma_x$ to the bits in positions given by the $n$-bit string $\alpha$, and similarly for $Z^\beta$. Next, define the inner product of two matrices, $M_1$ and $M_2$, as $Tr(M_1 M_2^\dagger)$. If the set of all $2^n \times 2^n$ matrices is seen as an inner product space (with respect to the preceding inner product), then one can easily verify that the set of $2^{2n}$ unitary matrices $\{X^\alpha Z^\beta\}$ forms an orthonormal basis. Expanding any message state, $\rho$, in this $X^\alpha Z^\beta$ basis gives:

$$\rho = \sum_{\alpha,\beta} a_{\alpha,\beta} X^\alpha Z^\beta , \tag{4.2}$$

where $a_{\alpha,\beta} = Tr(\rho Z^\beta X^\alpha)/2^n$. Using this formalism, it is clear that the given choice of $p_k$ and $U_k$ satisfies eqn. (4.1), and hence the underlying protocol is secure:

$$\begin{aligned}
\sum_k p(k) U_k \rho U_k^\dagger &= \frac{1}{2^{2n}} \sum_{\gamma,\delta} X^\gamma Z^\delta \rho Z^\delta X^\gamma \\
&= \frac{1}{2^{2n}} \sum_{\alpha,\beta} a_{\alpha,\beta} \sum_{\gamma,\delta} X^\gamma Z^\delta X^\alpha Z^\beta Z^\delta X^\gamma \\
&= \frac{1}{2^{2n}} \sum_{\alpha,\beta} a_{\alpha,\beta} \sum_{\gamma,\delta} (-1)^{\alpha \cdot \delta \oplus \gamma \cdot \beta} X^\alpha Z^\beta \\
&= \sum_{\alpha,\beta} a_{\alpha,\beta} \delta_{\alpha,0} \delta_{\beta,0} X^\alpha Z^\beta \\
&= a_{0,0} I = \frac{Tr(\rho)}{2^n} I = \frac{1}{2^n} I \tag{4.3}
\end{aligned}$$

## 4.4 An Equivalent Problem

Since there are a continuum of valid density matrices, the quantum security criterion (4.1) can be unwieldy to deal with. Here we introduce a modified condition that is necessary and sufficient for security.

**Lemma 4.4.1** An encryption set $\{p_k, U_k\}$ satisfies eqn. (4.1) if and only if it satisfies:

$$\sum_{k=1}^{M} p(k) U_k X^\alpha Z^\beta U_k^\dagger = \delta_{\alpha,0} \delta_{\beta,0} I . \tag{4.4}$$

**Proof:** To show that the above condition is sufficient, express $\rho$ in the $X^\alpha Z^\beta$ basis, as was done in eqn. (4.3) and apply the eqn. (4.4).

$$\begin{aligned}
\sum_{k=1}^{M} p(k) U_k \rho U_k^\dagger &= \sum_{k=1}^{M} p(k) U_k \left( \sum_{\alpha,\beta} a_{\alpha,\beta} X^\alpha Z^\beta \right) U_k^\dagger \\
&= \sum_{\alpha,\beta} a_{\alpha,\beta} \sum_{k=1}^{M} p(k) U_k X^\alpha Z^\beta U_k^\dagger \\
&= \sum_{\alpha,\beta} a_{\alpha,\beta} \delta_{\alpha,0} \delta_{\beta,0} I \\
&= a_{0,0} I = \frac{Tr(\rho)}{2^n} I = \frac{1}{2^n} I
\end{aligned}$$

To show that the modified condition eqn. (4.4), is necessary is somewhat more involved. First let us introduce some new notations:

$$\rho_i = \frac{I + \sigma_i}{2} \quad \text{and} \quad \rho_{mix} = \frac{I}{2} .$$

The proof may be obtained by induction. Suppose all $X^\alpha$ with $|\alpha| \leq k$ are mapped to zero by the encryption process. Now consider the following product state of $n - k - 1$ mixed states, with exactly $k + 1$ pure states $\rho_x$:

$$\rho = \rho_{mix} \otimes \rho_{mix} \otimes \ldots \otimes \rho_{mix} \otimes \rho_x \otimes \rho_x \otimes \ldots \otimes \rho_x$$

By expanding the above becomes:

$$\rho = \frac{I}{2^n} + \frac{1}{2^n} \sum_{\alpha=1}^{2^k - 1} X^\alpha + \frac{1}{2^n} X^{2^{k+1} - 1}$$

In the above we use decimal numbers where before we defined $X^\alpha$ with $\alpha$ in binary; hence $X^3 = X^{00\ldots011}$. When the above $\rho$ is encrypted we know that $\frac{I}{2^n}$ is mapped to itself. By assumption $X^\alpha$ with $|\alpha| \leq k$ is mapped to zero, hence the sum in the expansion of $\rho$ disappears. Since $\rho$ must be mapped to $\frac{I}{2^n}$, then the last term in the above, which is $X^\alpha$ with $|\alpha| = k + 1$, must be mapped to zero. By permuting the initial input states, all $X^\alpha$ with $|\alpha| = k + 1$ must be mapped to zero. The case where $k = 1$ is our base case. By induction all $X^\alpha$ are mapped to zero.

If $x$ is replaced by $z$ in the above, then all $Z^\beta$ are mapped to zero also. If $x$ is replaced by $y$ and using the fact that all $X^\alpha$ and $Z^\beta$ are mapped to zero, one sees that all $X^\alpha Z^\beta$ are mapped to zero, which proves the lemma.

∎

Thus, by using a basis for the set of $2^n \times 2^n$ matrices, the condition for security becomes discrete, and only $2^{2n}$ equations need to be satisfied by the set $\{p_k, U_k\}$. The above lemma will be useful for showing necessary conditions on encryption sets.

## 4.5 Characterization and Optimality of Quantum One-Time Pads

So far, we have provided one quantum encryption protocol based on bit-wise Pauli rotations, which uses $2n$ random classical bits in order to encrypt $n$ quantum bits. In this section we explore the following questions: (1) What are some of the other choices of $\{p_k, U_k\}$ that can be used to perform quantum encryption? In general, can one precisely characterize all possible valid choices of $\{p_k, U_k\}$? and (2) Is the simple quantum one time pad protocol optimal? That is, can one encrypt $n$-bit quantum states using less than $2n$ random secret classical bits? First, we prove a sufficient condition for choosing a secure encryption protocol, and then provide a corresponding necessary condition as well. In particular, we show that one *cannot* perform secure encryption of $n$-bit quantum states using less than $2n$ random classical bits.

**Lemma 4.5.1** Any unitary orthonormal basis for the $2^n \times 2^n$ matrices uniformly ap-plied encrypts $n$ quantum bits.

**Proof:** We can always write the matrices, $U_k$, in terms of the $X^\alpha Z^\beta$ basis as

$$U_k = \sum_{\alpha,\beta} C^k_{\alpha,\beta} X^\alpha Z^\beta . \tag{4.5}$$

Since these $U_k$'s form an orthonormal basis, the $2^{2n} \times 2^{2n}$ transformation matrix $C$, comprising of the transformation coefficients, is a unitary matrix. Hence, the rows and columns of $C$ are orthonormal:

$$\sum_{k=1}^{M} C^k_{\alpha,\beta}(C^k_{\gamma,\delta})^* = \delta_{\alpha,\gamma}\delta_{\beta,\delta} \text{ and } \sum_{\alpha,\beta} C^k_{\alpha,\beta}(C^l_{\alpha,\beta})^* = \delta_{k,l} . \tag{4.6}$$

By substitution of $U_k$ in (4.1) the lemma is obtained:

$$\begin{aligned}
\frac{1}{2^{2n}} \sum_k U_k \rho U_k^\dagger &= \frac{1}{2^{2n}} \sum_k \left( \sum_{\alpha,\beta} C^k_{\alpha,\beta} X^\alpha Z^\beta \right) \rho \left( \sum_{\gamma,\delta} C^{k\,*}_{\gamma,\delta} Z^\delta X^\gamma \right) \\
&= \frac{1}{2^{2n}} \sum_k \sum_{\alpha,\beta} \sum_{\gamma,\delta} C^k_{\alpha,\beta} C^{k\,*}_{\gamma,\delta} X^\alpha Z^\beta \rho Z^\delta X^\gamma \\
&= \frac{1}{2^{2n}} \sum_{\alpha,\beta} \sum_{\gamma,\delta} \left( \sum_k C^k_{\alpha,\beta} C^{k\,*}_{\gamma,\delta} \right) X^\alpha Z^\beta \rho Z^\delta X^\gamma \\
&= \frac{1}{2^{2n}} \sum_{\alpha,\beta} \sum_{\gamma,\delta} \delta_{\alpha,\gamma}\delta_{\beta,\delta} X^\alpha Z^\beta \rho Z^\delta X^\gamma \\
&= \frac{1}{2^{2n}} \sum_{\alpha,\beta} X^\alpha Z^\beta \rho Z^\beta X^\alpha \\
&= \frac{1}{2^n} I
\end{aligned}$$

∎

**Lemma 4.5.2** Given any quantum encryption set, $\{p_k, U_k\}, k = 1, \cdots, M$, (i.e., $\sum_k p_k = 1$, $U_k$ is unitary, and eqns. (4.1) and (4.4) are satisfied), let $\tilde{U}_k = \sqrt{p_k} U_k = \sum_{\alpha,\beta} \tilde{C}^k_{\alpha,\beta} X^\alpha Z^\beta$, and let $\tilde{C}$ be the $M \times 2^{2n}$ transformation matrix, comprising of the transformation coefficients $\tilde{C}_{\alpha,\beta}^k$. Then $M \geq 2^{2n}$, and

$$\tilde{C}^\dagger \tilde{C} = \frac{1}{2^{2n}} I_{2^{2n} \times 2^{2n}} .$$

**Proof:** $\{p_k, U_k\}$ satisfies eqns. (4.1) and (4.4). Hence, for every $\ell, m \in \{0,1\}^n$,

$$\begin{aligned}
\delta_{\ell,0}\delta_{m,0} I &= \sum_{k=1}^M p(k) U_k X^\ell Z^m U_k^\dagger \\
&= \sum_{k=1}^M \tilde{U}_k X^\ell Z^m \tilde{U}_k^\dagger \\
&= \sum_{k=1}^M \sum_{\alpha,\beta} \sum_{\gamma,\delta} \tilde{C}_{\alpha,\beta}^k (\tilde{C}_{\gamma,\delta}^k)^* X^\alpha Z^\beta X^\ell Z^m Z^\delta X^\gamma \\
&= \sum_{\alpha,\beta} \sum_{\gamma,\delta} (-1)^{\beta\cdot\ell+\gamma\cdot(\beta+\delta+m)} \left( \sum_{k=1}^M \tilde{C}_{\alpha,\beta}^k (\tilde{C}_{\gamma,\delta}^k)^* \right) X^{\alpha+\gamma+\ell} Z^{\beta+\delta+m} \\
&= \sum_{p,q} \left( \sum_{\alpha,\beta} (-1)^{\beta\cdot\ell+(p+\ell+\alpha)\cdot q} \left( \sum_{k=1}^M \tilde{C}_{\alpha,\beta}^k (\tilde{C}_{\alpha+p+\ell,\beta+q+m}^k)^* \right) \right) X^p Z^q .
\end{aligned}$$

Using the linear independence of the $X^p Z^q$, only the identity component is non-zero. Hence security implies:

$$\begin{aligned}
\delta_{\ell,0}\delta_{m,0}\delta_{p,0}\delta_{q,0} &= \sum_{\alpha,\beta} (-1)^{\beta\cdot\ell+\alpha\cdot q} \left( \sum_{k=1}^M \tilde{C}_{\alpha,\beta}^k (\tilde{C}_{\alpha+p+\ell,\beta+q+m}^k)^* \right) \\
&= \sum_{\alpha,\beta,\gamma,\delta} (-1)^{\beta\cdot\ell+\alpha\cdot q} \delta_{\gamma,\alpha+p+\ell}\delta_{\delta,\beta+q+m} \left( \sum_{k=1}^M \tilde{C}_{\alpha,\beta}^k (\tilde{C}_{\gamma,\delta}^k)^* \right) \quad (4.7)
\end{aligned}$$

As it will be evident, the second step in the above equation will be used to introduce a linear algebra formulation of the problem. Now, let

$$\Psi_{(\alpha,\beta),(\gamma,\delta)} = \sum_{k=1}^M \tilde{C}_{\alpha,\beta}^k (\tilde{C}_{\gamma,\delta}^k)^* ,$$

which is the standard inner product of the $(\alpha,\beta)^{th}$ and the $(\gamma,\delta)^{th}$ columns of $\tilde{C}$ or $\left(\tilde{C}^\dagger \tilde{C}\right)_{(\alpha,\beta),(\gamma,\delta)}$, and let

$$\mathbf{M}_{(\ell,m,p,q),(\alpha,\beta,\gamma,\delta)} = (-1)^{\beta\cdot\ell+\alpha\cdot q} \delta_{\gamma,\alpha+p+\ell}\delta_{\delta,\beta+q+m} .$$

Eqn. (4.7) can now be written as a set of $2^{4n}$ linear equations: $\mathbf{M}\mathbf{\Psi} = [1\ 0\ \cdots 0\ ]^T$, where $\mathbf{\Psi}$ is the $2^{4n} \times 1$ vector consisting of all the possible inner products of pairs of columns of $\tilde{C}$, and $\mathbf{M}$ is a $2^{4n} \times 2^{4n}$ matrix with elements from the set $1, 0, -1$. Next we observe that a matrix $\mathbf{A}$ is orthogonal if and only if $\sum_j A_{i,j} A_{i',j} = A_i^2 \delta_{i,i'}$, where $A_i$ is the norm of the $i^{th}$ row (which must be greater than zero). One can easily verify that $\mathbf{M}$ is an orthogonal matrix:

$$\sum_{\alpha,\beta,\gamma,\delta} \mathbf{M}_{(\ell,m,p,q),(\alpha,\beta,\gamma,\delta)} \mathbf{M}_{(\ell',m',p',q'),(\alpha,\beta,\gamma,\delta)}$$
$$= \sum_{\alpha,\beta,\gamma,\delta} (-1)^{\beta\cdot\ell+\alpha\cdot q} \delta_{\gamma,\alpha+p+l} \delta_{\delta,\beta+q+m} (-1)^{\beta\cdot\ell'+\alpha\cdot q'} \delta_{\gamma,\alpha+p'+l'} \delta_{\delta,\beta+q'+m'}$$
$$= \sum_{\alpha,\beta,\gamma,\delta} (-1)^{\beta\cdot(\ell+\ell')+\alpha\cdot(q+q')} \delta_{\gamma,\alpha+p+l} \delta_{\delta,\beta+q+m} \delta_{\gamma,\alpha+p'+l'} \delta_{\delta,\beta+q'+m'}$$
$$= \sum_{\alpha,\beta} (-1)^{\beta\cdot(\ell+\ell')+\alpha\cdot(q+q')} \delta_{p+l,p'+l'} \delta_{q+m,q'+m'}$$
$$= 2^{2n} \delta_{l,l'} \delta_{q,q'} \delta_{p+l,p'+l'} \delta_{q+m,q'+m'}$$
$$= 2^{2n} \delta_{l,l'} \delta_{q,q'} \delta_{p,p'} \delta_{m,m'}\ .$$

In showing the above we have also found the inverse of $\mathbf{M}$. The orthonormality of $\mathbf{M}$ means that $\mathbf{M}\mathbf{M}^T = 2^{2n} I$, and hence $\mathbf{M}^{-1} = \mathbf{M}^T/2^{2n}$. Therefore, $\mathbf{\Psi} = \frac{\mathbf{M}^T [1\ 0\ \cdots 0\ ]^T}{2^{2n}}$, which means $\mathbf{\Psi}$ is the first row of $\mathbf{M}$ renormalized:

$$\mathbf{\Psi}_{(\alpha,\beta),(\gamma,\delta)} = \frac{\mathbf{M}_{(0,0,0,0)(\alpha,\beta,\gamma,\delta)}}{2^{2n}} = \frac{1}{2^{2n}} \delta_{\alpha,\gamma} \delta_{\beta,\delta}\ .$$

Since $\left(\tilde{C}^\dagger \tilde{C}\right)_{(\alpha,\beta),(\gamma,\delta)} = \mathbf{\Psi}_{(\alpha,\beta),(\gamma,\delta)}$ we have

$$\tilde{C}^\dagger \tilde{C} = \frac{1}{2^{2n}} I_{2^{2n} \times 2^{2n}}\ .$$

Since $I_{2^{2n} \times 2^{2n}}$ is a full rank matrix, then $\tilde{C}$ must have at least as many rows as columns. $\tilde{C}$ has $2^{2n}$ columns so $M \geq 2^{2n}$. ∎

**Theorem 4.5.3** Any given quantum encryption set, $\{p_k, U_k\}$, $k = 1, \cdots, M$, (i.e., $\sum_k p_k = 1$, $U_k$ is unitary, and eqns. (4.1) and (4.4) are satisfied) has:

$$H(p_1, \cdots, p_M) = \sum_{i=1}^{M} p_i \log \frac{1}{p_i} \geq 2n.$$

Hence, one must use at least $2n$ random classical bits for any quantum encryption. Additionally, if $M = 2^{2n}$, then $p_k = \frac{1}{2^{2n}}$ and $U_k$'s form an orthonormal basis. Hence, a set $\{p_k, U_k\}$ involving only $2n$ secret classical bits is a quantum encryption set if and only if the unitary matrix elements form an orthonormal basis, and they are all equally likely.

**P**roof: By Lemma 4.5.2 we have that

$$\tilde{C}^\dagger \tilde{C} = \frac{1}{2^{2n}} I_{2^{2n} \times 2^{2n}}.$$

Using a singular value decomposition[GL89] of $\tilde{C}$, we have the following relationships:

$$\tilde{C} = W \Lambda V^\dagger, \quad \tilde{C}^\dagger \tilde{C} = V(\Lambda^\dagger \Lambda) V^\dagger, \text{ and } \tilde{C} \tilde{C}^\dagger = W(\Lambda \Lambda^\dagger) W^\dagger,$$

where $W$ and $V$ are $M \times M$ and $2^{2n} \times 2^{2n}$ unitary matrices, respectively, and $\Lambda$ is an $M \times 2^{2n}$ diagonal rectangular matrix: $\Lambda(i,j) = \lambda_i \delta_{i,j}$. Note that $\Lambda^\dagger \Lambda$ and $\Lambda \Lambda^\dagger$ are real diagonal matrices and have the same non-zero elements; hence, $\tilde{C}^\dagger \tilde{C}$ and $\tilde{C} \tilde{C}^\dagger$ have the *same* non-zero eigenvalues. Since $\tilde{C}^\dagger \tilde{C}$ has $2^{2n}$ repeated eigenvalues ($= \frac{1}{2^{2n}}$) and $M \geq 2^{2n}$, $\tilde{C} \tilde{C}^\dagger$ has $2^{2n}$ repeated eigenvalues ($= \frac{1}{2^{2n}}$) and the rest of its $M - 2^{2n}$ eigenvalues are 0. Also note that the diagonal entries of $\tilde{C} \tilde{C}^\dagger$ are the probabilities $p_k$'s and hence,

$$p_k = \frac{Tr(\tilde{U}_k \tilde{U}_k^\dagger)}{2^n} = (\tilde{C} \tilde{C}^\dagger)_{k,k} = \frac{1}{2^{2n}} \sum_{i=1}^{2^{2n}} |W_{i,k}|^2 \leq \frac{1}{2^{2n}}.$$

The above uses the facts that since $W$ is unitary, $\sum_{i=1}^{M} |W_{i,k}|^2 = 1$ and that $M \geq 2^{2n}$. Hence,

$$H(p_1, \cdots, p_M) = \sum_{i=1}^{M} p_i \log \frac{1}{p_i} \geq 2n \sum_{i=1}^{M} p_i = 2n \ .$$

In the particular case where $M = 2^{2n}$, we have $\tilde{C}\tilde{C}^\dagger = \tilde{C}^\dagger \tilde{C} = \frac{1}{2^{2n}} I_{2^{2n} \times 2^{2n}}$. Hence

$$\frac{Tr(\tilde{U}_k \tilde{U}_j^\dagger)}{2^n} = \delta_{k,j} \frac{1}{2^{2n}} ,$$

which gives $p_k = \frac{1}{2^{2n}}$, and that the set $\{U_k\}$ necessarily forms an orthonormal basis. The proof is completed by observing that by lemma 4.5.1 any unitary orthonormal basis applied uniformly is sufficient. ∎

## 4.6 Encryption vs. Teleportation and Superdense Coding

One of the most interesting results in quantum information theory is the teleportation of quantum bits by shared EPR pairs and classical channels[BBC93a]. The quantum one time pad described in Section 4.3 could be implemented using the usual teleportation scheme by encrypting the classical communications with a one time pad. Hence, teleportation gives one example of a quantum encryption algorithm. In the original teleportation paper[BBC93a] a proof that two classical bits are required to teleport is given. The proof is based on a construction that gives superluminal communication if teleportation can be done with less than two bits. This proof however does not imply that all quantum encryption sets require $2n$ bits. To do so would require one to prove that all quantum encryption sets correspond to a teleportation protocol. On the other hand, as we show next, all teleportation protocols correspond to a quantum encryption set; hence, Theorem 4.5.3 provides a new proof of optimality of teleportation.

A general teleportation scheme can be described as follows: Alice and Bob share a pure state comprising $2n$ qubits, $\rho_{AB}$, such that the traced out $n$-bit states of Alice and Bob satisfy: $\rho_A = \rho_B = \frac{1}{2^n}I$. Next, Alice receives an unknown $n$-bit quantum state $\rho$, and performs a joint measurement (i.e., on $\rho$ and $\rho_A$), which produces one of a fixed set of outcomes $m_k$, $k = 1, \ldots, M$, each with probability $p_k$. The particular outcome $m_k$ is classically communicated to Bob using $H(p_1, \ldots, p_M)$ bits. Bob performs a corresponding unitary operation $U_k$ on his state to retrieve $\rho$. Hence, after Alice's measurement (and before Bob learns the outcome), Bob's state can be expressed as $\rho_B = \frac{1}{2^n}I = \sum_{k=1}^{M} p(k) U_k \rho U_k^\dagger$, which is exactly the encrypted state of the message, $\rho$, defined in Eqn. (4.1). Hence, every teleportation scheme corresponds to an encryption protocol $\{p_k, U_k\}$. Since we prove that all quantum encryption sets require $2n$ classical bits, then all teleportation schemes must also require $2n$ classical bits. Note that our proof only relies on the properties of the underlying vector spaces.

Superdense coding[BW93] also has a connection to quantum encryption. Consider the case where Alice asks Bob to encrypt something and then Alice wishes to learn the key that Bob used to encrypt. In the case of the classical one time pad [Ver26] $c = m \oplus k$, and so given a message and it's accompanying ciphertext, one learns the key: $k = m \oplus c$. Quantumly, each quantum bit has two classical key bits to learn. Due to Holevo's theorem[Hol73] it may seem that this implies that there is no way to learn the classical key exactly. This intuition is not correct. Alice can learn Bob's key in the following way. Alice prepares $n$ singlets and gives half of each singlet to Bob. Bob encrypts them using the simple quantum one time pad and returns them to Alice. Alice can learn the key exactly by measuring each former singlet in the bell basis. The outcome would tell Alice exactly which transformation Bob applied. This protocol corresponds exactly to the superdense coding scheme[BW93].

Interestingly, some insight is gained as to where the factor of two between the number of classical and quantum bits comes from in both encryption and teleportation. In the case of classical bits, $\rho$ is diagonal. A basis for all diagonal matrices is $Z^\beta$.

Hence, for encryption of classical bits there are only $2^n$ equations. In the quantum case, by lemma 4.4.1, there are $2^{2n}$ equations to satisfy, so it is not too surprising that there are twice as many classical bits needed. Equivalently, the $\log$ of the size of the space is twice as large quantumly as opposed to classically. The *proof given here could be particularized to give a new proof of Shannon's original result* on informationally secure classical encryption[Ver26].

## 4.7  Discussion

We have presented an algorithm for using $2n$ secret classical bits to secure $n$ quantum bits. These encrypted quantum bits may now be held by an untrusted party with no danger that information may be learned from these bits. Any number of applications may be imagined for this algorithm, or class of algorithms $\{p_k, U_k\}$. For instance, rather than using random classical data of size $2n$, one could use a secret key ciphers[Sch96] or stream ciphers[Sch96] to keep a small finite classical key, for instance 256 bits, to generate pseudo-random bits to encrypt quantum data. In fact, these notions allow for straight-forward generalizations of many classical protocols to quantum data. Quantum secret sharing has been developed[CGL99] that may be used to share quantum secrets. Classical secret sharing schemes are known that are informationally secure[Sha79]. By encrypting a quantum state of $n$ bits with $2n$ classical bits, and then using classical secret sharing on the $2n$ bits, one may use these informationally secure classical methods in the quantum world. This protocol would allow users with only classical resources to perform secret sharing given an untrusted center to store the quantum data. One application independently suggested by Crépeau et. al.[CDM99] is to build quantum bit commitment schemes based on computationally secure classical bit commitment schemes.

# CHAPTER 5

# Mutually Unbiased Bases for Quantum States

## 5.1 Introduction

A $d$–level quantum system is described by a density operator $\rho$ that requires $d^2 - 1$ real numbers for its complete specification. A maximal orthogonal quantum test performed on such a system has, without degeneracy, $d$ possible outcomes, providing $d - 1$ independent probabilities. It follows that in principle one requires at least $d + 1$ different orthogonal measurements for complete state determination.

Since the quantum mechanical description of a physical system is characterized in terms of probabilities of outcomes of conceivable experiments consistent with quantum formalism, in order to obtain full information about the system under consideration we need to perform measurements on a large number of identically prepared copies of the system. The different measurements are performed on several subensembles. However, there may be redundancy in the measurement results as the probabilities will not, in general, be independent of each other unless a minimal set of measurements satisfying appropriate criteria is specified. This minimal set need not be necessarily optimal in the sense it may not serve the best way to ascertain the quantum state. However, intuitively speaking, a minimal set of measurements can be reasonably close to an optimal set if they mutually differ as much as possible, thereby ruling out possible overlaps in the results which become crucial in case of error prone measurements. The characterization and proving the existence of such a minimal set of measurements for complete quantum state determination is therefore of fundamental importance.

It has been shown that measurements in a special class of bases, i.e. mutually unbiased bases, not only form a minimal set but also provide the optimal way of determining a quantum state. Mutually unbiased measurements (MUM), loosely speaking, correspond to measurements that are as different as they can be so that each measurement gives as much new information as one can obtain from the system under consideration. In other words the MUM operators are maximally noncommuting among themselves. If the result of one MUM can be predicted with certainty, then all possible outcomes of every other measurement, unbiased to the previous one are equally likely.

As noted earlier mutually unbiased bases (MUB) have a special role in determining the state of a finite dimensional quantum system. Ivanovic [Iva81] first introduced the concept of MUB in the context of quantum state determination, where he proved the existence of such bases when the dimension is a prime by an explicit construction. Later Wootters and Fields [WF89] showed that measurements in MUB provide the minimal as well as optimal way of complete specification of the density matrix. The optimality is understood in the sense of minimization of statistical errors in the measurements. By explicit construction they showed the existence of MUB for prime power dimensions and proved that for any dimension $d$ there can be at most $d + 1$ MUB. However the existence of MUB for other composite dimensions which are not power of a prime still remains an open problem.

In this chapter we give a constructive proof of the results earlier obtained by Ivanovic, Wootters, and Fields [Iva81, WF89] with a totally different method. The two distinct features of our new proof are:

- Our approach is based on developing an interesting connection between maximal commuting bases of orthogonal unitary matrices and mutually unbiased bases, whereby we find a necessary condition for existence of MUB in any dimension. We then provide a constructive proof of existence of MUB in composite dimensions which are power of a prime. This allows us to connect encryption of quantum bits [BR00], which uses unitary bases of operators, to quantum key distribution, which uses mutually unbiased bases of quantum systems.

- Another advantage of our method is that we provide an explicit construction of the MUB observables (operators) as tensor product of the Pauli matrices for dimensions $d = 2^m$. This answers a critical related question: how can these mutually unbiased measurements be actually performed and what are the observables to which these measurements correspond to. When $d = 2$ the mutually unbiased operators are the three Pauli matrices, but unfortunately this observation cannot be generalized in a straightforward way to higher dimension. In addition to the obvious importance of mutually unbiased bases in the context of quantum state determination and foundations of quantum mechanics, recently it has also found useful applications in quantum cryptography where it has been demonstrated that using higher dimensional quantum systems for key distribution has possible advantages over qubits, and mutually unbiased bases play a key role in such a key distribution scheme [BT00b, BT00a]. Thus the fact that we provide an explicit construction of the MUB observables can turn out to be crucial in the application of MUB in quantum cryptography with systems with more than two states.

Before continuing it is useful to provide a formal definition of mutually unbiased bases.

**D**efinition. Let $B_1 = \{|\varphi_1\rangle, \ldots, |\varphi_d\rangle\}$ and $B_2 = \{|\psi_1\rangle, \ldots, |\psi_d\rangle\}$ be two orthonormal bases in the $d$ dimensional state space. They are said to be **m**utually unbiased bases

(MUB) if and only if $|\langle\varphi_i|\psi_j\rangle| = \frac{1}{\sqrt{d}}$, for every $i,j = 1,\ldots,d$. A set $\{\mathcal{B}_1,\ldots,\mathcal{B}_m\}$ of orthonormal bases in $\mathbb{C}^d$ is called a *set of mutually unbiased bases* (a set of MUB) if each pair of bases $\mathcal{B}_i$ and $\mathcal{B}_j$ are mutually unbiased.

The simplest example of a complete set of MUB is obtained in the case of spin 1/2 particle where each unbiased basis consists of the normalized eigenvectors of the three Pauli matrices respectively. However, the analysis of a set of MUB corresponding to a two level quantum system does not capture one of the basic features of MUB, i.e., its importance in determining the quantum state. In the case of two level systems, the density operator has three independent parameters and almost any choice of the three measurements is sufficient to have the complete knowledge of the system. This is not true in general for any other dimension greater than two, where the existence of MUB becomes more crucial in the context of minimal number of required measurements for quantum state determination.

In Section 5.2 we show the existence of $p+1$ MUB in the space $\mathbb{C}^p$, for any prime $p$. This result first shown by Ivanovic [Iva81] by explicitly defining the mutually unbiased bases. Here we show that these bases are in fact bases each consists of eigenvectors of the unitary operators
$$Z, X, XZ, \ldots, XZ^{d-1},$$
where $X$ and $Z$ are generalizations of Pauli operators to the quantum systems with more than two states (see, e.g., [Got99, GKP01]).

In Section 5.3 we show that there is a useful connection between mutually unbiased bases and special types of bases for the space of the square matrices. These bases consist of orthogonal unitary matrices which can be grouped in maximal classes of commuting matrices. As a result of this connection we show that every MUB over $\mathbb{C}^d$ consists of at most $d+1$ bases.

Finally, in Section 5.4 we present our construction of MUB over $\mathbb{C}^d$ when $d$ is a prime power. The basic idea of our construction is as follows. When $d = p^m$, imagine the system consists of $m$ subsystems each of dimension $p$. Then the total number of measurements on the whole system, viewed as performing measurement on every subsystem in their respective MUB is $(p+1)^m$. We show that these $(p+1)^m$ operators fall into $p^m + 1$ maximal noncommuting classes where members of each class commute among themselves. The bases formed by eigenvectors of each such mutually noncommuting class are mutually unbiased. It should be mentioned that the operators in each maximal commuting class have the same structure as the stabilizers of *additive quantum error correcting codes* (see, e.g., [CRS97, CRS98, Got99]).

There is a close connection between the MUB problem and the problem of determining arrangements of lines in the Grassmannian spaces so that they are as far apart as possible [CCK97] (see also [CHR99]). This problem (and some other combinatorial problems discussed in [CCK97]) can be related to the problem of finding the maximum number of lines through the origin of $\mathbb{C}^d$ that are either perpendicular or are at angle $\theta$, where $\cos\theta = 1/\sqrt{d}$. Any MUB $\mathcal{M}$ defines such a line–set: consider all lines through the origin defined by all vectors in the bases of $\mathcal{M}$. In [CCK97], for the case of $d = 2^m$, with an approach similar to the one presented in this chapter, such line–sets are constructed.

**Notation.** Let $\mathbb{M}_d(\mathbb{C})$ be the set of $d \times d$ complex matrices. In a natural way, the set $\mathbb{M}_d(\mathbb{C})$ is a $d^2$–dimensional linear space. Each matrix $A$ in $\mathbb{M}_d(\mathbb{C})$ can be also naturally considered as a $d^2$–dimensional complex vector $|v_A\rangle$, where the entries of the matrix $A$ being regarded as the components of the vector $|v_A\rangle$. In this way, for matrices $A, B \in \mathbb{M}_d(\mathbb{C})$ we can define the inner product $\langle A, B \rangle$ of matrices as the inner product $\langle v_A | v_B \rangle$ of vectors. It is easy to check that

$$\langle A, B \rangle = \text{Tr}(A^\dagger B).$$

We say the matrices $A, B \in \mathbb{M}_d(\mathbb{C})$ are **o**rthogonal if and only if $\langle A, B \rangle = 0$.

## 5.2 Construction of Sets of MUB for Prime Dimensions

Ivanovic [Iva81] for the first time showed that for any prime dimension $d$, there is a set of $d+1$ mutually unbiased bases. In that paper the bases are given explicitly. Here we show that there is a nice symmetrical structure behind these bases, and their existence can be derived as a consequence of properties of Pauli operators on $d$–state quantum systems. The core of our construction is the following theorem.

**Theorem 5.2.1** *Let $\mathcal{B}_1 = \{\,|\varphi_1\rangle, \ldots, |\varphi_d\rangle\,\}$ be an orthonormal basis in $\mathbb{C}^d$. Suppose that there is a unitary operator $V$ such that $V|\varphi_j\rangle = \beta_j|\varphi_{j+1}\rangle$, where $|\beta_j| = 1$ and $|\varphi_{d+1}\rangle = |\varphi_1\rangle$; i.e., $V$ applies a* cyclic shift modulo a phase *on the elements of the basis $\mathcal{B}_1$. Assume that the orthonormal basis $\mathcal{B}_2 = \{\,|\psi_1\rangle, \ldots, |\psi_d\rangle\,\}$ consists of eigenvectors of $V$. Then $\mathcal{B}_1$ and $\mathcal{B}_2$ are MUB.*

**P**roof. Assume that $V|\psi_k\rangle = \lambda_k |\psi_k\rangle$. Then $|\lambda_k| = 1$. Now, for every $k = 1, \ldots, d$, we have

$$\begin{aligned}
|\langle \psi_k | \varphi_1 \rangle| &= |\lambda_k^* \langle \psi_k | V | \varphi_1 \rangle| \\
&= |\beta_1 \langle \psi_k | \varphi_2 \rangle| \\
&= |\langle \psi_k | \varphi_2 \rangle|.
\end{aligned}$$

A similar argument shows

$$|\langle \psi_k | \varphi_1 \rangle| = |\langle \psi_k | \varphi_2 \rangle| = \cdots = |\langle \psi_k | \varphi_d \rangle|.$$

Therefore,

$$|\langle \psi_k | \varphi_j \rangle|^2 = \frac{1}{d}, \qquad 1 \leq j \leq d.$$

Thus $\mathcal{B}_1$ and $\mathcal{B}_2$ are MUB. ∎

Throughout this section, we suppose that $d$ is a prime number, and *all algebraic operations are modulo d*. We consider $\{\,|0\rangle, |1\rangle, \ldots, |d-1\rangle\,\}$ as the standard basis of $\mathbb{C}^d$. We define the unitary operators $X_d$ and $Z_d$ over $\mathbb{C}^d$, as a natural generalization of Pauli operators $\sigma_x$ and $\sigma_z$:

$$X_d|j\rangle = |j+1\rangle, \qquad (5.1)$$

$$Z_d|j\rangle = \omega^j|j\rangle, \qquad (5.2)$$

where $\omega$ is a $d^{\text{th}}$ root of unity; more specifically $\omega = \exp(2\pi i/d)$. We are interested in unitary operators of the form $X_d\,(Z_d)^k$. Note that

$$X_d\,(Z_d)^k\,|j\rangle = \left(\omega^k\right)^j |j+1\rangle.$$

**Theorem 5.2.2** *For $0 \leq k, \ell \leq d-1$, the eigenvectors of $X_d\,(Z_d)^k$ are cyclically shifted under the action of $X_d\,(Z_d)^\ell$.*

**P**roof. The eigenvectors of $X_d\,(Z_d)^k$ are

$$|\psi_t^k\rangle = \frac{1}{\sqrt{d}} \sum_{j=0}^{d-1} \left(\omega^t\right)^{d-j} \left(\omega^{-k}\right)^{s_j} |j\rangle, \qquad t = 0, \ldots, d-1, \qquad (5.3)$$

where $s_j = j + \cdots + (d-1)$. Then $|\psi_t^k\rangle$ is an eigenvector of $X_d\,(Z_d)^k$ with eigenvalue $\omega^t$, because

$$\begin{aligned}
X_d\,(Z_d)^k\,|\psi_t^k\rangle &= \frac{1}{\sqrt{d}} \sum_{j=0}^{d-1} \left(\omega^t\right)^{d-j} \left(\omega^{-k}\right)^{s_j} \left(\omega^k\right)^j |j+1\rangle \\
&= \frac{1}{\sqrt{d}} \sum_{j=0}^{d-1} \left(\omega^t\right)^{d-j} \left(\omega^{-k}\right)^{s_{j+1}} |j+1\rangle \\
&= \frac{1}{\sqrt{d}} \sum_{j=0}^{d-1} \left(\omega^t\right)^{d-j+1} \left(\omega^{-k}\right)^{s_j} |j\rangle \\
&= \omega^t |\psi_t^k\rangle.
\end{aligned}$$

The action of $X_d (Z_d)^\ell$ on $|\psi_t^k\rangle$ is as follows:

$$\begin{aligned} X_d (Z_d)^\ell |\psi_t^k\rangle &= \frac{1}{\sqrt{d}} \sum_{j=0}^{d-1} \left(\omega^t\right)^{d-j} \left(\omega^{-k}\right)^{s_j} \left(\omega^\ell\right)^j |j+1\rangle \\ &= \frac{1}{\sqrt{d}} \sum_{j=0}^{d-1} \left(\omega^t\right)^{d-j+1} \left(\omega^{-k}\right)^{s_{j-1}} \left(\omega^\ell\right)^{j-1} |j\rangle \\ &= \frac{\omega^{t-\ell}}{\sqrt{d}} \sum_{j=0}^{d-1} \left(\omega^t\right)^{d-j} \left(\omega^{-k}\right)^{s_j} \left(\omega^{-k}\right)^{j-1} \left(\omega^\ell\right)^j |j\rangle \\ &= \frac{\omega^{t+k-\ell}}{\sqrt{d}} \sum_{j=0}^{d-1} \left(\omega^t\right)^{d-j} \left(\omega^{-k}\right)^{s_j} \left(\omega^{\ell-k}\right)^j |j\rangle \\ &= \frac{\omega^{t+k-\ell}}{\sqrt{d}} \sum_{j=0}^{d-1} \left(\omega^{t+k-\ell}\right)^{d-j} \left(\omega^{-k}\right)^{s_j} |j\rangle \\ &= \omega^{t+k-\ell} |\psi_{t+k-\ell}^k\rangle. \quad \blacksquare \end{aligned}$$

Note that the standard basis $\{\,|0\rangle, |1\rangle, \ldots, |d-1\rangle\,\}$ is the set of the eigenvectors of $Z_d$. From (5.3) it follows that the $\left|\langle j|\psi_t^k\rangle\right|^2 = \frac{1}{d}$. Therefore, we have proved the following construction.

**Theorem 5.2.3** *For any prime d, the set of the bases each consisting of the eigenvectors of*

$$Z_d,\ X_d,\ X_d Z_d,\ X_d (Z_d)^2, \ldots, X_d (Z_d)^{d-1},$$

*form a set of $d+1$ mutually unbiased bases.*

**Example** $d = 2$. By Theorem 5.2.3, the eigenvectors of the operators $\sigma_z$, $\sigma_x$, and $\sigma_x \sigma_z$ form a set of mutually unbiased bases; i.e., the following set

$$\{|0\rangle, |1\rangle\},$$
$$\left\{\frac{|0\rangle+|1\rangle}{\sqrt{2}}, \frac{|0\rangle-|1\rangle}{\sqrt{2}}\right\},$$
$$\left\{\frac{|0\rangle+i|1\rangle}{\sqrt{2}}, \frac{|0\rangle-i|1\rangle}{\sqrt{2}}\right\}.$$

**Example** $d = 3$. The set of the eigenvectors of the following unitary matrices form a set of MUB (here $\omega = \exp(2\pi i/3)$):

$$\begin{pmatrix} 1 & 0 & 0 \\ 0 & 1 & 0 \\ 0 & 0 & 1 \end{pmatrix}, \quad \begin{pmatrix} 0 & 0 & 1 \\ 1 & 0 & 0 \\ 0 & 1 & 0 \end{pmatrix}, \quad \begin{pmatrix} 0 & 0 & \omega^2 \\ 1 & 0 & 0 \\ 0 & \omega & 0 \end{pmatrix}, \quad \begin{pmatrix} 0 & 0 & \omega \\ 1 & 0 & 0 \\ 0 & \omega^2 & 0 \end{pmatrix}.$$

## 5.3 Bases for Unitary Operators and MUB

In this section we study the close relation between MUB and a special type of bases for $\mathbb{M}_d(\mathbb{C})$. Here we are dealing with classes of commuting unitary matrices. The following lemma shows that the maximum size of such class is $d$.

**Lemma 5.3.1** *There are at most $d$ pairwise orthogonal commuting unitary matrices in $\mathbb{M}_d(\mathbb{C})$.*

**Proof.** Let $A_1, \ldots, A_m$ be pairwise orthogonal commuting unitary matrices in $\mathbb{M}_d(\mathbb{C})$. Then there is a unitary matrix $U$ such that the matrices $B_1, \ldots, B_m$, where $B_j = U A_j U^\dagger$, are diagonal. Moreover, $\langle B_j, B_k \rangle = \langle A_j, A_k \rangle$; so $B_j$ and $B_k$ are orthogonal for $j \neq k$. Let $|b_j\rangle \in \mathbb{C}^d$ be the diagonal of $B_j$. Then $\langle B_j, B_k \rangle = \langle b_j | b_k \rangle$. So the vectors $|b_1\rangle, \ldots, |b_m\rangle$ are mutually orthogonal; therefore, $m \leq d$. ∎

Let $\mathcal{B} = \{U_1, U_2, \ldots, U_{d^2}\}$ be a *basis of unitary* matrices for $\mathbb{M}_d(\mathbb{C})$. Without loss of generality, we can assume that $U_1 = \mathbb{1}_d$, the identity matrix of order $d$. We say that the basis $\mathcal{B}$ is a **m**aximal commuting basis for $\mathbb{M}_d(\mathbb{C})$ if $\mathcal{B}$ can be partitioned as

$$\mathcal{B} = \{\mathbb{1}_d\} \bigcup \mathcal{C}_1 \bigcup \cdots \bigcup \mathcal{C}_{d+1}, \tag{5.4}$$

where each class $\mathcal{C}_j$ contains exactly $d - 1$ commuting matrix from $\mathcal{B}$. Note that $\{\mathbb{1}_d\} \bigcup \mathcal{C}_j$ is a set of $d$ commuting orthogonal unitary matrices, which by Lemma 5.3.1 is maximal.

**Theorem 5.3.1** *If there is a maximal commuting basis of orthogonal unitary matrices in $\mathbb{M}_d(\mathbb{C})$, then there is a set of $d+1$ mutually unbiased bases.*

**P**roof. Let $\mathcal{B}$ be a maximal commuting basis of orthogonal unitary matrices in $\mathbb{M}_d(\mathbb{C})$, where (5.4) provides the decomposition of $\mathcal{B}$ into maximal classes of commuting matrices. For any $1 \leq j \leq d+1$, let

$$\mathcal{C}_j = \{U_{j,1}, U_{j,2}, \ldots, U_{j,d-1}\}.$$

We also define $U_{j,0} = \mathbb{1}_d$; then

$$\mathcal{C}'_j = \{U_{j,0}, U_{j,1}, U_{j,2}, \ldots, U_{j,d-1}\}$$

is a maximal set of commuting orthogonal unitary matrices. Thus for each $1 \leq j \leq d+1$, there is an orthonormal basis

$$\mathcal{T}_j = \{|\psi_1^j\rangle, |\psi_2^j\rangle, \ldots, |\psi_d^j\rangle\}$$

such that every matrix $U_{j,t}$ (for $0 \leq t \leq d-1$) relative to the basis $\mathcal{T}_j$ is diagonal. Let

$$U_{j,t} = \sum_{k=1}^{d} \lambda_{j,t,k} |\psi_k^j\rangle \langle \psi_k^j|. \tag{5.5}$$

Let $M_j$ be a $d \times d$ matrix whose $k^{\text{th}}$ row is the diagonal of the right-hand side matrix of (5.5); i.e.,

$$M_j = \begin{pmatrix} \lambda_{j,0,1} & \lambda_{j,0,2} & \cdots & \lambda_{j,0,d} \\ \lambda_{j,1,1} & \lambda_{j,1,2} & \cdots & \lambda_{j,1,d} \\ \vdots & \vdots & \ddots & \vdots \\ \lambda_{j,d-1,1} & \lambda_{j,d-1,2} & \cdots & \lambda_{j,d-1,d} \end{pmatrix}.$$

Then $M_j$ is a *unitary* matrix. Note that the first row of $M_j$ is the constant vector $(1,1,\ldots,1)$. We consider the classes $\mathcal{C}_1$ and $\mathcal{C}_2$. Then for $0 \leq s, t \leq d-1$, the orthogonality condition implies

$$\text{Tr}\left(U_{1,s}^\dagger U_{2,t}\right) = d\, \delta_{s,0}\, \delta_{t,0}.$$

But, since $\operatorname{Tr}(|\psi_k^1\rangle\langle\psi_\ell^2|) = \langle\psi_k^1|\psi_\ell^2\rangle^*$,

$$\begin{aligned}
\operatorname{Tr}\left(U_{1,s}{}^\dagger U_{2,t}\right) &= \operatorname{Tr}\left(\sum_{k=1}^d\sum_{\ell=1}^d \lambda_{1,s,k}{}^*\lambda_{2,t,\ell}|\psi_k^1\rangle\langle\psi_k^1|\psi_\ell^2\rangle\langle\psi_\ell^2|\right) \\
&= \sum_{k=1}^d\sum_{\ell=1}^d \lambda_{1,s,k}{}^*\lambda_{2,t,\ell}\langle\psi_k^1|\psi_\ell^2\rangle\operatorname{Tr}\left(|\psi_k^1\rangle\langle\psi_\ell^2|\right) \\
&= \sum_{k=1}^d\sum_{\ell=1}^d \lambda_{1,s,k}{}^*\lambda_{2,t,\ell}\left|\langle\psi_k^1|\psi_\ell^2\rangle\right|^2.
\end{aligned}$$

Therefore

$$\sum_{k=1}^d\sum_{\ell=1}^d \lambda_{1,s,k}{}^*\lambda_{2,t,\ell}\left|\langle\psi_k^1|\psi_\ell^2\rangle\right|^2 = d\,\delta_{s,0}\delta_{t,0}, \qquad 0 \leq s,t \leq d-1. \tag{5.6}$$

The system of equations (5.6) can be written in the following matrix form

$$A\,P = \Lambda,$$

where

$$\begin{aligned}
A &= M_1{}^* \otimes M_2, \\
P &= \left(|\langle\psi_1^1|\psi_1^2\rangle|^2, |\langle\psi_1^1|\psi_2^2\rangle|^2, \ldots, |\langle\psi_d^1|\psi_d^2\rangle|^2\right)^{\mathrm{T}}, \\
\Lambda &= (d, 0, 0, \ldots, 0)^{\mathrm{T}}.
\end{aligned}$$

Note that $A$ is a unitary matrix and its first row is the constant vector $(1, 1, \ldots, 1)$. Then from $P = A^{-1}\Lambda$ it follows

$$|\langle\psi_s^1|\psi_t^2\rangle|^2 = \tfrac{1}{d}, \qquad 1 \leq s,t \leq d.$$

By repeating the same argument for the classes $\mathcal{C}_j$ and $\mathcal{C}_k$, we conclude that

$$\{\mathcal{T}_1, \ldots, \mathcal{T}_{d+1}\}$$

is a set of MUB. ∎

Before we continue, we prove the following useful simple lemma.

**Lemma 5.3.2** *For any integers $m$ and $n$ such that $0 < m \leq n$ we have*

$$\sum_{k=1}^{n} e^{2\pi i \frac{mk}{n}} = 0.$$

**P**roof. We have

$$\sum_{k=1}^{n} \left(e^{2\pi i \frac{m}{n}}\right)^k = e^{2\pi i \frac{m}{n}} \frac{\left(e^{2\pi i \frac{m}{n}}\right)^n - 1}{e^{2\pi i \frac{m}{n}} - 1} = 0. \quad \blacksquare$$

The converse of Theorem 5.3.1, in the following sense, holds.

**Theorem 5.3.2** *Let $\mathcal{B}_1, \ldots, \mathcal{B}_m$ be a set of MUB in $\mathbb{C}^d$. Then there are $m$ classes $\mathcal{C}_1, \ldots, \mathcal{C}_m$ each consisting of $d$ commuting unitary matrices such that matrices in $\mathcal{C}_1 \bigcup \cdots \bigcup \mathcal{C}_m$ are pairwise orthogonal.*

**P**roof. Suppose that

$$\mathcal{B}_j = \left\{ |\psi_1^j\rangle, \ldots, |\psi_d^j\rangle \right\}.$$

Then

$$\langle \psi_s^j | \psi_t^j \rangle = \delta_{s,t}, \qquad 1 \leq s, t \leq d,$$

and

$$\left|\langle \psi_s^j | \psi_t^k \rangle\right|^2 = \frac{1}{d}, \qquad 1 \leq j < k \leq d, \ 1 \leq s, t \leq d.$$

We label the matrices in the class $\mathcal{C}_j$ as

$$\mathcal{C}_j = \{ U_{j,0}, U_{j,1}, \ldots, U_{j,d-1} \},$$

where

$$U_{j,t} = \sum_{k=1}^{d} e^{2\pi i \frac{tk}{d}} |\psi_k^j\rangle \langle \psi_k^j|, \qquad 0 \leq t \leq d - 1.$$

Note that $U_{j,0} = \mathbb{1}_d$. Then $U_{j,s}$ and $U_{j,t}$ are commuting, because both are diagonal relative to the basis $\mathcal{B}_j$. We now show that all these matrices are orthogonal. First we note that

$$\begin{aligned}
\langle U_{j,s}, U_{k,t}\rangle &= \text{Tr}\left(U_{j,s}{}^\dagger U_{k,t}\right) \\
&= \sum_{x=1}^{d}\sum_{y=1}^{d} e^{2\pi i \frac{ty-sx}{d}} \text{Tr}\left(|\psi_x^j\rangle\langle\psi_x^j|\psi_y^k\rangle\langle\psi_y^k|\right) \\
&= \sum_{x=1}^{d}\sum_{y=1}^{d} e^{2\pi i \frac{ty-sx}{d}} \left|\langle\psi_x^j|\psi_y^k\rangle\right|^2.
\end{aligned}$$

Thus, by Lemma 5.3.2, if $j = k$, then

$$\begin{aligned}
\langle U_{j,s}, U_{j,t}\rangle &= \sum_{x=1}^{d}\sum_{y=1}^{d} e^{2\pi i \frac{ty-sx}{d}} \delta_{x,y} \\
&= \sum_{x=1}^{d} e^{2\pi i \frac{x(t-s)}{d}} \\
&= d\,\delta_{s,t}.
\end{aligned}$$

If $j \neq k$ and $(s,t) \neq (0,0)$, then

$$\begin{aligned}
\langle U_{j,s}, U_{k,t}\rangle &= \sum_{x=1}^{d}\sum_{y=1}^{d} e^{2\pi i \frac{ty-sx}{d}} \frac{1}{d} \\
&= \frac{1}{d}\left(\sum_{x=1}^{d} e^{2\pi i \frac{sx}{d}}\right)^*\left(\sum_{y=1}^{d} e^{2\pi i \frac{ty}{d}}\right) \\
&= 0. \blacksquare
\end{aligned}$$

As an immediate corollary of the above theorem, we have the following upper bound on the size of a set of MUB.

**Theorem 5.3.3** *Any set of mutually unbiased bases in $\mathbb{C}^d$ contains at most $d+1$ bases.*

**P**roof. If a set of MUB contains $m$ bases, then by Theorem 5.3.2, there are at least $1 + m(d-1)$ pairwise orthogonal matrices in the $d^2$–dimensional space $\mathbb{M}_d(\mathbb{C})$. Therefore, $1 + m(d - 1) \leq d^2$, thus $m \leq d + 1$. $\blacksquare$

## 5.4 Construction of a Set of MUB for Prime Powers

### 5.4.1 The Pauli Group

To construct a maximal set of MUB in $\mathcal{H} = \mathbb{C}^{p^m}$, where $p$ is a prime number, we consider the Hilbert space $\mathcal{H}$ as tensor product of $m$ copies of $\mathbb{C}^p$; i.e.,

$$\mathcal{H} = \underbrace{\mathbb{C}^p \otimes \cdots \otimes \mathbb{C}^p}_{m \text{ times}}.$$

Like the case of $\mathbb{C}^p$, we build a set of MUB as the sets of eigenvectors of special types of unitary operators on the background space $\mathcal{H}$. On the space $\mathbb{C}^p$ we considered the generalized Pauli operators $X_p$ and $Z_p$, defined by equations (5.1) and (5.2). On the space $\mathcal{H}$, we consider the tensor products of operators $X_p$ and $Z_p$.

We denote the finite field $\{0, 1, \ldots, p-1\}$ by $\mathbb{F}_p$. Let $\omega = e^{2\pi i/d}$ be a primitive $p^{\text{th}}$ root of unity. Then

$$Z_p X_p = \omega X_p Z_p.$$

Therefore, if $U_1 = (X_p)^{k_1} (Z_p)^{\ell_1}$ and $U_2 = (X_p)^{k_2} (Z_p)^{\ell_2}$ then

$$U_2 U_1 = \omega^{k_1 \ell_2 - k_2 \ell_1} U_1 U_2. \tag{5.7}$$

We are interested on unitary operators on $\mathcal{H} = \mathbb{C}^p \otimes \cdots \otimes \mathbb{C}^p$ (the tensor product of $m$ copies of $\mathbb{C}^p$) of the form

$$U = M_1 \otimes \cdots \otimes M_m, \qquad \text{where } M_j = (X_p)^{k_j} (Z_p)^{\ell_j}, \ 0 \leq k_j, \ell_j \leq p - 1. \tag{5.8}$$

To describe an operator of the form (5.8) it is enough to specify the powers $k_j$ and $\ell_j$. So we represent an operator (5.8) by the following vector of length $2m$ over the field $\mathbb{F}_p$:

$$(k_1, \ldots, k_m \mid \ell_1, \ldots, \ell_m),$$

or equivalently as

$$X_p(k_1, \ldots, k_m) \, Z_p(\ell_1, \ldots, \ell_m).$$

If we let $\alpha = (k_1, \ldots, k_m)$ and $\beta = (\ell_1, \ldots, \ell_m)$, then $\alpha, \beta \in \mathbb{F}_p{}^m$ and we denote the corresponding operator by

$$X_p(\alpha)\, Z_p(\beta).$$

The **P**auli group $\mathbb{P}(p, m)$ is the group of all unitary operators on $\mathcal{H} = \mathbb{C}^p \otimes \cdots \otimes \mathbb{C}^p$ (the tensor product of $m$ copies of $\mathbb{C}^p$) of the form

$$\omega^j\, X_p(\alpha)\, Z_p(\beta), \tag{5.9}$$

for some integer $j \geq 0$ and vectors $\alpha, \beta \in \mathbb{F}_p{}^m$, where $\omega = \exp(2\pi i/p)$. In this section we are mainly interested in the subset $\mathbb{P}_0(p, m)$ of $\mathbb{P}(p, m)$ of the operators of the form (5.9) with $j = 0$. Note that $\mathbb{P}_0(p, m)$ is not a subgroup, but generators of subgroups of the Pauli group can always be considered as subsets of $\mathbb{P}_0(p, m)$.

If the operators $U$ and $U'$ in $\mathbb{P}_0(p, m)$ are represented by the vectors

$$(k_1, \ldots, k_m \mid \ell_1, \ldots, \ell_m) \quad \text{and} \quad (k'_1, \ldots, k'_m \mid \ell'_1, \ldots, \ell'_m),$$

respectively, then $U$ and $U'$ are commuting if and only if

$$\sum_{j=1}^m k_j \ell'_j - \sum_{j=1}^m k'_j \ell_j = 0 \mod p.$$

We can state this condition equivalently in the following form.

**Lemma 5.4.1** *If $U = X_p(\alpha)\, Z_p(\beta)$ and $U' = X_p(\alpha')\, Z_p(\beta')$, for $\alpha, \beta, \alpha', \beta' \in \mathbb{F}_p{}^m$, then $U$ and $U'$ are commuting if and only if*

$$\alpha \cdot \beta' - \alpha' \cdot \beta = 0 \mod p. \tag{5.10}$$

A set $X_p(\alpha_1)\, Z_p(\beta_1), \ldots, X_p(\alpha_t)\, Z_p(\beta_t)$ of operators in $\mathbb{P}_0(p, m)$ is represented by the $t \times (2m)$ matrix

$$\begin{pmatrix} \alpha_1 & \beta_1 \\ \vdots & \vdots \\ \alpha_t & \beta_t \end{pmatrix}.$$

Before we continue, we would like to get an explicit formula for the action of a $\mathbb{P}_0(p,m)$ operator $X_p(\alpha)\, Z_p(\beta)$. Let $\alpha = (\alpha_1, \ldots, \alpha_m)$ and $\beta = (\beta_1, \ldots, \beta_m)$. The standard basis of the Hilbert space $\mathcal{H} = \mathbb{C}^p \otimes \cdots \otimes \mathbb{C}^p$ consists of the vectors $|j_1 \cdots j_m\rangle$, where $(j_1, \ldots, j_m) \in \mathbb{F}_p{}^m$. Then

$$X_p(\alpha)\, Z_p(\beta)|j_1 \cdots j_m\rangle = \omega^{j_1 \beta_1 + \cdots + j_m \beta_m} |(j_1 + \alpha_1) \cdots (j_m + \alpha_m)\rangle.$$

Equivalently,

$$X_p(\alpha)\, Z_p(\beta)|a\rangle \;=\; \omega^{a \cdot \beta} |a + \alpha\rangle, \qquad a \in \mathbb{F}_p{}^m, \tag{5.11}$$

$$X_p(\alpha)\, Z_p(\beta) \;=\; \sum_{a \in \mathbb{F}_p{}^m} \omega^{a \cdot \beta} |a + \alpha\rangle \langle a|, \tag{5.12}$$

where the operations are in the field $\mathbb{F}_p$.

**Theorem 5.4.1** *Let $U = X_p(\alpha)\, Z_p(\beta)$ and $U' = X_p(\alpha')\, Z_p(\beta')$ be operators in $\mathbb{P}_0(p,m)$. If $U \neq U'$, i.e., $(\alpha, \beta) \neq (\alpha', \beta')$, then the operators $U$ and $U'$ are orthogonal.*

**Proof.** We have

$$\begin{aligned}
\langle U, U' \rangle &= \operatorname{Tr}\left( U^\dagger\, U' \right) \\
&= \operatorname{Tr}\left( \sum_{a \in \mathbb{F}_p{}^m} \sum_{b \in \mathbb{F}_p{}^m} \omega^{\beta' \cdot b - \beta \cdot a} |a\rangle \langle a + \alpha | b + \alpha' \rangle \langle b| \right) \\
&= \sum_{a \in \mathbb{F}_p{}^m} \omega^{\beta' \cdot b - \beta \cdot a} \langle a + \alpha | a + \alpha' \rangle.
\end{aligned}$$

If $\alpha \neq \alpha'$, then $\langle a + \alpha | a + \alpha' \rangle = 0$, for every $a \in \mathbb{F}_p{}^m$. Thus in this case $\langle U, U' \rangle = 0$. If $\alpha = \alpha'$ and $\beta \neq \beta'$ then, by Lemma 5.3.2,

$$\begin{aligned}
\langle U, U' \rangle &= \sum_{a \in \mathbb{F}_p{}^m} \omega^{(\beta' - \beta) \cdot a} \\
&= 0. \quad \blacksquare
\end{aligned}$$

### 5.4.2 The General Construction

Our scheme for constructing a set of MUB is based on Theorem 5.3.1. The maximal commuting orthogonal basis for $\mathbb{M}_{p^m}(\mathbb{C})$ with partition of the form (5.4) is such that each class $\{1\!\!1_p\} \bigcup \mathcal{C}_j$, in the following sense, is a *linear* space of operators in the Pauli group $\mathbb{P}(p, m)$. Let

$$X_p(\alpha_1)\, Z_p(\beta_1), \ldots, X_p(\alpha_{p^m})\, Z_p(\beta_{p^m})$$

be the operators in the class $\{1\!\!1_p\} \bigcup \mathcal{C}_j$. We say that this class is linear if the set of the vectors

$$\mathcal{E}_j = \{\, (\alpha_1|\beta_1), \ldots, (\alpha_{p^m}|\beta_{p^m}) \,\}$$

form an $m$–dimensional subspace of $\mathbb{F}_p^{2m}$. In this case, to specify a linear class, it is enough to present a basis for the subspace $\mathcal{E}_j$. Such a basis can be represented by an $m \times (2m)$ matrix. So instead of listing all operators in the classes $\mathcal{C}_1, \ldots, \mathcal{C}_{p^m+1}$, we could simply list the $p^m + 1$ matrices representing the bases of these classes.

More specifically, the bases of linear classes of operators in our construction are represented by the matrices

$$(0_m|1\!\!1_m), \quad (1\!\!1_m|A_1), \quad \ldots, \quad (1\!\!1_m|A_{p^m}),$$

where $0_m$ is the all–zero matrix of order $m$ and each $A_j$ is an $m \times m$ matrix over $\mathbb{F}_p$. It easy to see what conditions should be imposed on the matrices $A_j$ so that the requirements of Theorem 5.3.1 satisfied. The following lemma gives a simple necessary and sufficient condition for operators in each class commuting. Note that in a linear class of operators, if the basic operators are commuting then any pair of operators in these class will commute.

**Lemma 5.4.2** *Let $S$ be a set of $m$ operators in $\mathbb{P}_0(p, m)$, and $S$ be represented by the matrix $(1\!\!1_m|A)$, where $1\!\!1_m$ is the identity matrix of order $m$ and $A$ is an $m \times m$* matrix over $\mathbb{F}_p$. Then the operators in $S$ are pairwise commuting if and only if $A$ is a symmetric matrix.

**Proof.** Let $A = (a_{jk})$. Then, by (5.10), $S$ is a set of commuting operators if and only if $a_{jk} - a_{kj} = 0 \mod p$, for every $1 \leq j < k \leq m$. Since $a_{jk} \in \mathbb{F}_p$, $S$ is a set of commuting operators if and only if $A$ is symmetric. ∎

The other condition is that the classes $\mathcal{C}_j$ and $\mathcal{C}_k$ should be disjoint. This condition is met if the span of the matrices $(\mathbb{1}_m|A_j)$ and $(\mathbb{1}_m|A_k)$ are disjoint. The last condition is equivalent to $xA_j \neq xA_k$, for every non–zero $x \in \mathbb{F}_p{}^m$. The last condition is equivalent to $\det(A_j - A_k) \neq 0$. Thus we can summarize our construction in the following theorem.

**Theorem 5.4.2** *Let $\{A_1, \ldots, A_\ell\}$ be a set of symmetric $m \times m$ matrices over $\mathbb{F}_p$ such that $\det(A_j - A_k) \neq 0$, for every $1 \leq j < k \leq \ell$. Then there is a set of $\ell + 1$ mutually unbiased bases on $\mathbb{C}^{p^m}$.*

More specifically, the $\ell + 1$ bases of the above theorem are represented by the matrices
$$(0_m|\mathbb{1}_m), \quad (\mathbb{1}_m|A_1), \quad \ldots, \quad (\mathbb{1}_m|A_\ell).$$

**Example** $d = 4$. The four matrices (over $\mathbb{F}_2 = \{0, 1\}$) which satisfy the conditions of Theorem 5.4.2 are
$$\begin{pmatrix} 0 & 0 \\ 0 & 0 \end{pmatrix}, \quad \begin{pmatrix} 1 & 0 \\ 0 & 1 \end{pmatrix}, \quad \begin{pmatrix} 0 & 1 \\ 1 & 1 \end{pmatrix}, \quad \begin{pmatrix} 1 & 1 \\ 1 & 0 \end{pmatrix}. \tag{5.13}$$

Therefore the classes of maximal commuting operators are

$$\begin{aligned}
\mathcal{C}_0 &= \{Z \otimes I, I \otimes Z, Z \otimes Z\}, \\
\mathcal{C}_1 &= \{X \otimes I, I \otimes X, X \otimes X\}, \\
\mathcal{C}_2 &= \{Y \otimes I, I \otimes Y, Y \otimes Y\}, \\
\mathcal{C}_3 &= \{X \otimes Z, Z \otimes Y, Y \otimes X\}, \\
\mathcal{C}_4 &= \{Y \otimes Z, Z \otimes X, X \otimes Y\},
\end{aligned}$$

where

$$I = \begin{pmatrix} 1 & 0 \\ 0 & 1 \end{pmatrix}, \quad X = \begin{pmatrix} 0 & 1 \\ 1 & 0 \end{pmatrix}, \quad Y = \begin{pmatrix} 0 & -1 \\ 1 & 0 \end{pmatrix} = XZ, \quad Z = \begin{pmatrix} 1 & 0 \\ 0 & -1 \end{pmatrix}.$$

We represent this basis explicitly. To this end, we naturally represent each basis by a $4 \times 4$ matrix such that the $j^{\text{th}}$ row of this matrix is the components of the $j^{\text{th}}$ vector of the corresponding basis with respect to the standard basis $|00\rangle, |01\rangle, |10\rangle, |11\rangle$: the first matrix is $\mathcal{B}_0 = 1\!\!1_4$, and

$$\mathcal{B}_1 = \frac{1}{2}\begin{pmatrix} 1 & 1 & 1 & 1 \\ 1 & -1 & -1 & 1 \\ 1 & 1 & -1 & -1 \\ 1 & -1 & 1 & -1 \end{pmatrix}, \quad \mathcal{B}_2 = \frac{1}{2}\begin{pmatrix} 1 & i & i & -1 \\ 1 & -i & -i & -1 \\ 1 & i & -i & 1 \\ 1 & -i & i & 1 \end{pmatrix},$$

$$\mathcal{B}_3 = \frac{1}{2}\begin{pmatrix} 1 & 1 & -i & i \\ 1 & -1 & i & i \\ 1 & 1 & i & -i \\ 1 & -1 & -i & -i \end{pmatrix}, \quad \mathcal{B}_4 = \frac{1}{2}\begin{pmatrix} 1 & -i & 1 & i \\ 1 & i & -1 & i \\ 1 & i & 1 & -i \\ 1 & -i & -1 & -i \end{pmatrix}.$$

Note that, in this case, the mutually unbiasedness condition is equivalent to the condition that $\mathcal{B}_i \mathcal{B}_i^\dagger = 1\!\!1_4$, for every $0 \leq i \leq 4$, and each entry of $\mathcal{B}_i \mathcal{B}_j^\dagger$, for $0 \leq i < j \leq 4$, has absolute value equal to $\frac{1}{2}$.

### 5.4.3 Construction for $d = p^m$

By Theorem 5.4.2, to construct $p^m + 1$ mutually unbiased bases in $\mathbb{C}^{p^m}$, we only need to find $m$ symmetric nonsingular matrices $B_1, \ldots, B_m \in \mathbb{M}_m(\mathbb{C})$ such that the matrix $\sum_{j=1}^m b_j B_j$ is also nonsingular, for every *nonzero* vector $(b_1, \ldots, b_m) \in \mathbb{F}_p{}^m$. Because if this condition satisfied then the $p^m$ matrices

$$\sum_{j=1}^m a_j B_j, \qquad (a_1, \ldots, a_m) \in \mathbb{F}_p{}^m,$$

satisfy the condition of Theorem 5.4.2.

**Example** $d = 8$. The following eight $3 \times 3$ matrices determine a set 9 mutually unbiased bases on $\mathbb{C}^8$. Let $A_1 = \mathbf{0}_3$ (the zero matrix), $A_2 = \mathbb{1}_3$, and

$$A_3 = \begin{pmatrix} 0 & 1 & 0 \\ 1 & 1 & 1 \\ 0 & 1 & 1 \end{pmatrix} \qquad A_4 = \begin{pmatrix} 0 & 0 & 1 \\ 0 & 1 & 1 \\ 1 & 1 & 0 \end{pmatrix} \qquad A_5 = \begin{pmatrix} 1 & 1 & 0 \\ 1 & 0 & 1 \\ 0 & 1 & 0 \end{pmatrix}$$

$$A_6 = \begin{pmatrix} 1 & 0 & 1 \\ 0 & 0 & 1 \\ 1 & 1 & 1 \end{pmatrix} \qquad A_7 = \begin{pmatrix} 0 & 1 & 1 \\ 1 & 0 & 0 \\ 1 & 0 & 1 \end{pmatrix} \qquad A_8 = \begin{pmatrix} 1 & 1 & 1 \\ 1 & 1 & 0 \\ 1 & 0 & 0 \end{pmatrix}$$

Note that these matrices are of the following general form:

$$a_1 \begin{pmatrix} 1 & 0 & 0 \\ 0 & 1 & 0 \\ 0 & 0 & 1 \end{pmatrix} + a_2 \begin{pmatrix} 0 & 1 & 0 \\ 1 & 1 & 1 \\ 0 & 1 & 1 \end{pmatrix} + a_3 \begin{pmatrix} 0 & 0 & 1 \\ 0 & 1 & 1 \\ 1 & 1 & 0 \end{pmatrix}, \qquad a_1, a_2, a_3 \in \mathbb{F}_2.$$

Wootters and Fields [WF89] have found the following general construction for the matrices $B_1, \ldots, B_m$. Let $\gamma_1, \ldots, \gamma_m$ be a basis of $\mathbb{F}_{p^m}$ as a vector space over $\mathbb{F}_p$. Then any element $\gamma_i \gamma_j \in \mathbb{F}_{p^m}$ can be written uniquely as

$$\gamma_i \gamma_j = \sum_{\ell=1}^m b_{ij}^\ell \gamma_\ell.$$

Then $B_\ell = \left(b_{ij}^\ell\right)$; i.e., the $(i,j)^{\text{th}}$ entry of $B_\ell$ is $b_{ij}^\ell$.

### 5.4.3.1 A set of MUB for the case $d = p^2$

We would like to mention here that for the case $d = p^2$, there is a more explicit construction. We find $p^2$ matrices $A_1, \ldots, A_{p^2}$ over $\mathbb{F}_p$ which satisfy the conditions of Theorem 5.4.2. For this purpose, we let

$$A_j = \begin{pmatrix} a_j & b_j \\ b_j & sa_j + tb_j \end{pmatrix}, \quad a_j, b_j \in \mathbb{F}_p,$$

where $s, t \in \mathbb{F}_p$ are two constants which their value need to be determined. By construction, the matrix $A_j$ is symmetric, so we have to choose the values of the parameters $s$ and $t$ such that $\det(A_j - A_k) \neq 0$, for every $1 \leq j < k \leq p^2$. Let $\alpha = a_j - a_k$ and $\beta = b_j - b_k$. Then $(\alpha, \beta) \neq (0, 0)$, and we have

$$\det(A_j - A_k) = D(\alpha, \beta) = \begin{vmatrix} \alpha & \beta \\ \beta & s\alpha + t\beta \end{vmatrix} = s\alpha^2 + t\alpha\beta - \beta^2.$$

If $\alpha = 0$, then $D(\alpha, \beta) = -\beta^2 \neq 0$. Suppose now that $\alpha \neq 0$, and let $\beta/\alpha = \gamma$. Then

$$D(\alpha, \beta) = -\alpha^2(\gamma^2 - t\gamma - s).$$

Thus $D(\alpha, \beta) \neq 0$ if the quadratic polynomial $\gamma^2 - t\gamma - s$ is irreducible over $\mathbb{F}_p$. Since for every prime $p$ there is at least one irreducible quadratic polynomial over $\mathbb{F}_p$, it is possible to choose the parameters $s, t \in \mathbb{F}_p$ such that $D(\alpha, \beta) \neq 0$, for every $\alpha, \beta \in \mathbb{F}_p$.

**Example** $d = 4$. The four matrices (5.13) are obtained from the irreducible polynomial $x^2 + x + 1$ over $\mathbb{F}_2$. Therefore, all those matrices are of the following form

$$\begin{pmatrix} a & b \\ a & a+b \end{pmatrix}, \quad a, b \in \mathbb{F}_2.$$

**Example** $d = 9$. The polynomial $x^2 + x + 2$ is irreducible over $\mathbb{F}_3$. Therefore, the matrices $A_j$ are of the general form of

$$\begin{pmatrix} a & b \\ b & a + 2b \end{pmatrix}.$$

So the nine matrices are

$$\begin{pmatrix} 0 & 0 \\ 0 & 0 \end{pmatrix}, \quad \begin{pmatrix} 1 & 0 \\ 0 & 1 \end{pmatrix}, \quad \begin{pmatrix} 2 & 0 \\ 0 & 2 \end{pmatrix},$$

$$\begin{pmatrix} 0 & 1 \\ 1 & 2 \end{pmatrix}, \quad \begin{pmatrix} 1 & 1 \\ 1 & 0 \end{pmatrix}, \quad \begin{pmatrix} 0 & 2 \\ 2 & 1 \end{pmatrix},$$

$$\begin{pmatrix} 1 & 2 \\ 2 & 2 \end{pmatrix}, \quad \begin{pmatrix} 2 & 1 \\ 1 & 1 \end{pmatrix}, \quad \begin{pmatrix} 2 & 2 \\ 2 & 0 \end{pmatrix}.$$

## 5.5 Conclusion

In this chapter we partially solved the problem of existence of sets of MUB in composite dimensions. We formulated an interesting connection between maximal commuting basis of orthogonal unitary matrices and sets of MUB. We obtained the necessary condition for the existence of sets of MUB in any dimension. Using these we proved the existence of sets of MUB for dimensions which are prime power. We provided a sharp upper bound on the size of any MUB for any dimension. We expressed the sets of MUB observables as tensor products of Pauli matrices. However we could not apply this method when the dimension $d$ is a product of different primes instead of being a prime power (the simplest case that belongs to this category is when $d = 6$) because if we do so the convenient properties of the case $d = p^m$ no longer remain valid. For instance Theorem 5.4.2 does not hold in this case.

A useful application of our result is in secure key distribution using higher dimensional quantum systems. Specifically we note that the protocol suggested by Bechmann–Pasquinucci and Tittel [BT00a] using four dimensional quantum system will become more efficient if all the five mutually unbiased bases are used in the protocol instead of only two as suggested by the authors.

# CHAPTER 6

# A Quantum Protocol for Anonymous Broadcasts

## 6.1 Introduction

Quantum information has introduced many exciting new tools such as the qubit, teleportation of quantum information, ultra-fast quantum computation, and quantum cryptography protocols. In most cases, these exciting developments are beyond what is possible in the classical information theoretic setting. In particular, using quantum cryptography, a secret key can be generated over a channel by two participants that is informationally secure. This is impossible in classical cryptography. In this paper, the basic tools of quantum cryptography are used to build a multi-participant protocol which gives the participants the ability to anonymously announce classical information. This protocol is shown to be secure against any and all attacks. By security, we mean the following: an eavesdropper learns nothing about who sent a particular message by eavesdropping on the channel. This is not to say that an eavesdropper has no idea who sent a particular message. Given the content of the message, the eavesdropper may be able to guess which of the players might have sent it; however, no protocol could ever change that.

There does exist a classical protocol for anonymous broadcasts[Cha88]. However, the existing classical protocols have two problems. First, they require perfectly secret channels, which is only possible classically with a one time pad[Ver26]. Second, if there are $N$ participants, there must be a factor of $N$ more communication than for a non-anonymous broadcast. The current protocol increases the cost of anonymous broadcasts by a constant factor over that of non-anonymous broadcasts. However, the factor is *independent* of the number of participants, unlike the protocol given in [Cha88], where the factor is the number of users, $N$.

This protocol has many obvious uses: for instance, it is known classically that a secure anonymous channel implies informationally secure key distribution [AS83]. Furthermore, the ability to make information known anonymously is important in a society that values the freedom of speech and information. For instance, whistle-blowing, criticizing the government, or tipping off the police are a few of the things that people may prefer to do anonymously. From the standpoint of cryptography, the anonymous channel is an interesting tool that may now be added to the tool box of the quantum cryptographer. This tool joins the ranks of quantum key distribution[BB84, Ben92], quantum oblivious transfer given bit commitment[Yao95, May96], and, more recently, quantum gambling[GVW99], all of which have been shown to be secure. It should be noted that quantum bit commitment and coin tossing have been shown to be impossible[May97, LC98]. Bit committment and coin tossing are important primitives, and are not possible using only the tools of quantum information. This raises the question: which protocols can be made secure with quantum information, and which cannot? In this regard, our contribution shows that a protocol for an anonymous channel can be constructed using only the tools of quantum information.

## 6.2  Protocol for a Quantum Anonymous Channel

The protocol is based on *quantum teleportation*[BBC93a]. There are $N$ users of the protocol. There are two types of channels we will speak of: first, the virtual anonymous channel that the protocol will create, and, second, the physical quantum channels that connect neighboring users. So, user $i$ ($U_i$) has a quantum channel to $U_{i-1}$ and $U_{i+1}$.

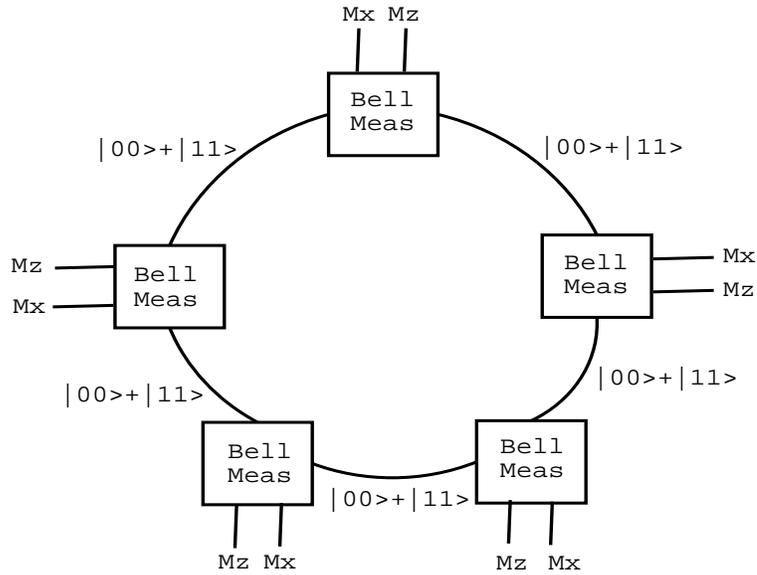

Figure 6.1: Anonymous channel with 5 users.

1. $U_i$ sends half of a Bell state $\frac{1}{\sqrt{2}}(|00\rangle + |11\rangle)$ to $U_{i+1}$. Now each user shares half a Bell state with each of his neighbors.

2. Each user tests these two bits with probability $p_{test}$. If the user tests, both bits should be measured in the $x$ basis. Otherwise, the user performs a Bell measurement on the two bits. The quantum part of the protocol is now over. The remaining steps involve classical processing of the measurements.

3. Each user makes an announcement: if $U_i$ tested, he announces the outcome of the measurement made on the bit received from $U_{i-1}$. If the user did not test, he announces the $z$ measurement on the Bell basis.

4. After all the above announcements have been made, each user announces whether or not he performed a test.

5. If no user tested, this bit is added to a string of bits $k'_{i_x}$ to be used for sending anonymous messages (*information bits*). If the bit is a test bit, all measurements that were not previously announced are now announced (the other $x$ basis measurement for testers, and the $x$ basis Bell measurement for those that did not test).

Note that it is crucial that step 3 happens before step 4. If step 4 occurs before 3 then a potential eavesdropper knows when to avoid attacking, and hence could never be detected. The point of step 3 is that the eavesdropper must commit to applying her attack *before* she knows if the bit will be used for testing.

### 6.2.1 Testing

In the case where there were some number of users that tested, the shared Bell states serve to create a teleportation channel between the testers. Hence, the testers verify that states sent in the $x$ basis have high fidelity along random sections of the network.

If each user tests with probability $p_{test}$, then the probability that someone tests is $1 - (1 - p_{test})^N$. The probability that no one tests, and hence that the channel is used, is given by $p_{use} = (1 - p_{test})^N$. If $p_{test} = \alpha/N$, note that $\lim_{N \to \infty} p_{use} = e^{-\alpha}$ and $p_{use} \leq e^{-\alpha}$. So, the probability that a bit is used for sending information is constant even as $N$ goes to infinity. The probability that a particular player tests a particular bit goes as $1/N$.

For each user, we can consider $2(N-1)$ channels: the channels for which he is the "starting node" for each of the other users (which provides $N-1$ channels), and the channel where he is the "ending node" for each of the other users (an additional $N-1$ channels). Since each user tests with equal probability, we can calculate the probability of testing each of these channels:

$$P_{test}(C_l^+) = (1 - \alpha/N)^{l-1}(\alpha/N) \tag{6.1}$$

For each channel, $1 \leq l \leq N - 1$, so

$$\alpha/N \geq P_{test}(C_l^+) \geq (1 - \alpha/N)^{N-1}(\alpha/N).$$

If we and look at the limit as $N \to \infty$, then:

$$\lim_{N \to \infty} P_{test}(C_l^+) \geq \lim_{N \to \infty} (1 - \frac{\alpha}{N})^{N-1} p_{test}$$
$$= e^{-\alpha} p_{test}$$

Hence, given that a user tested, he will test the channel between him and another user with probability $O(\frac{1}{N})$. Since each user tests with probability $\alpha/N$, if there are $O(N^2)$ test bits, then there are a constant number of tests of each channel. Note that this is different than the classical case[Cha88], which requires $N^2$ times as many bits to be sent. The reason for the $N^2$ factor in both cases is the fact that there are $N^2$ connections; however, in our case, the security exponent will be a function of the number of bits tested divided by $N^2$. This *does not* mean that we need to use $O(N^2)$ bits to send each message. It only means that the message should be long when compared to $N^2$ so that each user has many test bits for each message.

The fidelity of each of these channels in the $x$ basis is measured by the tests performed by the users. All of these fidelities should be high in order to ensure secrecy of the protocol. In section 6.4, we will see how high the fidelity needs to be.

### 6.2.2 Error Correction and Privacy Amplification

Given that the above test is passed, the remaining bits are used to send anonymous messages. The bits from the measurement done by the users in step 2 gives each user a string of bits ($k_i'$). If there were no errors on the channels used to share the EPR pairs, the parity of these measurements would be zero. Since there can be errors in the channel, the parity of all bits may not necessarily be zero. We denote the $z$ basis error between $U_i$ and $U_{i+1}$ as $e_i$. These errors would accumulate in the parity:

$$\vec{p} = \bigoplus_i \vec{k_i'} = \bigoplus_i \vec{e_i} = \vec{e} \tag{6.2}$$

These errors must be corrected. Additionally, Eve, due to her interaction, may have some knowledge about what each user measured. To combat this, a privacy amplification technique is employed: Eve will be required to learn the syndrome of $U_i$'s bits, $Hk_i'$.

To send information, two codes are used $C_1, C_2$, with

$$\{0\} \subset C_2 \subset C_1 \subset GF(2^n). \tag{6.3}$$

where $GF(2^n)$ is the binary vector space on $n$ bits. One code ($C_1$) corrects any errors on the information bits. $C_2$ can be thought of as a "privacy amplification" code. $C_1$ has some parity check matrix associated with it $H_1$, and all code words $w \in C_1$ have the relation: $H_1 \cdot w = 0$. $C_2$ has some parity check matrix $H_2$, and all code words $v \in C_2$ have the relation: $H_2 \cdot v = 0$. Since $C_2 \subset C_1$, all $v \in C_2$ also satisfy: $H_1 \cdot v = 0$, and, for $w \notin C_2$, $H_2 \cdot w \neq 0$.

The error correction is used in the following way: a code word has some errors on it $u \oplus e$. The message is the coset of $C_2$ in $C_1$ that also includes $u$. The vector $u \oplus e$ is error-corrected with $C_1$ to $u$, and the parity of $u$ is computed:

$$\begin{aligned} H_1 \cdot (u \oplus e) &= H_1 \cdot u \oplus H_1 \cdot e \\ &= 0 \oplus H_1 \cdot e \end{aligned}$$

If the error is small enough to be corrected, $e$ is now obtained. Consider $u = v \oplus s$

where $s \in C_1$ and $v \in C_2$. The message is:

$$\begin{aligned} H_2 \cdot u &= H_2 \cdot (v \oplus s) \\ &= H_2 \cdot v \oplus H_2 \cdot s \\ &= 0 \oplus H_2 \cdot s \\ &= m \end{aligned}$$

For the anonymous channel, each user $U_i$ announces $a_i = k'_i \oplus u_i$, with $H_2 \cdot u_i = m_i$, where $m_i$ is the message $U_i$ wants to announce. With this public information, anyone can compute:

$$\begin{aligned} H_1 \cdot (\bigoplus_i a_i) &= H_1 \cdot (\bigoplus_i (k'_i \oplus u_i)) \\ &= H_1 \cdot (\bigoplus_i e_i \oplus u_i) \\ &= H_1 \cdot \bigoplus_i e_i \oplus H_1 \cdot \bigoplus_i u_i \\ &= H_1 \cdot e \end{aligned}$$

Making use of the error correcting code, $e$ is computed. We may now compute the output of the channel:

$$\begin{aligned} H_2 \cdot (e \oplus \bigoplus_i a_i) &= H_2 \cdot (\bigoplus_i u_i) \\ &= \bigoplus_i (H_2 \cdot u_i) \\ &= \bigoplus_i m_i \end{aligned}$$

We reserve the message 0 for the $NULL$ message. We say a user $U_i$ uses a channel if $m_i \neq 0$. If each player uses the channel with probability $\gamma/N$, then the probability that one or less persons used the channel is close to $e^{-\gamma}$, providing a constant fraction of the total capacity. In the case of a collision (i.e. two users speaking at one time), each user will notice that their intended message was not broadcast successfully. They then simply resend the message with probability $\gamma/N$ until successful.

## 6.3 A Protocol Based on Quantum Error Correction Codes

Recently a method of proving security for quantum protocols that relies on quantum error correction codes was developed [SP00]. For a given protocol, one gives an alternate protocol based on quantum error correction codes that looks the same to Eve. What we mean is that all the classical announcements and quantum states would look exactly the same to Eve in this QECC based scheme, and therefore Eve cannot even know which protocol the participants are using. Since in the QECC case, the players could have corrected all the errors that Eve causes, the security result needs only to consider cases where the fidelity is exponentially close to unity, without having to worry about the announcements done for the error correction.

The basic idea is that the original protocol was equivalent to teleporting a qubit around a ring. Instead, we can consider just passing a qubit around a channel, with each user randomly applying a Pauli matrix to the qubit. In the end, the person that "sent" the qubit could measure it and see how the output differs from what they sent. Each user could then announce the exclusive OR of whether they applied $\sigma_x$ and his message bit. The user which "sent" the qubit would announce the parity of what he sent, what he measured, and his message bit. Testing would proceed in the same fashion. Each player would measure the bit with some probability in the $x$ basis and send a random bit in the $x$ basis. At the end of each round, they announce if each bit was a test bit.

In the following we denote $U_i$ to be user number $i$.

1. $U_0$ initiates a round by preparing $n$ bit state $|0_L\rangle$ in the CSS code given by $0 \subset C_2 \subset C_1$: $|0_L\rangle = \frac{1}{|C_2|^{1/2}} \sum_{w \in C_2} |w\rangle$. $U_0$ chooses $l_z$ tests bits with the value $|0_z\rangle$, and $l_x$ bits with the value $|0_x\rangle$. He concatenates these test bits to the end of the code and applies a random permutation, $\pi$. Next, he encrypts the entire state (see chapter 4) by choosing two random $n + l_x + l_z$ bit strings, $k_{0z}$ and $k_{0x}$, and applying $\sigma_z^{[k_{1z}]}$ and $\sigma_x^{[k_{1x}]}$. This produces the state: $\frac{1}{|C_2|^{1/2}} \sum_{w \in C_2} (-1)^{w \cdot k_{1z}} |w \oplus k_{1x}\rangle$.

2. $U_0$ sends the state to $U_1$. Note that, from Eve's point of view, since the state is encrypted, it is totally mixed.

3. $U_1$ measures each bit in the $x$ basis with probability $p_{test,x}$. $U_1$ replaces the measured bits $|0_x\rangle$.

4. $U_1$ chooses new random values $k_{1z}, k_{1x}$. He encrypts the state to obtain the new keys.

5. The above 3 steps are repeated for each user until the state gets back to $U_0$.

6. $U_0$ applies the inverse permutation $\pi^{-1}$ and measures the $l_x$ $x$-basis test bits and the $l_z$ $z$-basis test bits.

7. Each bit may or may not have been tested by player $i$. For each tested bit, the user announces the value he measured. For each untested bit, he announces the value of $k_{i_z}$ for that bit. The user does not disclose whether or not the bit was a test bit.

8. After all players complete step 7, they each announce which bits they tested. They also announce the values for $k_{i_z}$ for these bits. At this point, the fidelity in the $x$ basis between any two players can be determined. If any fidelity is too low, the protocol is aborted. If any user tested any one of the code bits or $z$-basis test bits (the first $n + l_z$ bits in $U_0$'s unpermuted ordering), the protocol is aborted.

9. $U_0$ discloses $\pi$. All users announce $k_{i_x}$ for the $l_z$ test bits in the $z$-basis. If the fidelity is too low, the protocol is aborted.

10. Defining $k'_{i_x}$ to be the values of $k_{i_x}$ which have not yet been announced, each user announces $a_i = k'_{i_x} \oplus m_i$, where $m_i$ is a codeword from $C_1$. $U_0$ XORs this $a_i$ with his state.

11. The state $U_0$ holds looks like:

$$\frac{1}{|C_2|^{1/2}} \sum_{w \in C_2} |w \oplus e \bigoplus_i k'_{i_x} \bigoplus_j a_j\rangle = \frac{1}{|C_2|^{1/2}} \sum_{w \in C_2} |w \oplus e \oplus \bigoplus_i m_{i_x}\rangle$$

where $e$ is the error. Correcting the error, $U_0$ obtains:

$$\frac{1}{|C_2|^{1/2}} \sum_{w \in C_2} |w \oplus \bigoplus_i s_{i_x}\rangle$$

$U_0$ computes the $H_2$ parity of this state and announces it; this is the output of the channel.

In the next section, we will show that the above protocol is secure. This protocol will only succeed a small number of times due to the probability that some user tests one of the code bits. To remedy this, $U_0$ could have made an alternate state preparation: he could have simply prepared $n + l_z + l_x$ EPR pairs and sent them into the channel. Only after step 8 would $U_0$ project some $n$ untested bits onto a quantum code word. From Eve's point of view, this is equivalent to the above case. Instead of each user applying a random set of Pauli matrices to encrypt the state in step 4, a user could prepare an EPR pair and measure the incoming bit in the Bell basis. This act of teleporting the state will have the effect of applying a random Pauli matrix to the state[BBC93a]. $U_0$ only makes use of the quantum code in one basis (namely the $z$-basis), so instead of projecting his state into a quantum code word, he could wait to receive the final state from $U_{N-1}$, measure that state with the one he holds, and perform the error correction and syndrome measurements on the collapsed states. At that point, the protocol is identical to one in the first part of the paper.

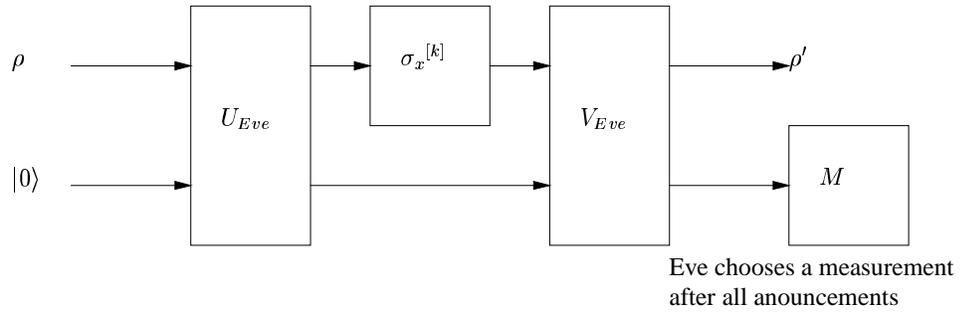

Figure 6.2: Eve attacks the state before and after a user applies $\sigma_x^{[k]}$.

## 6.4 Security of the QECC protocol

Since the announcement from the anonymous channel is public, Eve is concerned with *who* is responsible for a given announcement. Each user announces $k'_i \oplus s_i$, and Eve wants to know if $H_2 s_i$ is equal to the outcome of the channel.

Since the protocol is aborted if the $x$-basis error rate between any two parties exceeds what the code can correct, if we show that the protocol is secure for the case without error correction or privacy amplification, but with $x$-basis fidelity exponentially close to 1, then the security result will apply to the QECC protocol. We call the case with no error correction or privacy amplification the *simple case*.

In the simple case, each user $U_i$ receives a state and XORs the state with a random $k_i$. Subsequently, $U_i$ announces $k_i \oplus m_i$. Eve's job is learn if $m_i = 0$. The most general attack is depicted in figure 6.2. Eve does not, in general, know the state that is coming into $U_i$, and wishes only to learn the change that $U_i$ effects in the $z$-basis. We can assume that she applies some unitary operation $U$ before $U_i$ and $V$ after $U_i$ on her ancillary states. She waits until the end of the protocol, when all announcements have been, made in order to try to learn if what $U_i$ announced was $k_i$.

In the following section, we show that the bounds derived in chapter 2 can apply (with slight modifications) to the above two-sided attack to obtain:

$$I(E; F(M)|A) \leq (4 + 4\sqrt{2})H(F(K))\sqrt{1 - F_{min}} \tag{6.4}$$

where $F_{min}$ is the minimum fidelity in the $x$ basis of any channel that passes through or terminates with the user. For the simple case, we assume this to be exponentially close to unity, and hence we have security. In the error correction code protocol, the code provides a fidelity exponentially close to one; hence, we have security. Finally, since we showed that if the error-correction-based protocol was secure, then the original protocol is secure; therefore, the original protocol is secure. This proof follows the reduction technique introduced in [SP00].

## 6.5 Reducing Two-Sided Attacks to One-Sided Attacks

In chapter 2, we considered the case that Eve does a normal attack on qubit with an unknown value in the $z$-basis. In the anonymous channel protocol, Eve has the opportunity to attack the qubit before it arrives to the user, and again afterwards. Intuitively, if the fidelity in the $x$-basis is high for all Eve's attacks, then her attacks should commute with $\sigma_x$, which is the operator that the user may or may not apply. In this section, we formalize this result. It will very closely mirror theorem 2.4.2.

**Theorem 6.5.1** *If an eavesdropper performs the most general two-sided attack on a user who applies $\sigma_x^{[k]}$ to his incoming state, with $p_k = \frac{1}{2^n}$, and subsequently announces $a = m \oplus k$, then the most information the eavesdropper can learn about any* *function of the message is bounded by the minimum fidelity the attack would give in the x-basis:*

$$I(F(M); E|A) \leq (4 + 4\sqrt{2})H(F(K))\sqrt{1 - F_{min}}$$

**P**roof. The proof will follow closely theorem 2.4.2. The major difference will be that we will consider attacks of the form depicted in figure 6.2. The input into the attack will be a random state is the $z$-basis.

The input state that comes towards the user is a uniformly selected state in the $z$-basis. This variable will be denoted as $X$. $X$ is by definition independent of $K$, $M$, and $A$ (the key, message, and announcement, respectively). Using the above facts, we see that $H(F(M)|A, X) = H(F(M)|A)$, hence:

$$\begin{aligned} I(F(M); E|A, X) &= H(F(M)|A, X) - H(F(M)|E, A, X) \\ &= H(F(M)|A) - H(F(M)|E, A, X) \\ &= H(F(M)|A) - H(F(M)|E, A) \\ &\quad + H(F(M)|E, A) - H(F(M)|E, A, X) \\ &= I(F(M); E|A) + I(F(M); X|A, E) \\ &\geq I(F(M); E|A) \end{aligned}$$

So, if we want to bound $I(F(M); E|A)$, it is sufficient to bound $I(F(M); E|A, X)$.

The states that Eve has consistent with $f(m) = i$ given a particular announcement $a$ and input state $x$ are:

$$\sigma^{a,x}{}_i \equiv \frac{1}{q_i} \sum_{k: f(k \oplus a) = i} p_k \rho_{x,k}$$

where $q_i \equiv \sum_{k:f(k \oplus a)=i} p_k$, and $\rho_{x,k}$ is the state that Eve would have if the user applies $\sigma_x^{[k]}$ to an input state $|x\rangle\langle x|$. Since $p_k$ is independent of $k$, we see that $q_i$ is independent of $a$. The average of such states is:

$$\begin{aligned}\sigma^{a,x} &\equiv \sum_i q_i \sigma^{a,x}{}_i \\ &= \sum_i \sum_{k: f(k\oplus a)=i} p_k \rho_{x,k}\end{aligned}$$

Since $p_k$ is independent of $k$, and since each input to the function $f$ has one output, we obtain:

$$\sigma^{a,x} = \sum_k p_k \rho_{x,k}$$

From lemma 2.2.3 we have:

$$\begin{aligned}I(F(M); E|A, X) &= \sum_{a,x} p_a p_x I(F(M); E|a, x) \\ &\leq \sum_{a,x} p_a p_x \sum_i q_i \log \frac{1}{q_i} |\sigma^{a,x}{}_i - \sigma^{a,x}|\end{aligned}$$

Making use of the triangle inequality, we can obtain the following:

$$|\sigma^{a,x}{}_i - \sigma^{a,x}| \leq |\sigma^{a,x}{}_i{}' - \sigma^{a,x\prime}| + |\sigma^{a,x}{}_i - \sigma^{a,x}{}_i{}'| + |\sigma^{a,x} - \sigma^{a,x\prime}| \qquad (6.5)$$

for any choice of $\sigma^{a,x\prime}$ and $\sigma^{a,x\prime}_i$. Our approach will be to choose those states to be the ones that Eve would hold if, instead of the two-sided attack, the $\sigma_x{}^{[k]}$ had been applied to the input state $|x\rangle\langle x|$, and Eve had then applied her two-sided attack, $U$ then $V$.

We will use the sum notation for the first part of Eve's attack, $U$:

$$U(|0\rangle \otimes |i\rangle) = \sum_j |E_{i,j}\rangle |j\rangle$$

So, if we look at the user's and Eve's state together, after everything depicted in figure 6.2, we would have:

$$|\psi_{x,k}\rangle = V(\sum_j |E_{x,j\oplus k}\rangle_1 |j\rangle_2)$$

Thus, $\rho_{x,k} = Tr_2(|\psi_{x,k}\rangle\langle\psi_{x,k}|)$. The $primed$ states will be derived from the case where the user applies $\sigma_x{}^{[k]}$ $before$ Eve does any attack:

$$|\psi'_{x,k}\rangle \;=\; V(\sum_j |E_{x\oplus k,j}\rangle_1 |j\rangle_2)$$

Thus, $\rho'_{x,k} = Tr_2(|\psi'_{x,k}\rangle\langle\psi'_{x,k}|)$.

There are three terms in the right-hand side of equation 6.5. In theorem 2.4.2, we saw how to bound the first term. We now will handle the second and third terms from equation 6.5 in turn.

$$\sum_{a,x} p_a p_x \sum_i q_i \log \frac{1}{q_i} |\sigma^{a,x}{}_i - \sigma^{a,x}{}_i'|$$

$$= \sum_{a,x} p_a p_x \sum_i \log \frac{1}{q_i} |q_i \sigma^{a,x}{}_i - q_i \sigma^{a,x}{}_i'|$$

$$= \sum_{a,x} p_a p_x \sum_i \log \frac{1}{q_i} |\sum_{k:f(k\oplus a)=i} \frac{1}{2^n}(\rho_{x,k} - \rho'_{x,k})|$$

$$= \sum_{a,x} p_a p_x \sum_i \log \frac{1}{q_i} |\sum_{k':f(k')=i} \frac{1}{2^n}(\rho_{x,k'\oplus a} - \rho'_{x,k'\oplus a})|$$

$$\leq \sum_{a,x} p_a p_x \sum_i \log \frac{1}{q_i} \sum_{k':f(k')=i} \frac{1}{2^n} |\rho_{x,k'\oplus a} - \rho'_{x,k'\oplus a}|$$

$$\leq \sum_{a,x} p_a p_x \sum_i \log \frac{1}{q_i} \sum_{k':f(k')=i} \frac{1}{2^n} ||\psi_{x,k'\oplus a}\rangle\langle\psi_{x,k'\oplus a}| - |\psi'_{x,k'\oplus a}\rangle\langle\psi'_{x,k'\oplus a}||$$

$$= \sum_{a,x} p_a p_x \sum_i \log \frac{1}{q_i} \sum_{k':f(k')=i} \frac{1}{2^n} 2\sqrt{1 - |\langle\psi_{x,k'\oplus a}|\psi'_{x,k'\oplus a}\rangle|^2}$$

$$\leq \sum_{a,x} p_a p_x \sum_i \log \frac{1}{q_i} \sum_{k':f(k')=i} \frac{1}{2^n} 2\sqrt{2}\sqrt{1 - |\langle\psi_{x,k'\oplus a}|\psi'_{x,k'\oplus a}\rangle|}$$

$$\leq \sum_i \log \frac{1}{q_i} \sum_{k':f(k')=i} \frac{1}{2^n} 2\sqrt{2}\sqrt{1 - \sum_{a,x} p_a p_x |\langle\psi_{x,k'\oplus a}|\psi'_{x,k'\oplus a}\rangle|}$$

$$= \sum_i \log \frac{1}{q_i} (\sum_{k':f(k')=i} \frac{1}{2^n}) 2\sqrt{2}\sqrt{1 - \sum_{a,x} \frac{1}{2^{2n}} |\langle\psi_{x,a}|\psi'_{x,a}\rangle|}$$

$$= H(F(K)) 2\sqrt{2}\sqrt{1 - \sum_{a,x} \frac{1}{2^{2n}} |\langle\psi_{x,a}|\psi'_{x,a}\rangle|}$$

$$\leq H(F(K)) 2\sqrt{2}\sqrt{1 - \left|\sum_{a,x} \frac{1}{2^{2n}} \langle\psi_{x,a}|\psi'_{x,a}\rangle\right|}$$

All of the above is obtained merely by applying the triangle inequality and results from chapter 2. We still need to look at the term under the square root in the last line of the

above:

$$\sum_{a,x} \frac{1}{2^{2n}} \langle \psi_{x,a} | \psi'_{x,a} \rangle = \sum_{a,x} \frac{1}{2^{2n}} \sum_j \langle E_{x,j\oplus a} | E_{x\oplus a,j} \rangle$$
$$= F_U.$$

where $F_U$ is the probability that the attack would cause no error in the $x$-basis[1].

We must finally look at the third term in equation 6.5.

$$\sum_{a,x} p_a p_x \sum_i q_i \log \frac{1}{q_i} |\sigma^{a,x} - \sigma^{a,x'}|$$
$$= H(F(K)) \sum_{a,x} p_a p_x |\sigma^{a,x} - \sigma^{a,x'}|$$
$$= H(F(K)) \sum_{a,x} p_a p_x |\sum_k \frac{1}{2^n}(\rho_{x,k} - \rho'_{x,k})|$$
$$= H(F(K)) \sum_x p_x |\sum_k \frac{1}{2^n}(\rho_{x,k} - \rho'_{x,k})|$$
$$\leq H(F(K)) \sum_{x,k} p_x \frac{1}{2^n} |\rho_{x,k} - \rho'_{x,k}|$$
$$\leq H(F(K)) \sum_{x,k} p_x \frac{1}{2^n} ||\psi_{x,k}\rangle\langle\psi_{x,k}| - |\psi'_{x,k}\rangle\langle\psi'_{x,k}||$$
$$= H(F(K)) \sum_{x,k} \frac{1}{2^n} \frac{1}{2^n} 2\sqrt{1 - |\langle\psi_{x,k}|\psi'_{x,k}\rangle^2|}$$
$$\leq H(F(K)) \sum_{x,k} p_x \frac{1}{2^n} 2\sqrt{2}\sqrt{1 - |\langle\psi_{x,k}|\psi'_{x,k}\rangle|}$$
$$\leq H(F(K)) 2\sqrt{2} \sqrt{1 - \left|\sum_{x,k} \frac{1}{2^{2n}} \langle\psi_{x,k}|\psi'_{x,k}\rangle\right|}$$
$$= H(F(K)) 2\sqrt{2} \sqrt{1 - F_U}$$

Where we made use of the same connection to error rates as in the previous equations.

Putting these two bounds together with theorem 2.4.2, we have:

$$I(F(M); E|A) \leq H(F(M))(4\sqrt{1 - F_{UV}} + 4\sqrt{2}\sqrt{1 - F_U})$$

---
[1] refer back to chapter 2, theorem 2.4.1 to see this worked out in detail

Note that the second and third terms in equation 6.5 yeilded bounds that only depend on the fidelity of one of the sides, but the first term depends on the fidelity of the two sides considered as one attack. Of course, all the fidelities are greater than the minimum fidelity measured[2], so $F_{UV} \geq F_{min}$ and $F_U \geq F_{min}$, and hence we prove the theorem. ∎

## 6.6  Summary

We have described the first fundamentally multi-participant quantum cryptography protocol. In this multi-participant protocol, a new type of attack is possible. In section 6.5, we saw how to reduce these new attacks to the types we considered in chapter 2. Finally, this gives security as long as the error rate is low enough to allow a code with properties discussed at the end of chapter 3.

---

[2]In the protocol, both fidelities are estimated by test measurements

# CHAPTER 7

# An Experimental Realization of a Quantum Anonymous Channel

## 7.1 Introduction

In chapter 6, we introduced a new quantum protocol and also proved that this protocol is secure. Both the protocol as well as the proof of security are stated in a quantum information theoretic language. In this chapter, we will give a variant of the protocol in the language of quantum optics. We will see that this protocol is not beyond the reach of current quantum optical technology.

## 7.2 Tools and Terminology

The polarization state of a photon is, to many physicists, the most common example of a two level quantum system. We will discuss two polarization bases (see figure 7.1), $\{|H\rangle, |V\rangle\}$ and $\{|F\rangle, |S\rangle\}$. The second basis can be written in terms of the first:

$$|F\rangle = \frac{1}{\sqrt{2}}(|H\rangle + |V\rangle)$$
$$|S\rangle = \frac{1}{\sqrt{2}}(|H\rangle - |V\rangle)$$

We will now describe an anonymous channel protocol in terms of physical processes. We will forgo discussion of two aspects for now, namely production of sin-glet states: $\frac{1}{\sqrt{2}}(|H\rangle|V\rangle - |V\rangle|H\rangle)$, and performing polarization measurements in both bases, $M_{HV}, M_{FS}$. We will discuss these aspects in sections 7.6 and 7.7.

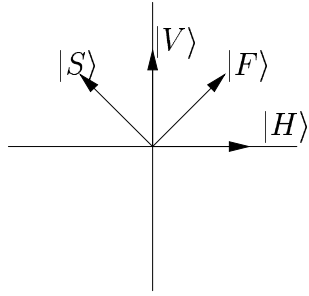

Figure 7.1: Two polarization bases

In addition to polarization measurements and singlet production, we also need an optical phase shifter ($P_\theta$):

$$P_\theta|H\rangle = |H\rangle \qquad P_\theta|V\rangle = e^{i\theta}|V\rangle$$

Note that $P_0$ is identity operation. We will make use of the shifter at either $\theta = 0$ or $\theta = \pi$. We can see how it looks with $\theta = \pi$:

$$P_\pi|H\rangle = |H\rangle \qquad P_\pi|V\rangle = -|V\rangle$$
$$P_\pi|F\rangle = |S\rangle \qquad P_\pi|S\rangle = |F\rangle$$

With these primitives, we can assemble a system which will allow us to perform the anonymous channel protocol.

## 7.3 Experimental Procedure

We will assume that there are $N$ participants in the protocol. As we saw in chapter 6, the quantum phase of the protocol is really quite simple. We will describe it slightly differently here.

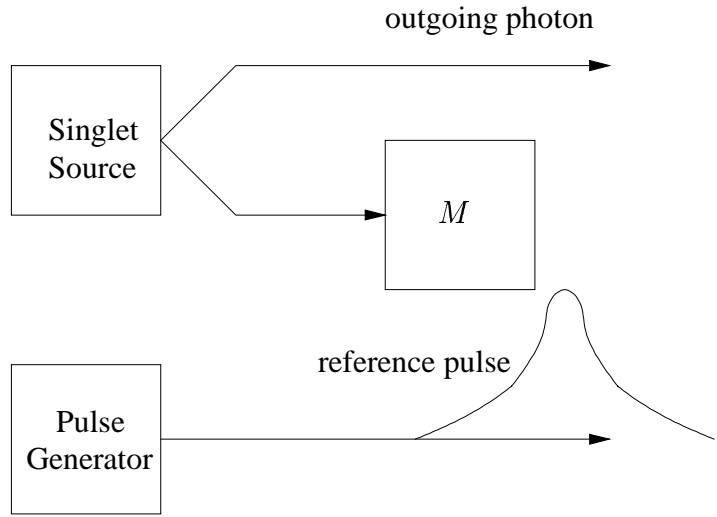

Figure 7.2: Starting a round of the experiment

Each participant, generally speaking, will make one of two actions: measurement of the photon to perform a test (see figure 7.3), or phase shifting the photon to generate a key (see figure 7.4). If no participant measures the photon in the $\{|H\rangle, |V\rangle\}$ basis, the photon will be used to generate a key. One participant will be special in that he will initiate and terminate each round. All the other participants will behave by the same rules (see figure 7.6).

### 7.3.1 The First Participant

The first participant will start and finish each round. He will produce a singlet and a reference pulse to trigger the other participants (see figure 7.2). He will measure the photon in the $\{|F\rangle, |S\rangle\}$ basis with probability $1 - \frac{1}{N}$. With probability $\frac{1}{N}$, he will measure in the $\{|H\rangle, |V\rangle\}$ basis. The optical fiber is arranged in a ring so it will eventually loop back to the first participant. When the reference pulse arrives back at his lab, he will measure in the same basis that he did when he emitted the reference pulse.

If he measures in the $\{|F\rangle, |S\rangle\}$ basis, and none of the other participants performs a test measurement, the result of the measurement is the bit he will use for this round of key generation. We now look at how the other participants behave.

### 7.3.2 Performing a Test Measurement

In addition to the polarized photon which will arrive via optical fiber, there is also a bright reference pulse. The bright reference pulse does not make use of any of the quantum principles; therefore, the signal may be amplified and measured by eavesdroppers. In order to reduce the effects of dark counts, the photo-detector is triggered by the reference pulse. Examining figure 7.3, we see that, in addition to measuring the incoming photon, a second photon is emitted by creating a singlet and measuring one of the photons in the singlet pair. Of course, since the next participant will use the reference pulse to time his measurement (should he make one), each participant must be careful to make sure that the his photon output is properly timed with the reference pulse.

### 7.3.3 Generating a Key

If no one performs a test measurement, the photon is used to form the shared key which will be used in the protocol. If the participant does not perform a test measurement, he applies one of two phase shifts with equal probability. If the participant applied the phase shift $P_0$, he counts the value zero as his bit for this round of the key. If he applied $P_\pi$, he counts the bit as one. One such device that can be electrically triggered to produce exactly such a phase shift is a Pockels Cell.

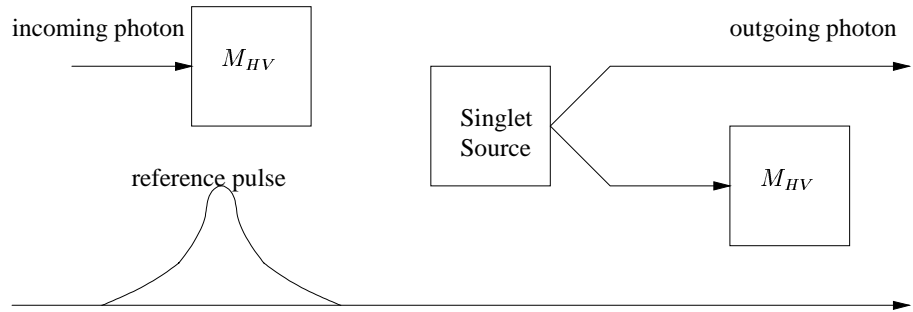

Figure 7.3: Test measurement procedure

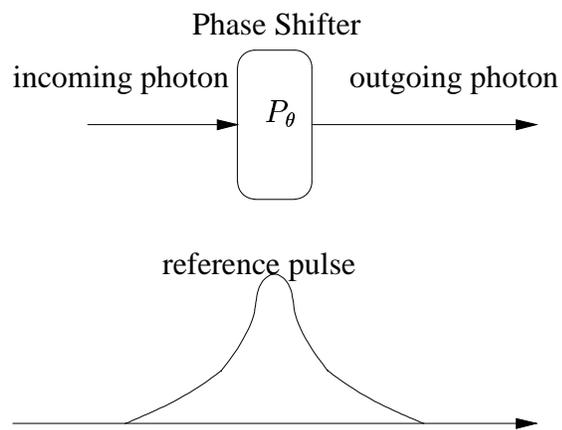

Figure 7.4: Phase shift ($P_0$,$P_\pi$) to generate key

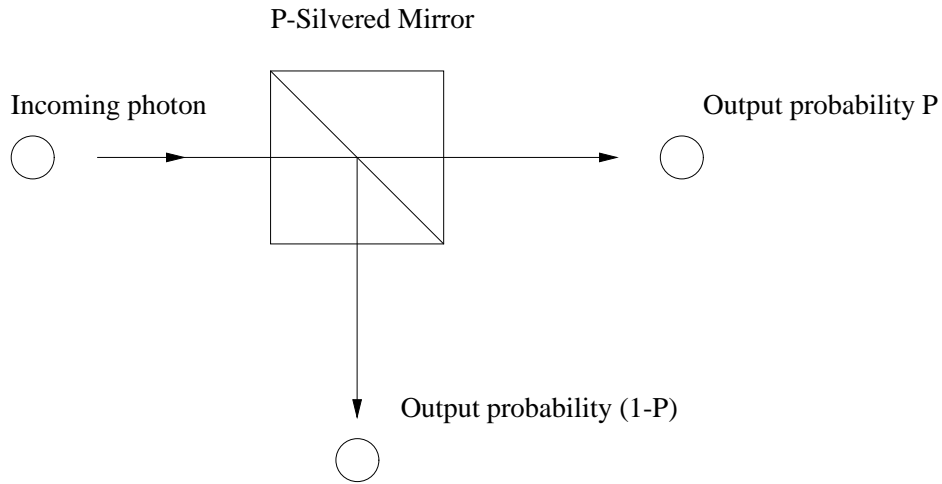

Figure 7.5: P-silvered mirror to randomly select test photons

## 7.4 Handling the Random Testing

As discussed in section 7.3.2, each photon needs to be tested with probability $1/N$. How can this be accomplished? One may use an electro-optical switch which can route photons between one of two positions. One position would reflect the beam toward the test measurement device; the other position would reflect the beam into the phase shifter (see section 7.3.3). A simpler method would be to use a partially silvered mirror for the incoming beam (see figure 7.5). In this case, the partially silvered mirror could be produced to reflect the photon with probability $P$, and to pass the photon with probability $1 - P$. If the photon is reflected into the test apparatus (see figure 7.3), the detectors would detect a photon. If no photon is detected, we can assume that it was not a test. This setup could resolve two difficulties. First, there is no need for a random number generator to select photons for testing. Second, there is no need for a mechanical routing of the photon into the testing or keying arms of the experiment. However, the routing difficulty is not completely resolved, since recombining the beam as the photon is leaving the experiment (see figure 7.6) would still require some kind

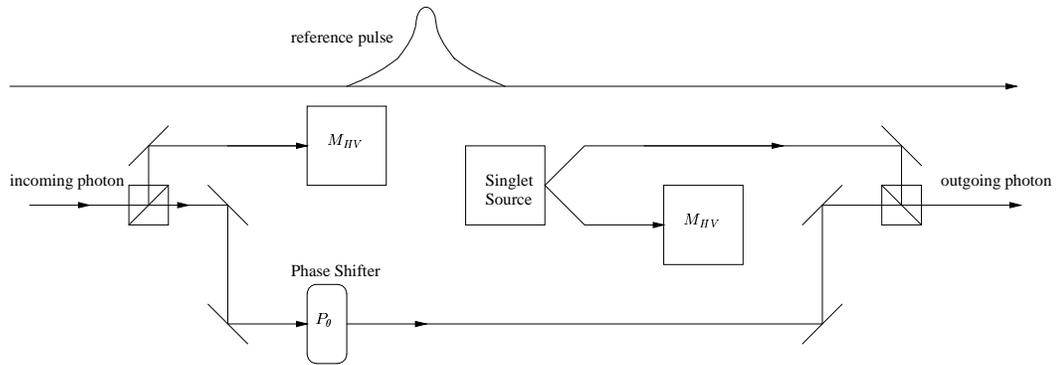

Figure 7.6: Full experimental setup for each participant

of electro-optical switch.

## 7.5 Processing the Results

After the experimental phase is complete, the participants are left with data. This data is discussed according to the protocol to allow the participants to send anonymous messages. First, each participant announces whether or not he tested. After all the participants have made this announcement, each participant that performed a test announces the value of the the first measurement (see figure 7.3). After all the above announcements are made, each participant that tested announces the result of their second measurement. From these announcements, everyone can compute the error rate between any two players. If the error rate between any two players is more than $7.5\%$[1], the protocol is aborted. Otherwise, they check the error rate of the non-test results. They then take a random sampling of the non-test information (specifically, which phase shifters each participant used and what measurements the first participant made). Using this information, the error rate of the key material can be estimated. This needs to be less than $7.5\%$ as well. If this is the case, the participants are ready to make

---
[1] the details of the security result are given in chapter 6

anonymous announcements.

As we saw in chapter 6, we use two linear codes, $C_1$ and $C_2$, such that $\{0\} \subset C_2 \subset C_1 \subset GF(2^n)$. These codes have parity check matrices $H_2, H_1$. The code $C_2$ needs to able to correct all the errors in the key material. The anonymous message is encoded in $C_1$ by selecting a random element $a$ of $C_1$ such that $H_2 \cdot a = m$, where $m$ is the anonymous message. Now each participant announces $a \oplus k$ where $k$ was the string of his key material. By decoding the parity of all these announcements, one will learn the parity of the messages. If only one person uses the channel, the result will be his message. If a participant uses the channel, but the outcome does not decode to his message, he waits a random time and tries again.

## 7.6 Producing Singlets Via Parametric Down Conversion

The singlet state $\frac{1}{\sqrt{2}}(|H\rangle|V\rangle - |V\rangle|H\rangle)$ is often referred to as an EPR pair, named after Einstein, Podolsky, and Rosen [EPR35]. It was the seemingly odd properties of this state that Einstein called "spooky action at a distance" and caused him to doubt quantum mechanics. In fact, the strange properties of such states, which we call entangled, are fundamental to many quantum information processing tasks such as quantum key distribution[Eke91], quantum state teleportation[BBC93a] and super-dense coding[BW93]. Therefore, the production of such states is of great importance to experimental efforts in quantum information processing.

The basic experimental procedure for parametric down conversion is given in figure 7.7. A laser sends a beam of photons into a non-linear crystal. Two beams come out of the crystal. By tuning various parameters (such as the wavelength and intensity the of laser light, the cut of the crystal, etc...), the quantum entanglement of the two emitted beams can be controlled. Energy and momentum are certainly conserved in this process, and so the wavelength of the emitted beam is larger. Recently, these sorts have experiments have been carried out by many groups[KWW99, ASS01, KCK01].

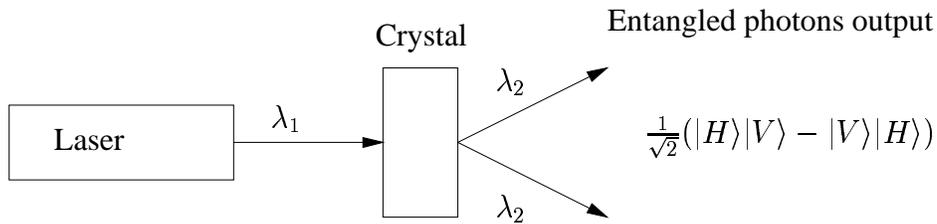

Figure 7.7: Parametric down conversion process

PDC is not, as of yet, a "push-button" process which can produce singlets precisely on demand. In the current protocol, we are measuring the second photon and using PDC only to produce photons with random, but known, polarization. A system has been recently demonstrated which produces photons "pseudo-on-demand"[PJF], meaning that they can be made to appear in known polarizations at prearranged times. This sort of system would be suitable for implementation of the current protocol.

It should be noted that, if the input beam does not contain exactly one photon, the output will have more than one photon in each beam. In order to fit the protocol, we need exactly one photon in each beam. This deviation from the ideal protocol has been partially analyzed in the case of BB84[BLM00, L99], and some results have been obtained by assuming that Eve may only do single-photon interactions. Full security results for such practical schemes is an open problem.

## 7.7 Performing Polarization Measurements

In the experimental procedure we have described in this chapter, we make use of measurements of photon polarization. It may not be immediately clear how this is to be performed. It is, in fact, quite simple.

The photon to be measured is sent through a polarization beam splitter. This device reflects photons of one polarization, and lets photons of the orthogonal polarization through. If photo-detectors are positioned in the paths that the photon might take in each case, then when a detector measures a photon, the polarization is known by which detector was activated (see figure 7.8). In the figure, we give an example of measuring the photon in the $\{|H\rangle, |V\rangle\}$ basis; however, by changing the configuration of the polarization beam splitter, this method can be adapted to any basis.

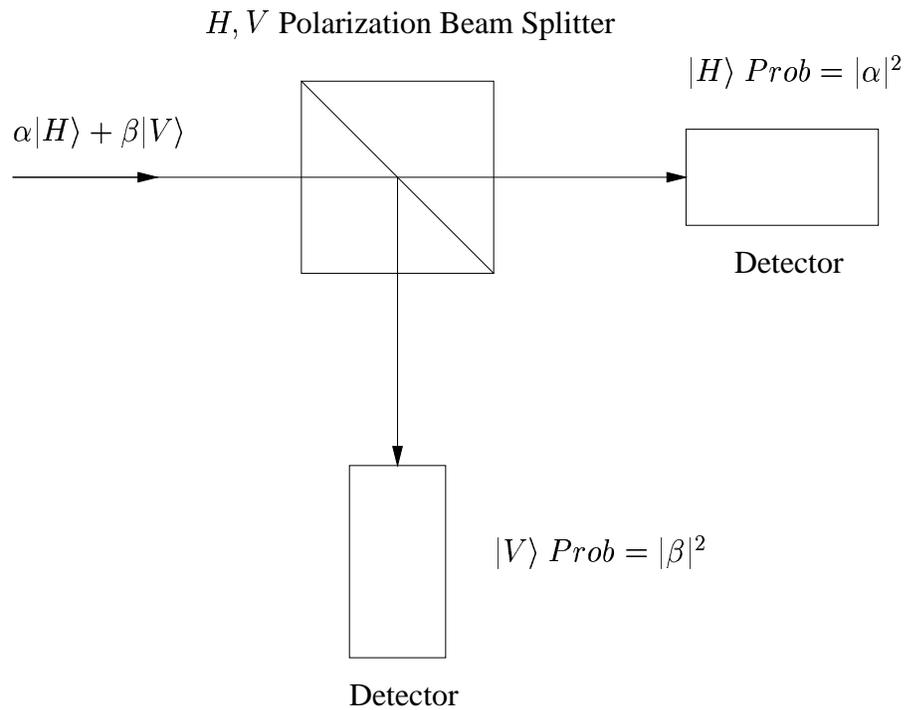

Figure 7.8: Measurement in the H,V basis: $M_{HV}$

## 7.8 Summary

We have given a prescription for how the protocol described in chapter 6 might be realized with current experimental techniques. While there have been many experimental implementations of quantum key distribution, this protocol is the first quantum security protocol to use more than two participants. As such, demonstration of this protocol will present interesting experimental challenges. The rate at which the protocol could proceed is limited by the rate of the processes described in sections 7.3.3 and 7.3.2.

# REFERENCES

[AS83]  Bowen Alpern and Fred B. Schneider. "Key exchange using 'keyless cryptography'." *Information Processing Letters*, **16**(2), 1983.

[ASS01]  Mete Atatüre, Alexander V. Sergienko, Bahaa E.A. Saleh, and Malvin C. Teich. "Entanglement in Cascaded-Crystal Parametric Down-Conversion." *Phys. Rev. Lett.*, **86**(18):4013–4016, 2001.

[BB84]  Charles H. Bennett and Gilles Brassard. "Quantum cryptography: Public key distribution and coin tossing." In *Proc. of IEEE Int. Conf. on Computers, Systems and Signal Processing*, pp. 175–179, Bangalore, India, December 1984.

[BBB92]  Charles H. Bennett, Francois Bessette, Gilles Brassard, Louis Salvail, and John Smolin. "Experimental quantum cryptography." *Jour. of Cryptology*, **5**:3–28, 1992.

[BBB98]  Eli Biham, Michel Boyer, Gilles Brassard, Jeroen van de Graaf, and Tal Mor. "Security of Quantum Key Distribution Against All Collective Attacks." QPH/9801022, 1998.

[BBB99]  Eli Biham, Michel Boyer, Patrick Oscar Boykin, Tal Mor, and Vwani Roychowdhury. "A Proof of Security of Quantum Key Distribution Against Any Attack." in preperation, 1999.

[BBC93a]  Charles H. Bennett, Gilles Brassard, Claude Crépeau, Richard Jozsa, Asher Peres, and William K. Wooters. "Teleporting an Unknown Quantum State via Dual Classical and Einstein-Podolsky-Rosen Channels." *Phys. Rev. Lett.*, **70**(13):1895–1899, 1993.

[BBC93b]  Charles H. Bennett, Gilles Brassard, Claude Crépeau, and Denis Langlois. "A quantum bit commitment scheme provably unbreakable by both parties." In *Proc. of 34th Ann. Symp. on Found. of Comp. Sc.*, pp. 362–371, Palo Alto, Ca., 1993.

[Ben92]  Charles H. Bennett. "Quantum Cryptography Using Any Two Nonorthogonal States." *Phys. Rev. Lett.*, **68**(21):3121–3124, 1992.

[BHM96]  Eli Biham, Bruno Huttner, and Tal Mor. "Quantum cryptographic network based on quantum memories." *Phys. Rev. A*, **54**(4):2651–2658, 1996.

[BLM00]  Gilles Brassard, Norbert Lütkenhaus, Tal Mor, and Barry C. Sanders. "Security aspects of practical quantum cryptography." *Phys. Rev. Lett.*, **85**(6):1330–1333, 2000.

[BM97a]   Eli Biham and Tal Mor. "Bounds on information and the security of quantum cryptography." *Phys. Rev. Lett.*, **79**(20):4034–4037, 1997.

[BM97b]   Eli Biham and Tal Mor. "Security of quantum cryptography against collective attacks." *Phys. Rev. Lett.*, **78**:2256–2259, 1997.

[BMS]   Gilles Brassard, Tal Mor, and B. C. Sanders. "Quantum cryptography via parametric downconversion." quant-ph/9906074. To appear in *Proceedings of the Quantum Communication, Computing, and Measurement 2 (QCM'98)* conference, Evanston, Ill., USA, Aug. 1998.

[BMS96]   Charles H. Bennett, Tal Mor, and John A. Smolin. "Parity bit in quantum cryptography." *Phys. Rev. A*, **54**(3):2675–2684, 1996.

[BR00]   P. O. Boykin and V. Roychowdhury. "Optimal encryption of quantum bits." QPH/0003059, 2000.

[BT00a]   H. Bechmann-Pasquinucci and W. Tittel. "Quantum cryptography using larger alphabets." *Phys. Rev. A*, **61**(6):062308/1–6, 2000.

[BT00b]   H. Bechmann-Pasquinucci and W. Tittel. "Quantum cryptography with 3–state systems." *Phys. Rev. Lett.*, **85**(15):3313–3316, 2000.

[BW93]   Charles H. Bennett and Stephen J. Wiesner. "Communication via One- and Two-Particle Operators on Einstein-Podolsky-Rosen States." *Phys. Rev. Lett.*, **69**(20):2881–2884, 1993.

[CCK97]   A. R. Calderbank, P. J. Cameron, W. M. Kantor, and J. J. Seidel. "Kerdock codes, orthogonal spreads, and extremal Euclidean line–sets." *Proc. London Math. Soc.*, **3**:436–480, 1997.

[CDM99]   Claude Crépeau, Paul Dumais, and Julien Marcil, December 1999. Talk given in NEC workshop on quantum cryptography.

[CGL99]   Richard Cleve, Daniel Gottesman, and Hoi-Kwong Lo. "How to share a quantum secret." *Phys. Rev. Lett.*, **83**:648–651, 1999. quant-ph/9901025.

[Cha88]   D. Chaum. "The Dining Cryptographers Problem: Unconditional Sender and Receiver Untraceability." *Journal of Cryptology*, **1**(1):65–75, 1988.

[CHR99]   A. R. Calderbank, R. H. Hardin, E. M. Rains, P. W. Shor, and N. J. A. Sloane. "A group–theoretic framework for the construction of packings in Grassmannian spaces." *J. Algebraic Combinatorics*, **9**:129–140, 1999.

[CRS97] A. R. Calderbank, E. M. Rains, P. W. Shor, and N. J. A. Sloane. "Quantum error correction and orthogonal geometry." *Phys. Rev. Lett.*, **78**(3):405–408, 1997.

[CRS98] A. R. Calderbank, E. M. Rains, P. W. Shor, and N. J. A. Sloane. "Quantum error correction via codes over GF(4)." *IEEE Transactions on Information Theory*, **44**(4):1369–1387, 1998.

[CT91] Thomas M. Cover and Joy A. Thomas. *Elements of information theory*. John Wiley and Sons, New York, 1991.

[Eke91] Arthur Ekert. "Quantum cryptography based on Bell's theorem." *Phys. Rev. Lett.*, **67**(6):661–663, 1991.

[EPR35] A. Einstein, B. Podolsky, and N. Rosen. "Can Quantum-Mechanical Description of Physical Reality Be Considered Complete?" *Physical Review*, **47**(10):777, 1935.

[FG99] C. A. Fuchs and J. van de Graaf. "Cryptographic distinguishability measures for quantum-mechanical states." *IEEE Transactions on Information Theory*, **45**(4):1216–1227, May 1999. quant-ph/9712042.

[FGG97] Christopher A. Fuchs, Nicolas Gisin, Robert B. Griffiths, Chi-Sheng Niu, and Asher Peres. "Optimal eavesdropping in quantum cryptography. I. Information bound and optimal strategy." *Phys. Rev. A*, **56**(2):1163–1172, 1997.

[Gal63] Robert C. Gallagher. *Low-density parity-check codes*. The M.I.T. Press, Cambridge, Mass., 1963. Chapter 2.

[GKP01] D. Gottesman, A. Kitaev, and J. Preskill. "Encoding a qubit in an oscillator." *Phys. Rev. A*, **64**(1):012310/1–21, 2001.

[GL89] G. Golub and C. Van Loan. *Matrix computations*. Johns Hopkins Univ. Press, Baltimore, 1989.

[Got99] D. Gottesman. "Fault–tolerant quantum computation with higher–dimensional systems." *Chaos, Solitons and Fractals*, **10**(10):1749–1758, 1999.

[GVW99] Lior Goldenberg, Lev Vaidman, and Stephen Wiesner. "Quantum Gambling." *Phys. Rev. Lett.*, **82**(16):3356–3359, 1999.

[Hoe63] Wassily Hoeffding. "Probability inequalities for sums of bounded random variables." *J. Amer. Stat. Assoc.*, **58**:13–20, 1963.

[Hol73]    A. S. Holevo. "Bounds for the quantity of information transmitted by a quantum channel." *Probl. Inf. Transm.*, **9**(3):177–183, 1973.

[Iva81]    I. D. Ivanovic. "Geometrical description of quantum state determination." *Journal of Physics A*, **14**(12):3241–3245, 1981.

[KCK01]    Yoon-Ho Kim, Maria V. Chekhova, Sergei P. Kulik, Morton H. Rubin, and Yanhua Shih. "Interferometric Bell-state preparation using femtosecond-pulse-pumped spontaneous parametric down-conversion." *Phys. Rev. A*, **63**:062301/1–11, 2001.

[KWW99]    Paul G. Kwait, Edo Waks, Andrew G. White, Ian Appelbaum, and Philippe H. Eberhard. "Ultra-bright source of polarization-entangled photons." *Phys. Rev. A*, **60**(2):773–776, 1999.

[L99]    Norbert Lütkenhaus. "Estimates for practical quantum cryptography." *Phys. Rev. A*, **59**(5):3301–3319, 1999.

[Lan61]    R. Landauer. "Irreversibility and heat generation in the computing process." *IBM Journal of Research and Development*, **5**:183–191, 1961.

[LC98]    H. Lo and H. Chau. "Why quantum bit commitment and ideal quantum coin tossing are impossible." *Physica D*, **120**:177–187, 1998.

[LC99]    Hoi-Kwong Lo and H. F. Chau. "Unconditional security of quantum key distribution over arbitrarily long distances." *Science*, **283**:2050–2056, 1999.

[Lo]    Hoi-Kwong Lo. "A simple proof of the unconditional security of quantum key distribution." quant-ph/9904091.

[May]    Dominic Mayers. "Unconditional security in quantum cryptography." quant-ph/9802025.

[May96]    Dominic Mayers. "Quantum key distribution and string oblivious transfer in noisy channel." In *Advances in cryptology - CRYPTO'96*, LNCS 1109, pp. 343–357. Springer-Verlag, 1996.

[May97]    Dominic Mayers. "Unconditionally Secure Quantum Bit Commitment is Impossible." *Phys. Rev. Lett.*, **78**(17):3414–3417, 1997.

[Per93]    Asher Peres. *Quantum Theory: Concepts and Methods*. Kluwer Academik Publishers, Dordrecht, 1993.

[PJF]    T. B. Pittman, B. C. Jacobs, and J. D. Franson. "Single Photons on Pseudo-Demand from Stored Parametric Down-Conversion." quant-ph/0205103.

[Pre]     John Preskill. "Quantum Computing Lecture Notes." http://www.theory.caltech.edu/people/preskill/ph229/.

[Sch95]   B. Schumacher. "Quantum Coding." *Phys. Rev. A*, **51**(4):2738–2747, 1995.

[Sch96]   Bruce Schneier. *Applied Cryptography Second Edition*. John Wiley & Sons, Inc., New York, 1996.

[Sha49]   C. E. Shannon. "Communication theory of secrecy systems." *Bell Syst. Tech. J.*, **28**:656–715, 1949.

[Sha79]   A. Shamir. "How to share a secret." *Communications of the ACM*, **22**:612–613, 1979.

[Sho97]   Peter W. Shor. "Polynomial-time algorithms for prime factorization and discrete logarithms on a quantum computer." *SIAM J. Comput.*, **26**:1484–1509, 1997.

[SP00]    Peter W. Shor and John Preskill. "Simple Proof of Security of the BB84 Quantum Key Distribution Protocol." *Phys. Rev. Lett.*, **85**(2):441–444, 2000.

[Ver26]   G. S. Vernam. "Cipher printing telegraph systems for secret wire and radio telegraphic communications." *J. Amer. Inst. Elect. Eng.*, **55**:109–115, 1926.

[WF89]    W. k. Wootters and B. D. Fields. "Optimal state–determination by mutually unbiased measurements." *Annals of Physics*, **191**(2):363–381, 1989.

[Wie83]   S.J. Wiesner. "Conjugate coding." *Sigact News*, **15**:77–88, 1983.

[Yao95]   Andrew Yao. "Security of quantum protocols against coherent measurements." In *Proc. of the 26th ACM Symp. on the Theory of Computing*, pp. 67–75, June 1995.